\documentclass[fleqn,usenatbib]{mnras}

\usepackage{newtxtext,newtxmath}

\usepackage[T1]{fontenc}
\usepackage{multirow}
\usepackage{rotating}
\usepackage{lscape}

\DeclareRobustCommand{\VAN}[3]{#2}
\let\VANthebibliography\thebibliography
\def\thebibliography{\DeclareRobustCommand{\VAN}[3]{##3}\VANthebibliography}

\usepackage{graphicx}	
\usepackage{amsmath}	

\title[Radio monitoring of transient BeXRBs]{Radio monitoring of transient Be/X-ray binaries and the inflow-outflow coupling of strongly-magnetized accreting neutron stars}

\author[Van den Eijnden et al.]{J. van den Eijnden,$^{1}$ N. Degenaar,$^{2}$ T. D. Russell,$^{2,3}$ J. C. A. Miller-Jones,$^{4}$ A. Rouco Escorial,$^{5}$
\newauthor R. Wijnands,$^{2}$ G. R. Sivakoff,$^{6}$ J. V. Hern\'andez Santisteban,$^{7}$
\\
$^{1}$Department of Physics, Astrophysics, University of Oxford, Denys Wilkinson Building, Keble Road, Oxford OX1 3RH, UK\\
$^{2}$Anton Pannekoek Institute for Astronomy, University of Amsterdam, Science Park 904, 1098 XH, Amsterdam, the Netherlands\\
$^{3}$INAF, Istituto di Astrofisica Spaziale e Fisica Cosmica, Via U. La Malfa 153, I-90146 Palermo, Italy\\
$^{4}$International Centre for Radio Astronomy Research, Curtin University, GPO Box U1987, Perth, WA 6845, Australia\\
$^{5}$Center for Interdisciplinary Exploration and Research in Astrophysics (CIERA) \\
and Department of Physics and Astronomy, Northwestern University, Evanston, IL 60208, USA\\
$^{6}$Department of Physics, CCIS 4-181, University of Alberta, Edmonton, AB, T6G 2E1, Canada\\
$^{7}$SUPA School of Physics and Astronomy, University of St Andrews, North Haugh, St Andrews KY16 9SS, Scotland, UK\\
}

\date{Accepted XXX. Received YYY; in original form ZZZ}

\pubyear{2021}

\begin{document}
\label{firstpage}
\pagerange{\pageref{firstpage}--\pageref{lastpage}}
\maketitle

\begin{abstract}
Strongly-magnetized ($B\geq10^{12}$ G) accreting neutron stars (NSs) are prime targets for studying the launching of jets by objects with a solid surface; while classical jet-launching models predict that such NSs cannot launch jets, recent observations and models argue otherwise. Transient Be/X-ray binaries (BeXRBs) are critical laboratories for probing this poorly-explored parameter space for jet formation. Here, we present the coordinated monitoring campaigns of three BeXRBs across four outbursts: giant outbursts of SAX 2103.5+4545, 1A 0535+262, and GRO J1008-57, as well as a Type-I outburst of the latter. We obtain radio detections of 1A 0535+262 during ten out of twenty observations, while the other targets remained undetected at typical limits of $20$--$50$ $\mu$Jy. The radio luminosity of 1A 0535+262 positively correlates with its evolving X-ray luminosity, and inhabits a region of the $L_X$--$L_R$ plane continuing the correlation observed previously for the BeXRB Swift J0243.6+6124. We measure a BeXRB $L_X$--$L_R$ coupling index of $\beta = 0.86 \pm 0.06$ ($L_R \propto L_X^\beta$), similar to the indices measured in NS and black hole low-mass X-ray binaries. Strikingly, the coupling's $L_R$ normalisation is $\sim 275$ and $\sim 6.2\times10^3$ times lower than in those two comparison samples, respectively. We conclude that jet emission likely dominates during the main peak of giant outbursts, but is only detectable for close-by or super-Eddington systems at current radio sensitivities. We discuss these results in the broader context of X-ray binary radio studies, concluding that our results suggest how supergiant X-ray binaries may host a currently unidentified additional radio emission mechanism.

\end{abstract}

\begin{keywords}
accretion, accretion discs -- stars: neutron -- X-rays: binaries -- radio continuum: transients
\end{keywords}

\section{Introduction}

What mechanism underlies the formation of jets remains an important but poorly-understood question across a wide range of astrophysical objects. Jets are observed in combination with accretion processes onto sub-Solar mass objects up to supermassive black holes, as well as during the end-phases of the lives of massive stars, mergers of compact objects, and formation of stars. Accretion-driven, stellar-mass jet sources, such as neutron stars (NSs) and black holes accreting in binary systems, can be studied to observe the jet launching process and its evolution over time-scales from weeks to months. In such systems, X-ray and radio observations probe distinct, but coupled components of the system; the inflow of matter in the accretion flow by the former, and the collimated jet outflow by the latter.

The coupling between inflows and outflows in accreting NS and black hole systems has predominantly been established for systems with low-mass donor stars. Among such systems, called low-mass X-ray binaries (LMXBs), black holes were first discovered to show a correlation between their X-ray and radio luminosity \citep{hannikainen98,corbel00,corbel03,merloni03,gallo03}, often spanning multiple orders of magnitude during the quiescent and hard accretion states of outbursts \citep{fender04}. Monitoring of a large sample of black holes has revealed a radio-bright (shallower) and radio-quiet (steeper) track of this correlation, merging as sources decay into quiescence. However, debate remains regarding the origin of these different correlations for black hole systems \citep{gallo2014,soleri11,dincer14,meyer14,drappeau15,espinasse18}.

The behavior of NS LMXB in the X-ray--radio (hereafter $L_X$--$L_R$) diagram is more complicated and varied. In individual NS systems, a correlation between the X-ray and radio luminosity has been observed \citep{migliari03,migliari06,tudor17,gusinskaia17,russell18}; however, such couplings across a range of X-ray luminosities have been seen in only a small number of sources, and, surprisingly, different couplings have been observed between different outbursts of the same target \citep[e.g.][]{gusinskaia20}. As a sample, as a sample, the NS LMXBs are significantly radio-fainter than the sample of black hole systems, complicating the study of their $L_X$--$L_R$ behaviour \citep{fender01,vandeneijnden2021}. The source class, as a sample, shows an overall correlation in the $L_X$--$L_R$ diagram whose coupling index is similar to the black hole sample \citep{gallo18}; however, individual sources have been observed to follow steeper indices \citep{migliari06,gusinskaia17,gusinskaia20}. Therefore, while the in- and outflow in individual NS LMXBs are thought to be coupled, it remains unclear whether a single correlation can describe the entire source class.

Similar monitoring in the X-ray and radio band has, to date, rarely been performed for high-mass X-ray binaries (HMXBs): binary systems wherein the compact object accretes from a massive, early-type O/B donor star, with a mass typically exceeding $10$ $M_{\odot}$. Based on the donor star type and the mode of accretion, HMXBs are often divided into three broad categories. Be/X-ray binaries (BeXRBs), Supergiant X-ray Binaries (SgXBs), and Superfast X-ray Transients (SFXTs). BeXRBs are systems with a Be-type donor star \citep{porter03}, characteristically showing optical emission lines and an IR excess due to the presence of a decretion disk around the star \citep{reig11}. The great majority of BeXRBs are transient systems, with typically long and eccentric orbits, . Accretion outbursts can occur close to periastron passages, as the compact object moves through the stellar decretion disk, in what is called a Type-I outburst \citep{okazaki01}. Alternatively, giant outbursts (also known as Type-II outbursts) can occur at any orbital phase, may last longer than the orbital period, and reach higher X-ray luminosities than Type-I outbursts. The trigger of giant outbursts currently remains debated, but may be related to the properties of the Be-star disk, increased Be-star activity, and instabilities driven by the interaction between the NS and decretion disk \citep{okazaki01,moritani13,martin14,monageng17,laplace17}.

The second and third general types of HMXBs both host supergiant O/B stars as the secondary, but differ in their X-ray properties. The SgXBs persistently emit at X-ray energies, typically between $10^{35}$ erg/s and several times $10^{37}$ erg/s, although they can vary in luminosity and accretion state \citep{sidoli2018}. SFXTs, on the other hand, are typically X-ray faint, with luminosities between $L_X \sim 10^{32}$--$10^{34}$ erg/s \citep{sidoli17}. However, as their name suggests, SFXTs show occasional flares lasting a number of kilo-seconds, reaching up to $\sim 10^{37}$ erg/s. The origin of the difference in X-ray properties between SgXBs and SFXTs remains debated, with possible explanations involving the quasi-spherical settling accretion regime \citep{shakura2012} or the influence of the NS magnetosphere \citep{bozzo2008}.

Alternatively, HMXBs can be divided based on their compact object. A small minority of known HMXBs host a black hole, with only few confirmed Galactic systems such as Cyg X-1 and MWC 656 (and, e.g. LMC X-1 and LMC X-3 in the Large Magellanic Cloud). The former persistently accretes from the  stellar wind of their donor star, showing strong variability in X-rays. MWC 656, on the other hand, is the only confirmed black hole BeXRB \citep{casares2014}, and has just one recorded X-ray outburst \citep{williams2010}. Both black hole systems have been detected at radio frequencies and monitored across limited ranges of X-ray luminosities \citep{ribo17}. The remainder of the HMXB class, instead, hosts a NS as its primary with a strong, $B>10^{12}$ G magnetic field and a slow, typically $P>1$ second, rotation period. However, despite their larger numbers, monitoring campaigns including multiple detections have only been obtained for two systems\footnote{Here, we ignore the famous systems Cir X-1 and SS 433. While both have been studied extensively in radio and X-rays, neither is confirmed to be a NS HMXB. For the former, the donor star nature remains debated \citep{johnston16}, while for the latter, the primary is likely to be a black hole \citep[see e.g.][]{fabrika04}. In addition, even if Cir X-1 is a HMXB, its inner accretion flow displays many characteristics of LMXBs instead.}.

Firstly, the persistent but X-ray variable NS HMXB GX 301-2 was monitored at radio frequencies at different orbital phases by \citet{pestalozzi09}. The system is (marginally) detected in most observations, regardless of orbital phase, with variable radio flux density and spectrum. The authors postulate that the emission is dominated by stellar emission, with a possible intermittent contribution from a short-lived and weak jet. Secondly, the BeXRB Swift J0243.6+6124 was monitored extensively in radio during its super-Eddington discovery outburst in 2017/2018. Comparing the coupled X-ray and radio properties during the main peak of the giant outburst, \citet{vandeneijnden2018_swj0243} attributed the radio emission to a relativistic jet. Surprisingly, during a later X-ray re-brightening at significantly lower X-ray luminosity, the radio emission re-brightened to similar levels as the main outburst peak \citep{vandeneijnden2019_reb}. Additionally, \citet{vandeneijnden2021} presented a sample of radio observations of active SgXBs and BeXRBs in quiescence, including several detections of the former class\footnote{As well as symbiotic X-ray binaries: strongly-magnetized NSs accreting from the stellar wind of a low-mass, evolved donor in a wide orbit.}. However, those samples did not include monitoring campaigns, but instead consisted of single observations.

In this work, we present coordinated radio and X-ray monitoring campaigns of three BeXRBs across four outbursts. Our main goal is to better understand the possibility and properties of jet launching in these systems, in order to constrain jet launching in general. This is critical as currently, no jet model can fully account for the launch and properties of jets by NSs with magnetic fields exceeding $\sim 10^{9}$--$10^{10}$ G \citep{massi08,parfrey16,vandeneijnden2021}. With just a single transient strongly-magnetized accreting NS detected and monitored at radio frequencies, few constraints on such models currently exist. Observing more BeXRBs, across their different outburst types, we aim to obtain more stringent constraints on jet properties across a wider range of source properties and X-ray luminosity. 

\subsection{Targets: GRO J1008-57, SAX J2103.5+4545, 1A 0535+262}
\label{sec:targets}
For this study, we obtained radio and X-ray monitoring of the 2019 Type-I outburst and the 2020 giant outburst of GRO J1008-57, the 2020 giant outburst of SAX J2103.5+4545, and the 2020--2021 giant outburst of 1A 0535+262.

GRO J1008-57 is a regularly-outbursting BeXRB, hosting a NS with a spin period of $\sim 93.7$ seconds in an orbit of $\sim 249.5$ days with an eccentricity of $e\sim0.68$ around the donor Be star \citep{kuhnel2013}. GRO J1008-57 shows a cyclotron resonance scattering feature (CRSF) at 78 keV \citep{shrader1999,yamamoto2014}, confirming directly that the NS is strongly magnetized --- the line energy implies a field strength around $B\sim(6$--$7)\times10^{12}$ G \citep[see e.g.][for a review]{staubert19}. GRO J1008-57 shows a Type-I outburst at almost every periastron passage, as well as less frequent giant outbursts \citep{kuhnel2013}. The remarkable consistency of its outbursts at each periastron passage allowed us to target its Type-I outburst in June 2019 with a pre-planned observing campaign. Just less than one year later, \textit{NICER} \citep{reynolds2020} and \textit{MAXI} \citep{nakajima2020}, on 2020 May 21 and 22 respectively, reported the onset of a giant outburst, which we subsequently monitored at radio frequencies as well. As the giant outbursts of GRO J1008-57 are typically preceded by an enhanced flux state, this outburst was expected and we were able to trigger early during the outburst rise.

SAX J2103.5+4545 is a BeXRB with a NS, spinning at a period of approximately $346$ seconds \citep{hulleman1998}\footnote{See also the \textit{Fermi}/GBM pulse frequency monitoring at \url{https://gammaray.msfc.nasa.gov/gbm/science/pulsars.html}.}). \citet{brumback2018} presented the possible detection of a cyclotron line at $12$ keV, consistent with a $\sim 10^{12}$ G magnetic field. While this detection has not been repeated independently \citep[see][]{staubert19}, we assume such a magnetic field in this work given the HMXB type and long spin period of the NS. SAX J2103.5+4545 is not a typical BeXRB. Compared to other systems of similarly long spin, it has a relatively short orbit of 12.7 days \citep{baykal2007,camero2007,corbet86}. Secondly, its super-orbital X-ray behaviour is atypical. As summarized by \citet{reig2014}, it displays low and high X-ray flux states lasting months \citep{reig2010}, typically around $\sim 1.5\times10^{35}$ erg/s in the former and one order of magnitude higher in the latter state. During the high flux state, which starts with a bright and short flare, SAX J2103.5+4545 shows outbursts, similar to other BeXRBs. In August 2020, \citet{grishina2020} reported the optical brightening of this BeXRB. We subsequently triggered radio and X-ray monitoring, which we halted after a single radio non-detection, as the outburst peak had already passed, making later radio detections unlikely. While not strictly a monitoring campaign, we include this radio observation in our study.

Finally, 1A 0535+262 is a BeXRB showing both Type-I and giant outbursts. Due to its proximity, the Type-I outbursts can typically reach fluxes close to one Crab, while the giant outbursts easily reach several Crab, even when the X-ray binary reaches only $\sim 10\%$ of the Eddington luminosity. Time scales of years separate the giant outbursts, with previous ones occurring in 1980, 1983, 1989, 1994, 2005, 2009, and 2011. The NS has a spin period of $\sim 103$ seconds, while the orbit has an eccentricity of $\sim 0.47$ \citep{finger1994} and a period of $\sim 111$ days \citep{motch1991}. Its CRSF energy suggests a magnetic field strength of $\sim 5\times10^{12}$ keV. In November 2020, \citet{mandal2020} reported the onset of a new giant outburst, which reached the highest peak X-ray flux observed from this source ($\sim 11$ Crab). During previous outbursts, \citet{tudose10} and \citet{migliari11} obtained radio observations of 1A 0535+262 but did not detect a radio counterpart, with 4.9-GHz flux density upper limits of $210$ $\mu$Jy and $160$ $\mu$Jy, respectively. At the start of the 2020-2021 outburst, we reported the detection of the radio counterpart at $39\pm4$ $\mu$Jy \citep{vandeneijnden2020_a0535atel}. In this paper, we report on the remainder of that observing campaign.

\section{Observations and data analysis}

In this section, we describe the radio and X-ray observations, data reduction, and analysis. Given the number of observations, targets, and observatories, we discuss these topics in relatively general terms here. The actual fitted parameters and fluxes/flux densities, as well as observation details such as ObsID, dates, and observing times, are tabulated in the Online Supplementary Materials.

\label{sec:obs}

\subsection{Radio observational campaign setup and analysis}

The radio observations of the three targets were taken with the Karl G. Jansky Very Large Array (hereafter VLA; for SAX J2103.5+4545 and 1A 0535+262) and the Australia Telescope Compact Array (ATCA; for GRO J1008-57). SAX J2103.5+4545 was observed once, 1A 0535+262 was observed over twenty epochs, and the giant and Type-I outbursts of GRO J1008-57 were targeted with five and six radio observations, respectively. All VLA observations were taken in 3-bit mode at C band, with a central frequency of $6.0$ GHz and a bandwidth of $4.0$ GHz. These observations span a range of different configurations, including several non-standard configurations in transitions (i.e. BnA$\rightarrow$A and A$\rightarrow$D)\footnote{\url{https://science.nrao.edu/facilities/vla/proposing/configpropdeadlines}}. The ATCA data were obtained simultaneously at two central frequencies of 5.5 GHz and 9.0 GHz, both with $2.0$ GHz of bandwidth. Again, multiple array configurations were used\footnote{\url{https://www.narrabri.atnf.csiro.au/operations/array\_configurations/configurations.html}}: 6A for the Type-I outburst monitoring of GRO J1008-57; and both 1.5C and H214 for its giant outburst.

For the observation of SAX J2103.5+4545, the primary and nearby secondary calibrator were 3C 286 and J2102+4702. For both campaigns of GRO J1008-57, 0939-608 was the secondary calibrator. In the Type-I outburst campaign, only 0823-500 was used as the primary calibrator; in the giant outburst observations, either 0823-500 or 1934-638 was used as primary calibrator, depending on their visibility. Similary, for the 1A 0535+262 observations, we used either 3C 286 or 3C 48 as the primary calibrator, depending on the time of the observation. For this final campaign, we used J0547+2721 as the secondary calibrator.

For all four observing campaigns, we used standard practices in \textsc{common astronomy software application} \citep[\textsc{casa};][]{mcmullin07} v5.4.1 to flag, calibrate, and image the radio data. We used a combination of manual inspection and automated routines to flag RFI and other data quality issues, before performing standard calibration steps. We then used the multi-scale, multi-frequency \textsc{tclean} task to image the field. For this final step, we use robust parameters of $1.0$ for 1A 0535+262, $1.0$ for SAX J2102.5+4545, and $-0.5$/$0.0$ (5.5/9 GHz) for GRO J1008-57, chosen per target and frequency to optimize the balance between sensitivity and imaging artefacts. If the target was detected, we used the \textsc{imfit} task to measure its flux density by fitting a 2D elliptical Gaussian profile with FWHMs and position angle equal to the synthesized beam's minor and major axis and angle. Using \textsc{imfit}, we also measured the source position. We measured the RMS sensitivity of the observation over a nearby region devoid of point sources. We then also re-imaged individual sub-bands of $1$-GHz width to determine the radio spectral index. If no radio emission was detected from the source, we determined the RMS sensitivity over the source position and tripled this value to obtain the $3-\sigma$ flux density upper limit.

\subsection{X-ray data reduction}

In order to assess the X-ray properties and measure the X-ray flux at or close to the time of the radio observations, we used publicly available observations from four instruments: the X-ray Telescope \citep[XRT;][]{burrows04} and Burst Alert Telescope \citep[BAT;][]{barthelmy05} aboard the \textit{Neil Gehrels Swift Observatory} \citep[hereafter \textit{Swift};][]{gehrels04}, as well as the \textit{Monitor of All-sky X-ray Image}/Gas-Slit Camera \citep[\textit{MAXI}/GSC;][]{matsuoka09} and \textit{Neutron Star Interior Composition ExploreR}/X-ray Timing Instrument \citep[\textit{NICER}/XTI;][]{gendreau2016} mounted on the International Space Station. The \textit{Swift}/BAT data were obtained via the Hard X-ray Transient Monitor webpage\footnote{\url{https://swift.gsfc.nasa.gov/results/transients/}}, which hosts light curves of the average daily and orbital X-ray flux in the $15$--$50$ keV band for known X-ray sources. We did not perform additional analysis of the \textit{Swift}/BAT data but instead employed it as long-term reference light curves of the three sources and as evenly-spaced monitoring observations for the targeted outbursts.

For the other three observatories, we extract and model the X-ray spectra to measure the X-ray flux. For the \textit{NICER} observations, only available for 1A 0535+262, we downloaded the observations from the \textsc{heasarc}\footnote{\url{https://heasarc.gsfc.nasa.gov}} and re-ran the level 2 data reduction tool \textsc{nicerl2} v1.6 to apply the latest version of the calibration, accessed via the online \textsc{caldb}. We then extracted the source spectrum using \textsc{xselect}, selecting all counts in the energy range $0.5$-$10$ keV across the entire field-of-view, since \textit{NICER} is a non-imaging instrument. Due to the very high count rates of the target \cite[i.e. often greatly exceeding $10^3$ ct/s, compared to a typical background rate $< 1$ ct/s;][]{remillard2021}, we did not generate a background spectrum. Finally, when fitting the spectra, we used the pre-calculated instrument response files \textsc{nixtiref20170601v002.rmf} and \textsc{nixtiaveonaxis20170601v004.arf} for each observation\footnote{As discussed in the Online Supplementary Materials, we tested whether using the recently developed \textsc{nicerarf}, \textsc{nicerrmf}, and \textsc{nicer\_bkg\_estimator} tools to determine observation-specific response and background files, resulted in significant differences in the fits. As it did not -- and, importantly, since they also left soft instrumental features remaining -- we instead used the pre-calculated response files in our analysis.}.

For the \textit{Swift}/XRT observations, we used the online data reduction pipeline \citep{evans09}\footnote{\url{https://www.swift.ac.uk/user\_objects/}} to extract X-ray source and background spectra, as well as instrument response files. This pipeline automatically corrects for pile-up at high count rates, which is particularly relevant for our analysis of 1A 0535+262. Finally, for \textit{MAXI}, we used the On-demand Process tool (\url{http://maxi.riken.jp/mxondem/}) to extract spectrum and response files. Given the lower sensitivity and spatial resolution of \textit{MAXI} compared to \textit{Swift} and \textit{NICER}, we only used \textit{MAXI} spectra for the giant outburst of GRO J1008-57, where no pointed observations by the latter two observatories are available.

\subsection{X-ray spectral fitting}

After extracting the X-ray spectra, we used \textsc{xspec} v12.10.1 to model the emission, using the \textsc{tbabs} model with abundances from \citet{wilms00} and cross-sections from \citet{verner96} to account for interstellar absorption. All errors and confidence intervals quoted in this paper, appendices, and supplementary material, are calculated at the $1-\sigma$ level.

\subsubsection{SAX J2103.5+4545 and GRO J1008-57}

For SAX J2103.5+4545, only two pointed X-ray observations are available close in time to the single VLA radio observation: \textit{Swift}/XRT observations were taken $5.2$ and $6.8$ days after the radio observation. Due to the low number of total counts in both Photon Counting-mode (PC) observations, we used C-statistics \citep{cash1979} in an energy range of $1$-$10$ keV to attempted fits with two simple, phenomenological models: an absorbed black body (\textsc{tbabs*bbody}) and an absorbed power law (\textsc{tbabs*powerlaw}). For both spectra, the latter model fitted significantly better and returned a non-zero absorption column, contrary to the black body model and consistent with earlier X-ray analyses of this target \citep[$N_H = (2.9-4.4)\times10^{22}$ cm$^{-2}$;][]{brumback2018}.

During the 2019 Type-I outburst of GRO J1008-57, \textit{Swift}/XRT observed the target in a cadence coordinated with the ATCA radio observations. Of these six observations, the first two were taken only in PC mode due to the low source flux at the start of the outburst. For the third and sixth observations, both PC and Window Timing (WT) mode data are available with a sufficient exposure time, while for observation four and five, only WT-mode data were taken. Similar to SAX J2103.5+4545, we use C-statistics in the $1$-$10$ keV range and attempt fits with the same two phenomenological models (an absorbed black body or power law). We find that, in all six spectra, the absorbed power law model provides a statistically better fit. Additionally, making use of the larger number of total counts compared to SAX J2103.5+4545, we also attempt to fit a combined model, \textsc{tbabs*(bbody+power law)}. However, in none of the six spectra does the addition of a second spectral component significantly improve the fit: in all cases, $\Delta C < 3$ with two additional parameters, which does not correspond to a significant improvement\footnote{For a single extra free parameter, a $3-\sigma$ improvement would require $\Delta C \geq 9$ \citep{cash1979}; for two extra free parameters, this change should be even larger.}.

During the 2020 giant outburst of GRO J1008-57, no pointed X-ray observations were taken. Therefore, we instead analysed the \textit{MAXI}/GSC spectra to measure the flux during or close to the five radio observations. For each MJD with available \textit{MAXI} data during the radio-monitored part of the outburst (i.e. MJD 58983--59025), we extracted a spectrum by combining all source counts from that MJD. During the final six days of the above period, the X-ray flux had dropped significantly; therefore, at those times, we instead generated two spectra by combining three days of observations (i.e. combining MJD 59020--59022 and MJD 59023--59025). We then fitted these spectra using $\chi^2$ statistics, as the \textit{MAXI} pipeline automatically bins the spectra to a sufficient number of counts for this approach. To constrain the absorption column despite the low number of counts and energy band of $2$--$10$ keV, we fit all seventeen spectra jointly, tying $N_H$ between the spectra. We find that the spectra are better described by an absorbed power law than a black body model ($\chi^2_\nu = 1.04$ vs $1.16$ for $466$ free parameters in both cases).

\subsubsection{1A 0535+262}
Finally, the 2020/2021 giant outburst of 1A 0535+262 was monitored in extensive detail by both \textit{NICER} and \textit{Swift}. We analyse observations up to MJD 59300, since the final radio observation was taken on MJD 59279. Given the extremely bright nature of the outburst (brighter than any previous outburst monitored by \textit{Swift}/BAT; cf. Figure \ref{fig:LC_A0535}), simple phenomenological models do not yield statistically acceptable fits. Instead, we follow the approach applied by \citet{jaisawal2019} to fit joint \textit{NICER} and \textit{NuSTAR} spectra of the bright giant outburst of Swift J0243.6+6124. As we solely intend to accurately measure the flux, and do not study the spectral evolution in this work, we employ their Model I, a phenomenological model defined as \textsc{tbabs*(bbodyrad + cutoffpl + gauss + gauss + gauss)}. The three Gaussian components in this model correspond to three narrow iron lines in the range $6.4$--$7$ keV\footnote{Note that we leave out the iron edge included in the original model by \citet{jaisawal2019} as we do not find significant evidence for its necessary inclusion in our spectral model.}.

\begin{figure}
    \centering
    \includegraphics[width=\columnwidth]{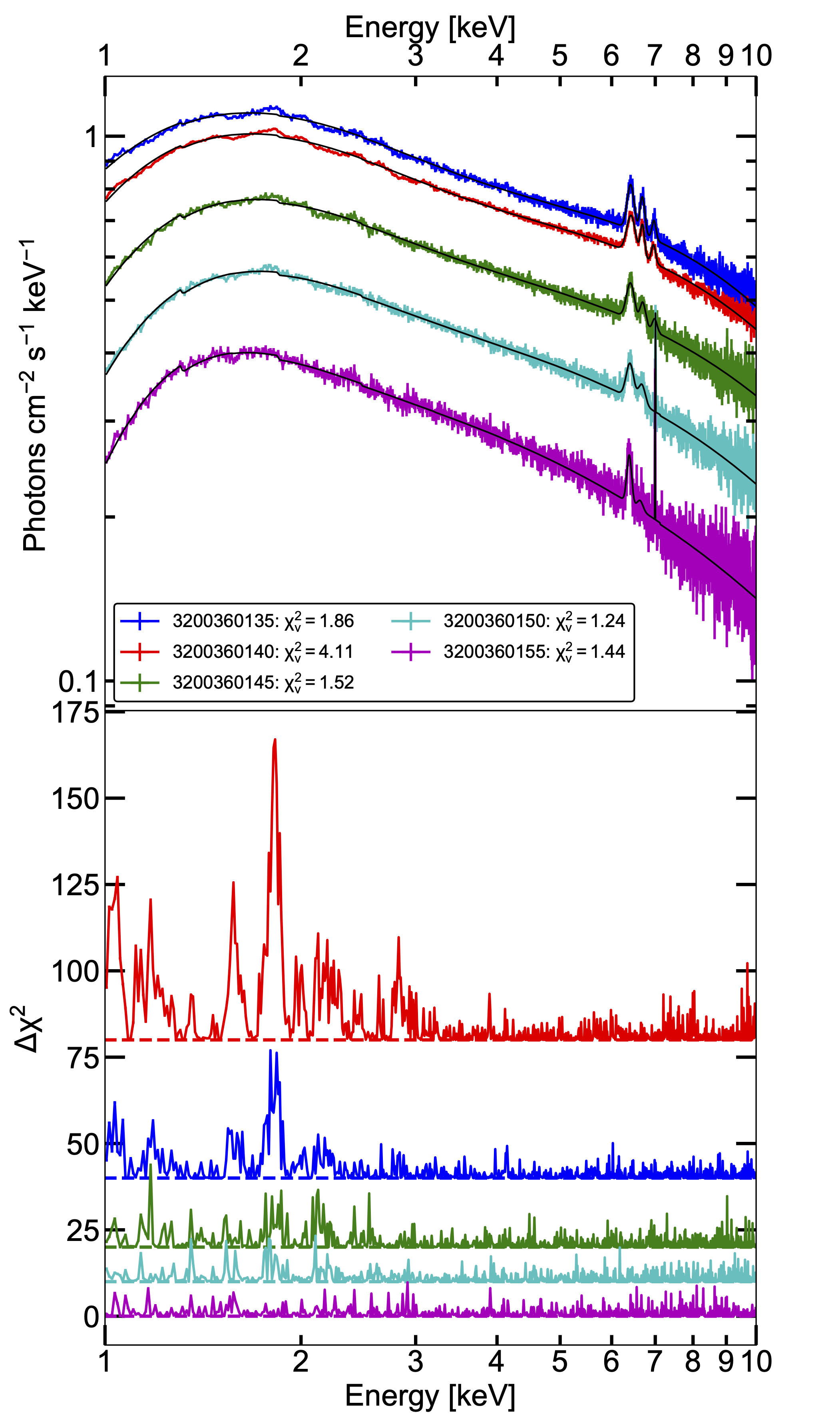}
    \caption{\textit{Top:} unfolded \textit{NICER} spectra for five observations of 1A 0535+262 (ObsIDs listed in legend) around the outburst peak. The decay in flux is evident in the overall decrease of the normalization. The black line indicates each of the model fits. \textit{Bottom:} the $\Delta \chi^2$ fit residuals with the same colour-coding as the top panel. Each observation is vertically offset for clarity but plotted on the same vertical scale. The presence of large and significant instrumental residuals below $3$ keV is clear at high flux (e.g. red and blue) but disappears into the noise towards lower fluxes (e.g. purple). Note that the large narrow peak in the bottom spectrum is caused by unfolding the spectrum around the model containing a very narrow Gaussian line.}
    \label{fig:nicerfits}
\end{figure}

However, fitting this, or any other model, to the \textit{NICER} spectra, runs into issues for the observations at the highest flux. At those fluxes, instrumental response residuals appear below $3$ keV, particularly between $1$ and $2$ keV. One can see this effect in the five \textit{NICER} spectra, taken around the peak of the 1A 0535+262 giant outburst, shown in the top panel of Figure \ref{fig:nicerfits}. The bottom panel shows the $\Delta \chi^2$ residuals compared to a model fit between $1$ and $10$ keV, vertically offset for clarity, highlighting how at higher fluxes, the instrumental residuals are present at high significance. While the model fit, as shown in the top panel, clearly describes the broadband spectral shape and can provide a reliable flux estimate, it is not formally statistically acceptable. Moreover, for poor fits with $\chi^2_\nu \geq 2$, \textsc{xspec} does not allow for the calculation of errors on parameters and fluxes. Therefore, we employ the following approach to fitting the \textit{NICER} spectra: using the above model as a starting point, we run an automated fitting script written in \textsc{tcl}\footnote{Available with the data reproduction notebook for this paper; see the Data Availability Section.} to first fit the model in the $1$--$10$ keV range. If no decent broad-band fit is obtained due to the presence of instrumental residuals (i.e. $\chi_\nu^2 > 2$) the energy range is limited instead to $2$--$10$ keV. We note again that using \textit{NICER}-observation-specific response files and backgrounds does not alleviate this issue. For the \textit{Swift} spectra, we employ the same fitting script, but find that the fitting range does not require restriction.

With this automated approach, further inspired by the large number of both \textit{Swift} and \textit{NICER} observations, two issues require careful attention. Firstly, we confirmed that all poor fits are indeed driven by instrumental residuals, instead of an incorrect or incomplete model, by searching for asymmetric residual structures and unphysical parameters. This conclusion is further confirmed by comparing the \textit{NICER} flux measurement with \textit{Swift} results, revealing that both instruments show a consistent flux evolution and therefore that instrument-specific effects do not significantly change the measured fluxes. Secondly, at the highest fluxes, all three narrow Gaussian lines are clearly present and fitted in the \textit{NICER} spectra. However, the inclusion of three such components may cause \textsc{xspec} to diverge at lower fluxes, as such narrow features can then also be fitted to any local noise deviation. Therefore, when we notice that a fit does not converge, we instead apply a similar scripted fit, first trying a single Gaussian instead, and if the issue persists, no Gaussian at all. For this reason, we ended up not including a Gaussian component in any of the \textit{Swift} spectra. The number of Gaussian components fitted to each \textit{NICER} spectrum is listed in Table 1 in the Online Supplementary Materials.

\subsubsection{Flux measurements and cross-checks}

After finishing each of the aforementioned spectral fits, we use the \textsc{cflux} convolution model to measure the unabsorbed flux and its error in the standard energy band of $0.5$--$10$ keV. As stated earlier, all relevant remaining details regarding the X-ray analysis, such as the exact ObsID, date and length of the observations, instrument setup, fitted model parameters with errors and quality of the final fit, as well as all the measured fluxes and errors, are presented in the Online Supplementary Materials. In addition, we calculate the $1$-$10$ keV and $2$-$10$ keV fluxes, which we include as machine-readable Online Supplementary Data tables to allow comparison with other works using these different energy ranges.

Since \textit{MAXI} is not a pointed instrument, we explicitly checked whether fitting the $2$--$10$ keV \textit{MAXI} spectrum introduces biases in the flux determination for the giant outburst of GRO J1008-57. Therefore, we also extracted a \textit{MAXI}/GSC spectrum on the same day as the brightest Swift observation of the Type-I outburst of this source (ObsID 31030152; MJD 58652). We then jointly fit the \textit{Swift}/XRT PC and WT mode $1$-$10$ keV spectra and the \textit{MAX}/GSC $2$-$10$ keV spectrum with an absorbed power law model. The inferred power law indices are consistent at $1-\sigma$ between the two instruments. However, the \textit{MAXI} fit systematically overestimates the flux by $70$\% compared to \textit{Swift}, in each of the three considered flux energy bands. Given the pointed nature of the \textit{Swift} observations, and the significantly lower spatial resolution of \textit{MAXI}, we expect the \textit{Swift} flux measurement to be more accurate. Since the giant and Type-I outburst of GRO J1008-57 traverse similar X-ray regimes (with peak BAT rates within a factor $\sim 2$), we correct the giant outburst \textit{MAXI} fluxes by dividing by a factor 1.7.

\subsection{Matching X-ray and radio observations}

To place the observations on the X-ray binary $L_X$--$L_R$ diagram, the radio observations need to be matched up with the best estimate for the quasi-simultaneous X-ray flux. For the Type-I outburst of GRO J1008-57, this is straighforward, since each radio observation has an associated \textit{Swift}/XRT pointing, coordinated to be taken within a day. For the giant outburst of this same source, we use the \textit{MAXI} flux from the spectrum extracted from the day of (or the range of days covering) the radio observation. The only exception is the first radio observation, which was not covered by \textit{MAXI} data. Therefore, we instead performed a linear interpolation between the measured logarithmic fluxes measured as close as possible before and after the radio observation. For 1A 0535+262, the majority of radio observations were obtained on a day where either \textit{Swift} or \textit{NICER} (or both) observed the target. Then, we associated the radio observation with the closest X-ray observation. Otherwise, we again performed a linear interpolation of the logarithmic flux measured one day earlier and later. The final case, of SAX J2103.5+4545, is somewhat more complicated, as X-ray measurements are only available after the radio observation, preventing interpolation. Therefore, we instead take the X-ray flux measured by \textit{Swift} closest to the radio observation, and scale it by the ratio of \textit{Swift}/BAT daily count rates between those dates. This implies an increase in flux of $\sim 66$\%, under the assumption that the X-ray spectrum did not significantly change shape.

\subsection{A note on distances}

For all three targets considered in this work, several distances estimates exist in the literature. For consistency, we calculate distances based on the Gaia eDR3 parallax measurements \citep{bailerjones2020}. Following \citet{atri2019}, we apply the Galactic distribution of LMXBs as a prior when converting parallaxes to distances\footnote{We find consistent results when using the more standard exponentially decreasing space density prior instead.} and apply a zero-point correction. At a $68$\% confidence level, we find distances of $1.79^{+0.08}_{-0.07}$ kpc for 1A 0535+262, $3.55^{+0.17}_{-0.15}$ kpc for GRO 1008-57, and $6.23^{+0.55}_{-0.47}$ kpc for SAX J2103.5+4545. Finally, we find a distance of $5.21^{+0.32}_{-0.28}$ kpc for the BeXRB Swift J0243.6+6124, which will be relevant in Section \ref{sec:results}. These derived distances can be included in our modelling while taking their uncertainties fully into account. We note that they are consistent with the aforementioned literature estimates based on other techniques: $\sim 2$ kpc \citep{bailerjones18} for 1A 0535+262; $3.6^{+0.4}_{-0.5}$ kpc \citep{arnason2021} and $\sim 5$ kpc \citep{coe1994} for GRO J1008-57; and $\sim 6.5$ kpc \citep{reig2004} and $\sim 4.5$ kpc \citep{baykal2007} for SAX J2103.5+4545.

\section{Results}
\label{sec:results}

\subsection{Light curves}

\begin{figure*}
    \centering
    \includegraphics[width=\textwidth]{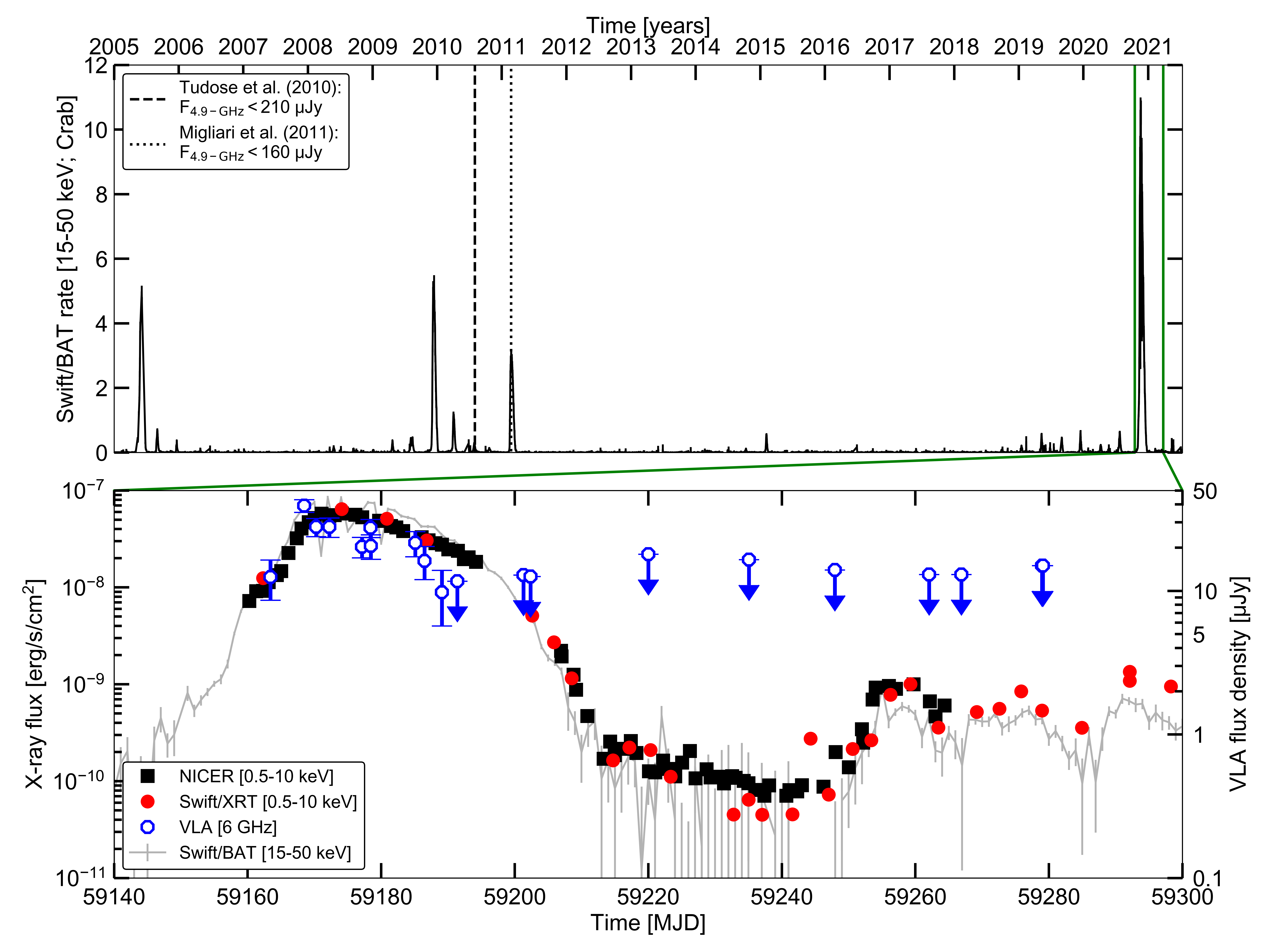}
    \caption{X-ray and radio light curves of 1A 0535+262. \textit{Top:} the long-term \textit{Swift}/BAT light curve, expressed in Crab units, covering Jan 2005 to June 2021. The black dashed and dotted lines indicate the times of the radio observations by Tudose et al. (2010) and Migliari et al. (2011), respectively, neither of which resulted in a detection of 1A 0535+262. The outburst starting in 2020, surrounded by the red lines, is the focus of this work. \textit{Bottom:} zoomed-in light curve of the 2020--2021 giant outburst of 1A 0535+262. The \textit{Swift}/BAT light curve is shown in grey, re-scaled to arbitrary units and shown in log scale. The black squares and red circles indicate the $0.5$--$10$ keV flux measured by \textit{Swift}/XRT and \textit{NICER}, respectively. The VLA radio monitoring is shown by the blue octagons. Note that we do not show two Swift observations (00013945008, 00013945025) due to their short exposures.}
    \label{fig:LC_A0535}
\end{figure*}

In Figures \ref{fig:LC_A0535}, \ref{fig:LC_J2103}, and \ref{fig:LC_J1008}, we show the X-ray and radio monitoring light curves of our three targets, 1A 0535+262, SAX J2103.5+4545, and GRO J1008-57, respectively. All plotted pointed observations, i.e. radio flux densities and X-ray fluxes, are listed in the Online Supplementary Materials. The top panel in Figure \ref{fig:LC_A0535} shows the long-term \textit{Swift}/BAT daily monitoring light curve of 1A 0535+262, re-scaled to Crab units, over the past sixteen-and-a-half years. The dashed and dotted black lines indicate the times of the earlier radio observations, by \citet{tudose10} and \citet{migliari11}, respectively (non-detections with $210$ and $160$ $\mu$Jy upper limits). The red lines indicate the time range plotted in the zoomed-in bottom light curve. The 2020-2021 giant outburst was clearly brighter than any observed previously with \textit{Swift}/BAT, reaching $\sim 11$ Crab, compared to $\sim 6$ Crab in 2009, during the second-brightest outburst. Type-I outbursts can be seen as the short-duration spikes, reaching up to $\sim 1$ Crab fluxes.

In the bottom panel, we plot the \textit{Swift}/XRT, \textit{NICER}, and VLA light curves of the 2020-2021 outburstof 1A 0535+262, in red, black, and blue, respectively. The \textit{Swift}/BAT light curve is shown as well, scaled by a single arbitrary factor to allow for the comparison of its shape to the light curve of pointed X-ray observations. While the outburst becomes visible in the \textit{Swift}/BAT monitoring between MJD 59140 and 59150, pointed X-ray observations start around MJD 59160, when the X-ray flux had already reached $\sim 7\times10^{-9}$ erg/s/cm$^2$. As shown by both the daily \textit{NICER} and the less-frequent \textit{Swift}/XRT observations, the outburst rise continued for two weeks, peaking at a flux of $\sim 5.8\times10^{-8}$ erg/s/cm$^2$ on MJD 59174. Subsequently, the initial outburst decayed gradually until MJD 59194, where the \textit{Swift}/BAT monitoring reveals an acceleration of the flux decrease during a gap in the \textit{NICER} monitoring. 1A 0535+262 later reaches a relatively stable, low-flux plateau, decaying the flux from $\sim 2.5\times10^{-10}$ to $8\times10^{-11}$ erg/s/cm$^2$ between MJD 59214 and 59246. Finally, as shown predominantly by the \textit{Swift} monitoring, the X-ray flux reaches a higher-flux plateau, stabilizing in the range of $(5$--$10)\times10^{-10}$ erg/s/cm$^2$. While the \textit{Swift}/XRT monitoring has continued after MJD 59300, we end the light curve due to the lack of radio data.

\begin{figure}
    \centering
    \includegraphics[width=\columnwidth]{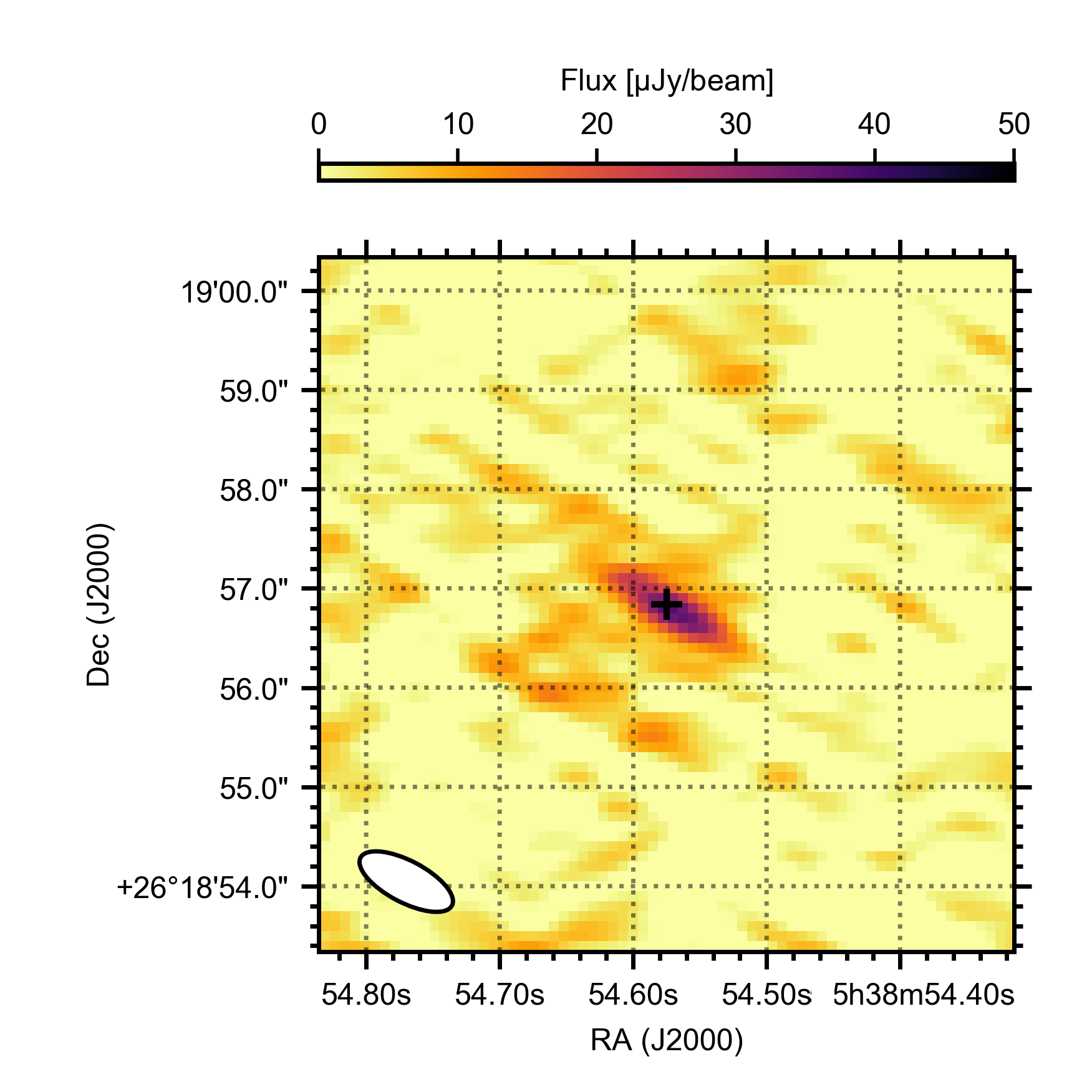}
    \caption{The 6-GHz radio counterpart of 1A 0535+262 during the second VLA observation, where it showed the highest observed radio flux. The black cross indicates the known position of the source; the synthesized beam is shown in bottom left corner.}
    \label{fig:image_A0535}
\end{figure}

In the VLA radio monitoring, 1A 0535+262 is detected in the first $9$ observations at $\geq 3-\sigma$ significance, with flux densities between $12.5\pm3.9$ $\mu$Jy and $39.2\pm4.0$ $\mu$Jy. In Figure \ref{fig:image_A0535}, we show the 6-GHz field around the target in the second observation, highlighting the faint but clear counterpart of the BeXRB. The best-fit source position in this image is:

\begin{align*}
    &\text{RA (J2000)} = 05\text{h } 38\text{m } 54.571\text{s } \pm 0.008\text{s} \\
    &\text{Dec (J2000)} = 26^{\rm o} 18\text{' } 56.79\text{" } \pm 0.09\text{"}
\end{align*}

\noindent The aforementioned flux densities, observed in the first and second radio observations, show how the radio flux density increased during the outburst rise, peaking close to the time of the X-ray peak. During radio observations 3--9, the flux density globally decayed, as the X-rays peaked and subsequently decayed as well. This radio flux density decrease is, however, very gradual, and the difference in peak times in X-rays and radio may be affected by a changing radio spectral shape. During the tenth observation, radio emission is still observed at the position of 1A 0535+262, albeit at less than $3\-\sigma$ significance: $9.8\pm4.1$ $\mu$Jy. We detect no radio emission from the source position in any of the remaining ten observations (i.e.,  starting on MJD 59191). The resulting upper limits on the radio flux density, below the typical radio levels during the outburst peak, show that the radio emission is enhanced during the accretion outburst and is not present (at detectable levels) for X-ray fluxes below $2.4\times10^{-8}$ erg/s/cm$^2$. Finally, based on our results, the earlier radio-non-detections by \citet{tudose10} and \citet{migliari11} can be attributed to the lower sensitivity in those observations.

To study the spectral shape, we divided the full 4-8 GHz observing band into four sub-bands of 1 GHz width, thereby roughly halving the sensitivity per band. During the brightest radio epoch (observation two), we obtain the best single-observation constraint on the spectral index $\alpha$ (where the flux density scales with frequency as $S_\nu \propto \nu^{\alpha}$). Even in this observation, however, we measure a relatively poor measurement of $\alpha = -0.9 \pm 1.1$, encompassing the range of expected indices for both optically-thin discrete ejecta (i.e. $\alpha = -0.7$) and unresolved, compact jets ($\alpha \geq 0$) \citep{fender04,russell13}. To increase signal-to-noise, we then repeated this procedure combining the three brightest observations (2, 3, and 4), all taken in the same array configuration. There, we measure a spectral index $\alpha = -0.1 \pm 0.1$, consistent with a flat-spectrum radio jet. The radio spectrum for both cases is shown in the Online Supplementary Materials.

\begin{figure}
    \centering
    \includegraphics[width=\columnwidth]{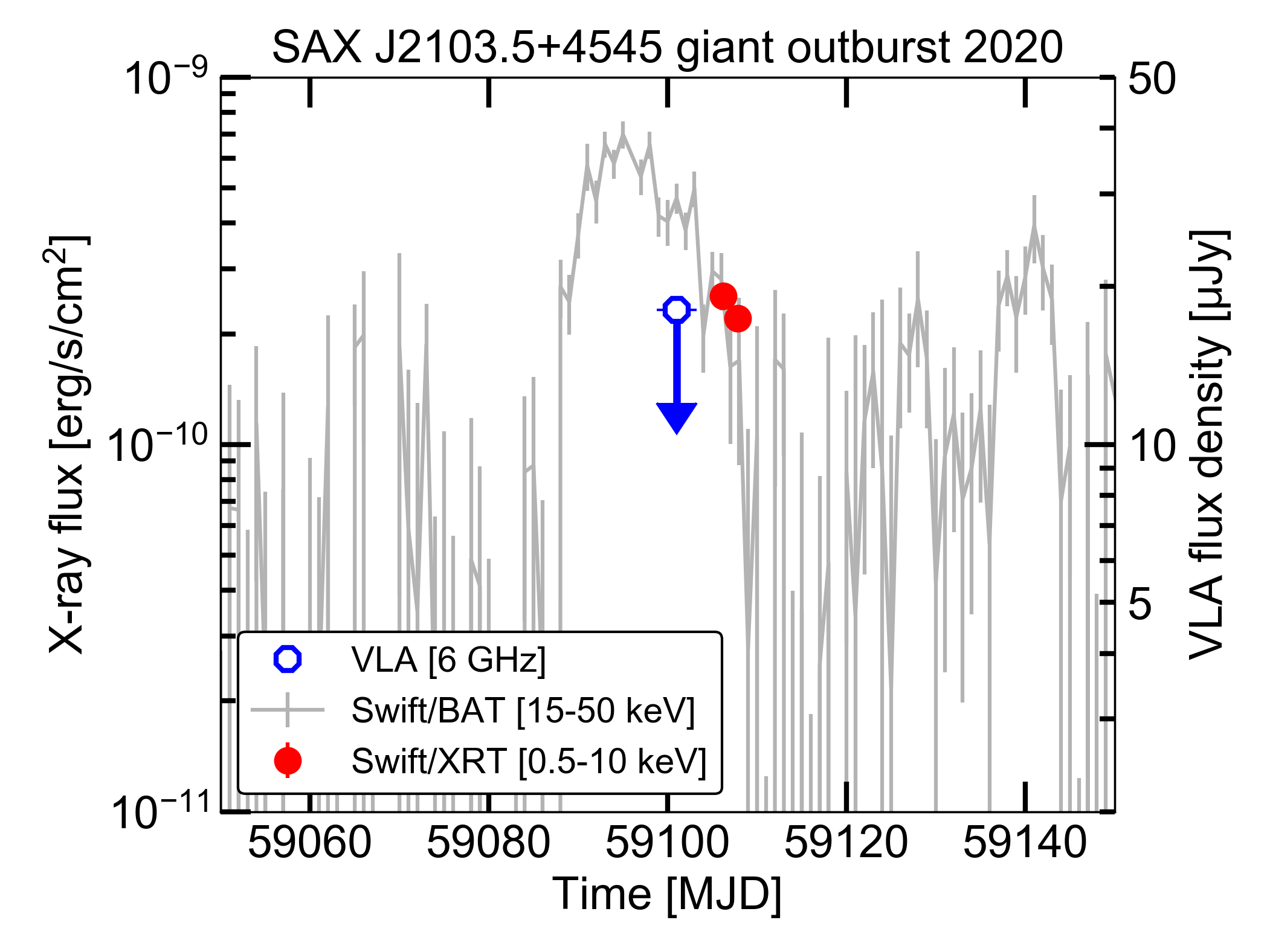}
    \caption{X-ray and radio light curves of SAX J2103.5+4545 during its 2020 giant outburst. X-ray monitoring/pointed and radio observations are shown in the same style as Figure \ref{fig:LC_A0535}.}
    \label{fig:LC_J2103}
\end{figure}

We now turn to SAX J2103.5+4545, shown in Figure \ref{fig:LC_J2103}. Here, we plot the \textit{Swift}/BAT light curve for the one hundred days around the VLA observation, scaled to arbitrary units similar to the bottom panel of Figure \ref{fig:LC_A0535}. The X-ray fluxes, measured from the two Swift/XRT pointed observations, are shown in red. These two observations clearly capture the behaviour during the decay of the outburst, just before the outburst cannot be clearly identified anymore in the \textit{Swift}/BAT light curve. The VLA observation occurs slightly earlier, albeit also during the outburst decay, and returns a non-detection. The 6-GHz image RMS is $6$ $\mu$Jy, resulting in a $3-\sigma$ upper limit on the flux density of $18$ $\mu$Jy. No further pointed observations, in the radio or X-ray band, were performed afterwards.

\begin{figure*}
    \centering
    \includegraphics[width=\textwidth]{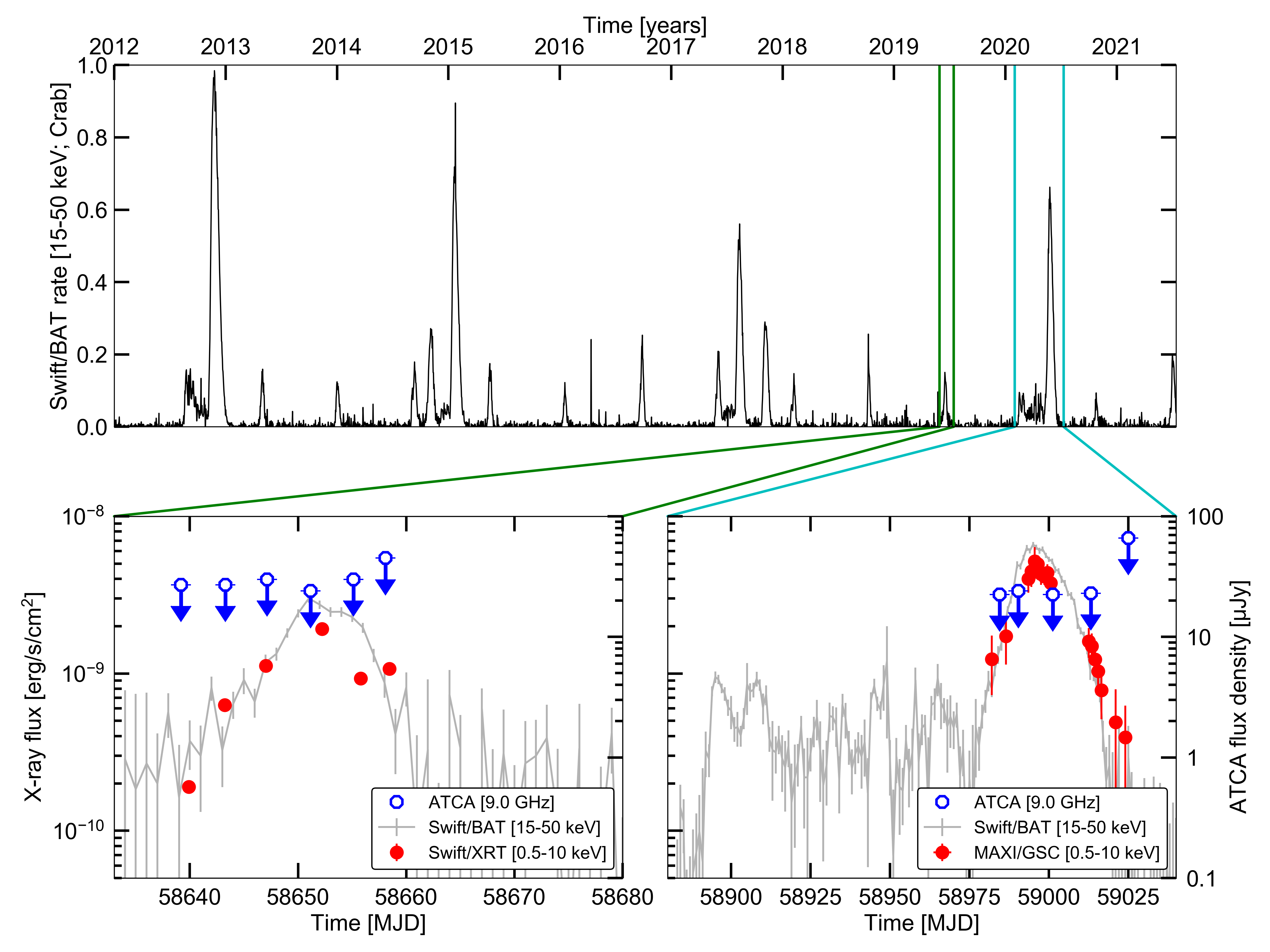}
    \caption{X-ray and radio light curves for GRO J1008-57. The setup of this Figure is the same as Figure \ref{fig:LC_A0535}. \textit{Top:} long-term \textit{Swift}/BAT monitoring, expressed in Crab units. The blue and red regions indicate the Type-I and giant outbursts targeted in radio, respectively. Before several giant outbursts -- i.e. in 2012, 2014--2015, 2017, 2020 -- an enhanced X-ray flux state is visible before the outburst onset. \textit{Bottom:} zoomed-in light curves of the Type-I (\textit{left}) and giant (\textit{right}) outburst. No radio counterpart of GRO J1008-57 is detected at any time. The jump in the radio upper limit in the final observation results from a change in ATCA array configuration.
    }
    \label{fig:LC_J1008}
\end{figure*}

Finally, in Figure \ref{fig:LC_J1008}, we show the long-term \textit{Swift}/BAT monitoring and our pointed X-ray and radio monitoring during two outbursts, of GRO J1008-57, in similar fashion to Figure \ref{fig:LC_A0535}. The upper light curve confirms that GRO J1008-57 is a prolific outbursting source, showing Type-I outbursts every orbital period (separated by $249.5$ days), as well as multiple giant outbursts. On two occasions since 2012, we can identify the occurrence of two giant outbursts in between successive Type-I outbursts: around the start of 2015 and in 2017. Moreover, in 2012 and 2020, the regular Type-I outburst is followed by a variable state of enhanced X-ray flux and then followed by a giant outburst. Evidence for this effect can also be seen before the giant burst in early 2015 and the brightest of the two giant bursts in 2017.

Our radio and X-ray monitoring campaigns targeted the 2019 Type-I outburst and the 2020 giant outburst, where we were able to catch the outburst rise early due to the the enhanced flux state before the latter outburst. In the two bottom panels for Figure \ref{fig:LC_J1008}, we show the zoomed in light curves of these two outbursts, with X-ray fluxes measured from the \textit{Swift}/XRT observations in the left panel, and those measured from the \textit{MAXI}/GSC spectra in the right. In both cases, the radio monitoring cadence samples the outburst evolution well. However, we note the difference in the scaling on the horizontal axis; the separation between the radio
observations in the Type-II outburst is longer and less regular, due to the triggered nature of the campaign. In none of the eleven ATCA observations, we detect any significant radio emission at either $5.5$ or $9$ GHz at RMS sensitivities usually ranging between $\sim 7$ and $10$ $\mu$Jy (at $9$ GHz), leading to typical $3-\sigma$ upper limits of $\sim 20$--$30$ $\mu$Jy. The higher upper limit in the final observation during the giant outburst ($66$ $\mu$Jy) results from a change in ATCA configuration to a compact H214 configuration. When we stack the four first giant outburst observations, or all Type-I observations (i.e. those taken in the same setup and configurations), we also do not detect a counterpart, down to slightly deeper levels than the individual observations (See the Online Supplementary Materials).

\subsection{The X-ray -- radio luminosity plane}

\begin{figure*}
    \centering
    \includegraphics[width=\textwidth]{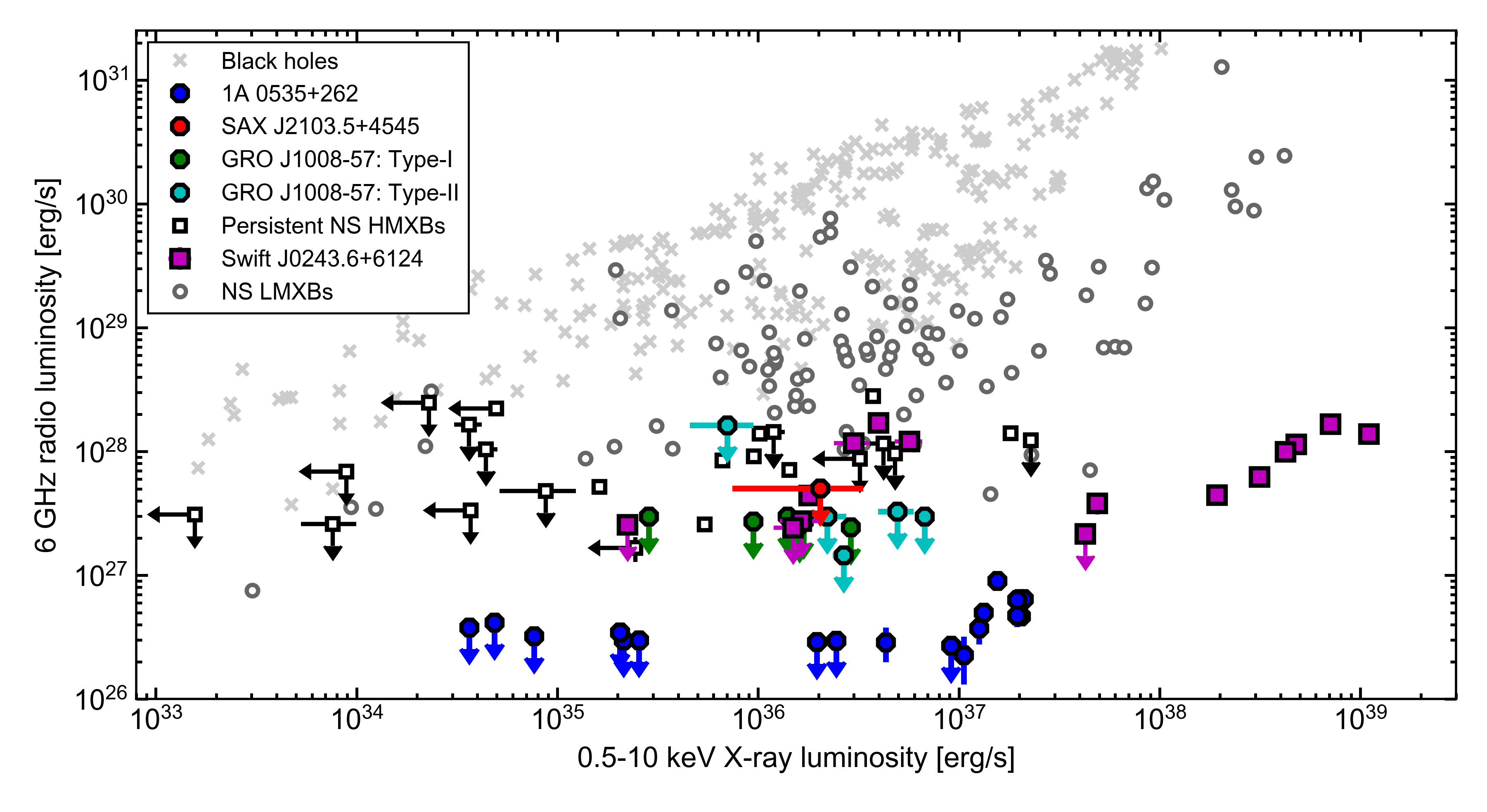}
    \caption{The X-ray -- radio luminosity plane for X-ray binaries.
    The colored and filled-in points indicate transient BeXRBs. Octagons indicate the three sources studied in this work, while the two outbursts of GRO J1008-57 are indicated with different colors. For comparison, Swift J0243.6+6124 is also shown, as the purple squares. The open black squares show persistently accreting NS HMXBs and quiescent BeXRBs, while the grey crosses and circles indicate archival observations of black hole and NS LMXBs, respectively. This comparison black/grey sample was compiled by \citet{vandeneijnden2021}.
    }
    \label{fig:lxlr_full}
\end{figure*}

Combining the X-ray and radio flux (density) measurements taken close in time (see Section \ref{sec:obs}), we can place our three targets on the X-ray binary $L_X$--$L_R$ plane. For this purpose, we calculated the radio luminosity (upper limit) at $6$ GHz, assuming a flat spectrum, and we assume the distances listed in Section \ref{sec:targets}. In Figure \ref{fig:lxlr_full}, we show the resulting luminosities alongside three comparison samples, all taken from \citet{vandeneijnden2021}: black hole LMXBs shown as the grey crosses, NS LMXBs shown as grey circles, and persistent NS HMXBs shown as black squares. The transient BeXRBs are shown as the colored and filled-in data points: SAX J2103.5+4545, GRO J1008-57, and 1A 0535+262 as octagons of different colors per source and outburst type, and archival Swift J0243.6+6124 data as purple squares. Below X-ray luminosities of $10^{37}$ erg/s, the BeXRB sample is dominated by radio non-detections. The only exceptions are a single detection of 1A 0535+262, as well as the four radio detections of Swift J0243.6+6124 obtained after its main giant outburst \citep{vandeneijnden2019_reb}.

\begin{figure}
    \centering
    \includegraphics[width=\columnwidth]{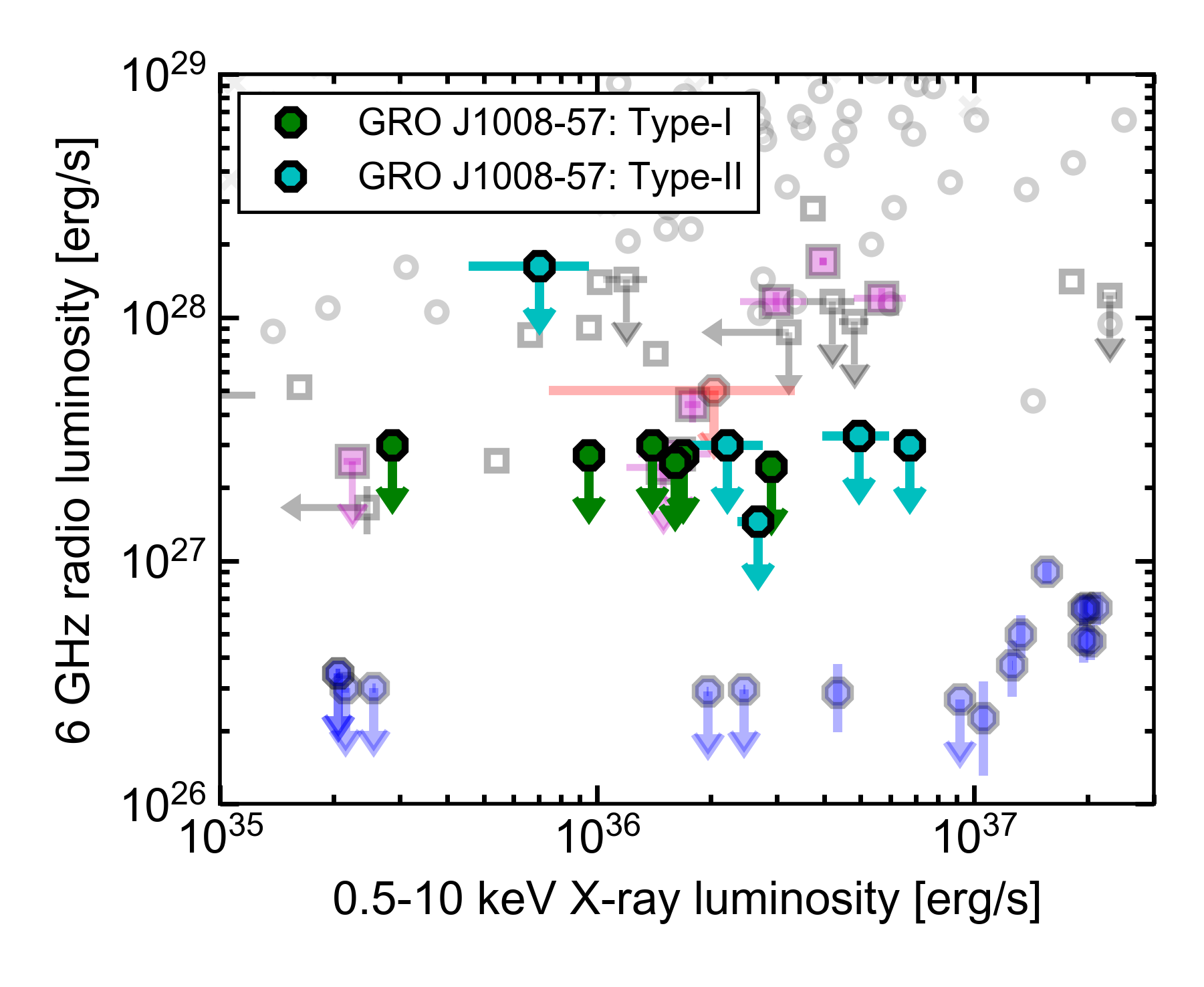}
    \caption{Zoomed in inset of Figure \ref{fig:lxlr_full}, focusing on the crowded region where the GRO J1008-57 data reside. To highlight the data from this BeXRB, all other data points have been faded.
    }
    \label{fig:lxlr_J1008}
\end{figure}

To highlight the observations of GRO J1008-57 within the cluster of data points surrounding it, we show a zoomed version of the radio -- X-ray luminosity plane in Figure \ref{fig:lxlr_J1008} with the other sources faded out. Between the Type-I and giant outbursts, our observations spanned a factor $\sim 25$ in X-ray luminosity. The two outbursts also overlap in X-ray luminosity. However, no radio emission is detected in either outburst. When we consider this crowded region of the radio -- X-ray luminosity plane for all targets, the archival radio detections of Swift J0243.6+6124 stand out. These data points lie above the majority of radio upper limits for the other three targets at similar X-ray luminosity. This discrepancy hints towards a difference in radio behaviour between main outbursts and X-ray re-flares, which we will discuss in more detail in Section \ref{sec:discussion}.

The sensitivity limits of current radio observatories in common monitoring observation lengths are clearly visible in Figure \ref{fig:lxlr_full}: approximately $2$--$3\times10^{27}$ erg/s for sources located at distances of the order of $\sim5$ kpc and approximately $3$--$5\times10^{26}$ erg/s for sources located at $\sim2$ kpc. The close distance to 1A 0535+262 has been essential in detecting its radio emission. From this sample of four BeXRBs, it appears that current radio telescopes may detect radio emission if: i) the source is located at relative close distances (i.e. $\leq 2$ kpc); ii) the source accretes close to, or at, super-Eddington luminosities; \textit{or} iii) the source is observed during re-flaring activity after a main giant outburst.

\subsection{A possible X-ray/radio correlation for transient BeXRBs}

\begin{figure*}
    \centering
    \includegraphics[width=\textwidth]{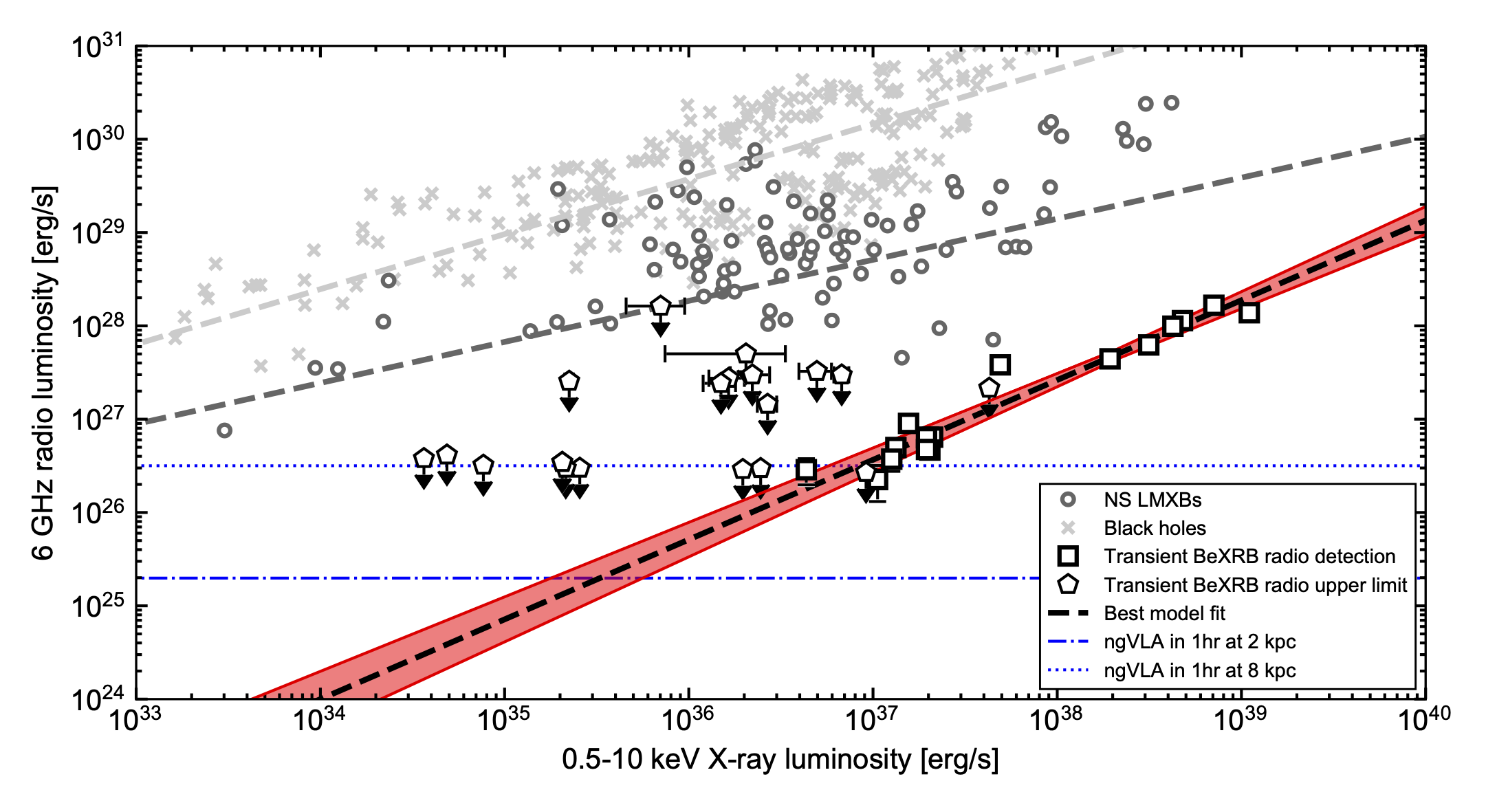}
    \caption{The fit to the BeXRB giant outburst X-ray -- radio luminosity correlation. The black open squares indicate radio detections of BeXRBs, while the black open hexagons show upper limits. The black dashed lines plot the best fit relation of the model $L_X \propto \xi L_R^\beta$, while the red lines show 100 random draws of the posterior distributions from the parameter distributions of $\xi$ and $\beta$. Finally, the blue dotted and dash-dotted lines indicate the luminosity sensitivity limits with the planned next-generation VLA (ngVLA), for distances of $8$ and $2$ kpc, respectively. The grey points show the LMXB comparison samples, as in Figure \ref{fig:lxlr_full}, and their $L_X$--$L_R$ correlation from \citet{gallo18}.
    }
    \label{fig:lxlr_fitted}
\end{figure*}

Based on the radio monitoring of the main outburst of Swift J0243.6+6124, \citet{vandeneijnden2018_swj0243} argued that its X-ray and radio luminosity display a global coupling during the outburst decay, measuring the relation $L_R \propto L_X^{0.54 \pm 0.16}$. This measurement, however, did not take into account the initial radio non-detection. In addition, the later observations of Swift J0243.6+6124, presented in \citet{vandeneijnden2019_reb}, were not taken into account. In our extended sample, that now includes four sources, we can expand on this analysis. From Figure \ref{fig:lxlr_full}, it appears that during the peak of its outburst, 1A 0535+262 also followed a coupling between its X-ray and radio luminosity. Moreover, this apparent correlation visually seems to continue the relation seen during the giant outburst decay in Swift J0243.6+6124. Under the simplest assumption that giant BeXRB outbursts follow a similar relation between their X-ray and radio luminosity, we can attempt to determine a global $L_X$--$L_R$ coupling index for this source class and outburst type.

To properly measure a giant outburst $L_X$--$L_R$ coupling, we combine the data from the main outburst of Swift J0243.6+6124, all data from 1A 0535+262 and SAX J2103.5+4545, and the giant outburst data from GRO J1008-57. While most radio upper limits for the latter two sources, as well as those for 1A 0535+262 at low X-ray luminosities, are likely unconstraining when assuming a single correlation, other radio limits will have a more significant effect. The early non-detection of Swift J0243.6+6124 and the three X-ray-brightest radio upper limits of 1A 0535+262 appear, by eye, to lie close to any global correlation. Therefore, improving upon \citet{vandeneijnden2018_swj0243}, it is important to properly account for all giant outburst radio non-detections. All radio detections and upper limits used in the fit, are shown in Figure \ref{fig:lxlr_fitted}.

We follow the approach originally developed by \citet{kelly2007}, introduced to the study of the $L_X$--$L_R$ plane by \citet{gallo2014}, in the \textsc{LinMix} method: a Bayesian Markov-Chain Monte-Carlo fit of a linear model, fully accounting for upper limits in the dependent variable. We specifically applied the \textsc{python}-version of this method\footnote{Publicly available via \href{https://github.com/jmeyers314/linmix}{https://github.com/jmeyers314/linmix}.}, to fit a model of the form
\begin{equation}
    \frac{L_R}{L_{R,0}} = \xi \left(\frac{L_X}{L_{X,0}}\right)^\beta\text{ ,}
\end{equation}
where $\xi$ is an arbitrary scaling factor and $\beta$ is the coupling index between the luminosities. Luminosities with a subscript $0$ denote the mean luminosity of the radio-detected observations: $L_{X,0} = 2\times10^{38}$ erg/s and $L_{R,0}=4.2\times10^{27}$ erg/s. To apply the \textsc{LinMix} method, we linearize the model as
\begin{equation}
    \log L_R - \log L_{R,0} = \log \xi + \beta\left(\log L_X - \log L_{X,0} \right) + \epsilon\text{ ,}
    \label{eq:linmodel}
\end{equation}
where $\epsilon$ is an additional parameter describing the intrinsic Gaussian scatter around the best fit correlation. To measure the fitted parameters $\log \xi$, $\beta$, and $\epsilon$, we followed \citet{gallo18} and calculate the mean parameter from $10^4$ draws from the posterior distribution (instead using the median returns equivalent results). We repeat this fit $500$ times, following \citet{gusinskaia20}, and report the mean of the 500 parameter estimates as the fitted values. To take into account distance uncertainties, we draw a random distance for each source from the Gaia distance distribution for each of these 500 iterations. However, we find the uncertainties in the fit are dominated by the low signal-to-noise of the radio observations. The $1-\sigma$ errors are calculated by taking, for each of the 500 runs, the $16^{\rm th}$ and $84^{\rm th}$ percentile from the $10^4$ parameter draws, and subsequently averaging those 500 values. We show an example distribution from a single run and the distributions after 500 runs, for $\log \xi$ and $\beta$ in the Online Supplementary Materials.

Following the above approach, we measure $\log \xi = 0.057 \pm 0.052$, $\beta = 0.86 \pm 0.06$, and $\epsilon = 0.17^{+0.05}_{-0.03}$. Figure \ref{fig:lxlr_fitted} shows the best fit version of Equation \ref{eq:linmodel} and its uncertainty range. The index $\beta$ is steeper than that measured for Swift J0243.6+6124 alone, although the exclusion of the radio upper limit may have pushed that earlier fit to shallower slopes. Comparing with \citet{gallo18}, we find that the measured index is steeper that that seen in black holes ($\beta = 0.59 \pm 0.02$) and the full sample of NS LMXBs ($\beta = 0.44^{+0.05}_{-0.04}$; however, we note that individual NS systems have been observed to show significantly different coupling indices). The scatter seen for the BeXRBs is smaller, compared to $\epsilon = 0.46 \pm 0.02$ and $\epsilon = 0.43^{+0.05}_{-0.04}$, in the other two samples, respectively. However, this is hardly surprising, as our sample includes only four BeXRBs and the fit was motivated specifically by the similarity in correlation between Swift J0243.6+6124 and 1A 0535+262. The scatter seen for these BeXRBs may, to some degree, be driven by short-time-scale ($< 1$ day) variability in the radio flux density, especially in combination with the association of non-simultaneous X-ray and radio observations separated by up to a day. 

We cannot directly compare the intercept, $\log \xi$, to \citet{gallo18}, due to the different values of $L_{X,0}$ and $L_{R,0}$. Re-scaling the measured $\log \xi$ value to their values, e.g. $L_{X,0,\rm G+18} = 2.00\times10^{36}$ erg/s and $L_{R,0,\rm G+18}=3.72\times10^{28}$ erg/s, we find that $\log \xi_{\rm G+18} = -2.61$. This value is significantly lower than measured for the black hole and NS LMXB populations, i.e $\log \xi = 1.18 \pm 0.03$ and $\log \xi = -0.17 \pm 0.05$, respectively. This makes the transient BeXRB population, while similar in its inferred coupling index, $\sim275$ times radio fainter than the NS LMXB population, at an X-ray luminosity of $2\times10^{36}$ erg/s. Due to the steeper index for the transient BeXRBs, this difference decreases towards higher X-ray luminosities, as is visible in Figure \ref{fig:lxlr_full}.

The two BeXRBs with radio detections, Swift J0243.6+6124 and 1A 0535+262, do not overlap in the X-ray luminosity during the radio monitoring of the main outburst. Therefore, physical differences or observational uncertainties can systematically affect the measured correlation for the entire BeXRB sample. For instance, an incorrect distance measurement for one of the sources may affect the measured slope of the correlation. Similarly, if the magnetic field strength or spin affects the radio luminosity (as discussed in Section \ref{sec:disc_BS}), this would affect the slope and normalization of the coupling: the spin period of Swift J0243.6+6124 is more than 10 times smaller than 1A 0535+262. While this issue plays a role in NS LMXBs as well, their overlap in X-ray luminosity and small differences in spin \citep{patruno17} implies that any distance, spin, and magnetic field effects, affect the scatter more than the slope.

With the above considerations in mind, it is an interesting exercise to assess Swift J0243.6+6124 and 1A 0535+262 separately (using $L_{X,0,\rm G+18}$ and $L_{R,0,\rm G+18}$). When we repeat our \textsc{LinMix} fits for 1A 0535+262 individually, we find $\beta = 0.80 \pm 0.26$, $\log \xi = -2.59 \pm 0.24$, and $\epsilon = 0.18\pm0.08$. For Swift J0243.6+6124, we find $\beta = 0.65^{+0.17}_{-0.14}$, $\log \xi = -2.13^{+0.32}_{-0.38}$, and $\epsilon = 0.26^{+0.16}_{-0.02}$. The slopes $\beta$ for the sources individually are consistent with each other, and with the NS LMXBs. However, we note that the reduced number of data points and the smaller range in X-ray luminosity contribute to significantly enhanced uncertainties on $\beta$ \citep[see][for a discussion on the effects of small ranges in X-ray luminosity]{corbel13}.

\section{Discussion}
\label{sec:discussion}

In this paper, we have presented coordinated radio and X-ray monitoring of three transient BeXRBs during four outbursts. No radio counterpart was detected in the eleven ATCA observations of GRO J1008-57 across two outbursts of different types, nor in the single VLA observation of SAX J2103.5+4545. 1A 0535+262, on the other hand, was detected in the first ten of twenty VLA giant outburst monitoring observations. Here, we will discuss the origin and properties of this detected radio emission and its apparent relation to the X-ray luminosity of the BeXRBs, before investigating the origin of their radio faintness. Finally, we will compare our results with the behaviour of Swift J0243.6+6124 during X-ray re-flares and with the radio properties of persistently accreting NS HMXBs.

\subsection{The origin of transient radio emission in BeXRBs}
\label{sec:whatdowesee}

The launch of a relativistic jet from the inner accretion flow can explain the observed radio properties of the three targets. As the only detected source, 1A 0535+262 naturally dominates this interpretation. Firstly, the set of ten non-detections in the tail of the outburst cannot be explained by a decrease in radio sensitivity. Instead, these non-detections appear to be linked to the decrease in X-ray luminosity, connecting the prior radio detections to the presence of accretion. In other words, we do not expect that the radio emission originates from either the NS or the donor star itself, or their interaction, via processes that also operate in quiescence. Secondly, the global correlation between the X-ray and radio luminosity of 1A 0535+262 is consistent with a coupling between an in- and outflow, as commonly observed in LMXBs. Finally, this coupling appears to extend the correlation observed during the Swift J0243.6+6124 giant outburst. A major difference, however, with that Swift J0243.6+6124 data set, is the lack of accurate spectral shape measurements for 1A 0535+262 across the outburst, due to its relative faintness; the only constraining measurement of $\alpha = -0.1 \pm 0.1$ could be obtained by combining the three brightest observations. This value is consistent with a flat-spectrum radio jet and the spectral shape that Swift J0243.6+6124 tended towards as it decreased in X-ray luminosity.

In a radio jet scenario, the non-detections of GRO J1008-57, SAX J2103.5+4545, and 1A 0535+262 below $L_X = 5\times10^{36}$ erg/s, can be explained by the limits of observational sensitivity in combination with a coupling between X-ray and radio luminosity. As can be seen in Figure \ref{fig:lxlr_fitted}, the distance to GRO J1008-57 and SAX J2103.5+4545 makes both BeXRBs undetectable in radio below $L_X \approx 0.5$--$1\times10^{38}$ erg/s, given their radio upper limits. These limits are representative of current radio sensitivities in standard monitoring observations (i.e. 4 hours with ATCA). Future observatories are therefore likely needed to detect giant outbursts below $\sim10$\% $L_{\rm Edd}$ for BeXRBs located beyond $\sim 4$ kpc. This statement assumes that all giant outbursts follow a single X-ray -- radio luminosity coupling (but see Section \ref{sec:disc_BS}). 

In addition to a jet origin, it is essential to consider alternative emission origins. We first consider the Be star in a BeXRB, itself. Isolated Be stars at distances of, typically, tens to hundreds of parsecs, have been observed extensively at radio wavelengths \citep[e.g.][]{taylor1987,taylor1990,drake1990,dougherty1991,clark1998}. The majority of these observations did not yield radio detections, while several sources were detected only in a subset of observations \citep{dougherty1991}. The detections of thermal radio emission of Be stars typically reveal specific radio luminosities between $\sim 1.3\times10^{15}$ to $2.7\times10^{16}$ erg/s/Hz, where the maximum was observed in $\beta$ Mon A by \citet{taylor1990}. Assuming a flat spectral shape, these specific luminosities correspond to $\sim 8\times10^{24}$ to $\sim 1.6\times10^{26}$ erg/s at $6$ GHz, firmly below the deepest upper limits and the radio detections for our four BeXRB targets. An interesting exception is EW Lac, which was observed by \citet{taylor1990} at a specific luminosity of $\sim 10^{17}$ erg/s/Hz, or a luminosity of $\sim 6\times10^{26}$ erg/s at $6$ GHz. While fainter than the majority of radio detections of 1A 0535+262, it is higher than the upper limits on the radio flux of this target during its late outburst decay. However, \citet{dougherty1991} did not detect EW Lac, with an upper limit $5$ times lower than the previous radio detection. This highlights the intrinsic radio variability of Be stars, and the need for coordinated X-ray and radio observations during BeXRB outbursts, to connect radio emission to the presence of accretion in BeXRBs. It similarly shows how, for BeXRBs at small ($<1$ kpc) distances at low X-ray luminosities, the radio flux and variability from the Be star may create a limit to our ability to track the relation between X-ray and radio luminosity into quiescence.

We next consider shock interactions between the Be-star's outflow or circumstellar disk and a pulsar wind, thought to be responsible for the radio emission in $\gamma$-ray binaries. These interactions can be ruled out on three grounds. Firstly, it is typically assumed that no pulsar wind is launched by actively accreting pulsars, which would imply that shock radio emission should become visible towards very low accretion rates; in 1A 0535+262, we observe the opposite \citep[although the recent detection of possibly-spin-powered optical/UV pulsations in the accreting millisecond X-ray pulsar SAX J1808.4-3658 could suggest that the pulsar mechanism may still operate during accretion episodes in some accreting NSs;][]{ambrosino2021}. Secondly, since the spin evolution of the pulsar in BeXRBs is (during outburst) regulated by the transfer of angular momentum between accretion flow and NS, we can estimate what their spin down energy $\dot{E}_{\rm spin}$ would be as isolated pulsars. The spin down energy is $\dot{E}_{\rm spin} = 4\pi^2 I \dot{P}/P^3$, for a NS with period $P$, period derivative $\dot{P}$, and moment of inertia $I$. $\dot{P}$ scales with the magnetic field as $B \propto \sqrt{P \dot{P}}$. Combined, this yields

\begin{equation}
\dot{E}_{\rm spin} \leq 4\times10^{31} \left( \frac{B}{10^{12} {\rm G}} \right)^2  \left( \frac{P}{1 {\rm s}} \right)^{-4} {\rm erg/s}\text{ .}
\end{equation}

\noindent Using magnetic field estimates from their cyclotron line measurements, we find upper limits for 1A 0535+262 and GRO J1008-57 of $\dot{E}_{\rm spin} \leq 5\times10^{24}$ erg/s and $\dot{E}_{\rm spin} \leq 3\times10^{25}$ erg/s, respectively. The radiative luminosity of a shock between a pulsar wind and the Be-disk will be a fraction of this spin down energy. Therefore, the energetics of a pulsar wind, even if launched, are not sufficient to account for the observed radio luminosities. For SAX J2103.5+454, no cyclotron lines have been detected. However, with its slow spin ($P \sim 346$ s), its upper limit will be even lower. Thirdly, given their (measured or assumed) magnetic field strengths and spin periods, all three sources fall beyond the pulsar death line, implying they are not expected to launch a pulsar wind even if isolated \citep[e.g.][]{ruderman1975,zhang2000}.

Recently, \citet{chatzis2021} formulated a new model for the radio emission from BeXRBs hosting a strongly-magnetized NS. In this shock-based model, the radio emission consists of a superposition of a thermal stellar-wind component and a non-thermal synchrotron component. In an X-ray binary analogy to colliding wind binaries, this shock takes place between the stellar wind from the Be star and a non-relativistic outflow launched from the accretion flow. Assuming spherical morphologies for both outflow and stellar wind, and a constant presence of disk outflows at all accretion rates (both sub- and super-Eddington, without any requirements on the exact mechanism), this shock-model derives the resulting radio luminosity, spectrum, and coupling to the X-ray luminosity. The latter is found to be steep in the sub-Eddington regime ($\beta = 12/7$), and dependent on the electron number density distribution in the super-Eddington regime --- $\beta = 2(p-1)/7$, where p is the the power-law index of the electron distribution. Interestingly, the analytical nature of this model allows for fits to the observed X-ray -- radio behaviour of BeXRBs. When fitting the full outburst behaviour of Swift J0243.6+6124, \citet{chatzis2021} find that the sub-Eddington behaviour can be consistently explained via this model for reasonable binary and outflow parameters, as we will return to in Section \ref{sec:biggerpicture}. The super-Eddington properties, on the other hand, cannot be explained in this model, as the inferred shock location is too close to the Be star (i.e. within $\sim 20$ Solar radii).

We can consider whether this new approach could account for the radio emission observed in 1A 0535+262. The most reliable method to assess this question, is to perform a full fit to the new data, similar to \citet{chatzis2021}. As such a fit is beyond the scope of this work, we will instead consider some qualitative lines of thought. There are a number of arguments suggesting that, similarly to the super-Eddington phase of Swift J0243.6+6124, these observations may be challenging to explain via such shocks.

The similarity in $L_X$--$L_R$ coupling between these two sources presents several issues. Firstly, this coupling does not fit with the predicted $\beta = 12/7$ for the sub-Eddington regime, and secondly, no large change in coupling is observed between the super- and sub-Eddington regime. Observing a similar coupling index in those two regimes requires $p \approx 7$, which is inconsistent with typical values for diffusive shock acceleration \citep[i.e. $p\approx2$--$2.2$;][]{bell1978,matthews2020}. Finally, such a single slope is significantly steeper than the observed $\beta = 0.84 \pm 0.06$. However, we reiterate that this inference is based on only two sources. 

Another challenge is that the apparently similar correlation between X-ray and radio luminosity is suggestive of a single underlying mechanism. If the super-Eddington Swift J0243.6+6124 data cannot be explained through such shocks, this line of reasoning would also argue against that origin in 1A 0535+262. However, we reiterate that a full fit will shed more light on this question.

\subsection{A BeXRB X-ray -- radio luminosity correlation: effects of magnetic field and spin?}
\label{sec:disc_BS}

Considering the X-ray -- radio luminosity plane, our observations confirm the inference in \citet{vandeneijnden2018_swj0243} that giant BeXRB outbursts are significantly radio underluminous compared to LMXB outbursts. Such a striking difference makes one naturally wonder about its origin. Remarkably, \citet{ribo17} showed how the only-known black hole BeXRB, MWC 656, falls on the black hole X-ray -- radio luminosity correlation at quiescent X-ray luminosities (where the two black hole tracks have converged). While it is rather speculative to extrapolate from a single source, this may suggest that the radio faintness of BeXRBs does not result from the binary or accretion flow properties. Instead, the compact object properties would then appear to play a more significant role in the low radio luminosity of NS BeXRBs. A similar conclusion follows from a comparison with NS LMXBs. During giant outbursts, the NS may accrete from an accretion disk in a similar fashion to accreting NSs in LMXBs. Therefore, the fundamental difference between BeXRB giant outbursts and LMXB hard states, appears to lie in the NS properties: the strong magnetic field in BeXRB truncates the accretion disk at hundreds to thousands of gravitational radii \citep[e.g.][]{tsygankov17}, compared to maximally a few to tens of $R_g$ in LMXBs
\citep{degenaar17,vandeneijnden2018_igr17379,ludlam16,ludlam17a,ludlam17b}. Moreover, the NSs in BeXRBs typically have orders of magnitude slower spins than those of NSs in LMXBs \citep{reig11,patruno17}.

To quantitatively compare NS jet launching between LMXBs and BeXRBs, we fundamentally assume a single jet launch mechanism underlies this process in both cases. This argues against Blandford-Payne-type jet launch models, i.e. classical magneto-centrifugal jet launch models, as those suggest a maximum NS magnetic field for jet formation that excludes BeXRB NSs \citep[i.e. $\sim 10^9$ G;][]{massi08,kylafis2012}. As an alternative, spin-powered jet launching models do not carry this restriction. From that category of models, we will consider the model proposed by \citet{parfrey16}, although we stress that currently, the main argument for assuming a single model for both source classes comes from Occam's razor instead of direct observational evidence.

In the jet-launching model by \citet{parfrey16}, the jet power is provided by magnetic field lines from the spinning NS, opened up by the accretion flow. Importantly, this model proposes that the jet power $L_J$ scales with three physical parameters: the NS spin period $P$, the NS magnetic field $B$, and the mass accretion rate $\dot{M}$:

\begin{equation}
    L_J \propto P^{-2} B^{6/7} \dot{M}^{4/7}\text{ .}
    \label{eq:L_J}
\end{equation}

\noindent This equation immediately reveals the fundamental difference with magneto-centrifugal models, as the jet power \textit{increases} with magnetic field strength. This may, however, still be reconciled with the radio faintness of strongly-magnetic BeXRBs, as their NSs spin slowly. More recently, \citet{das2022} presented GRMHD simulations of accreting NSs with a complex magnetic field morphology \citep[inspired by e.g.,][]{riley19}, finding the same relation between jet power, magnetic field, and spin period. In the remainder of this discussion, we will consider this class of models as \textit{magneto-rotational} models.

To review these models' scaling with NS and accretion parameters more quantitatively, we can introduce an additional piece of information and an assumption. Firstly, out of the three parameters setting the jet power in Equation \ref{eq:L_J}, only the accretion rate is related to the X-ray luminosity; $L_X \propto \dot{M}$ if we consider a range in mass accretion rate where the accretion flow does not transition between, e.g., radiatively efficient and inefficient. Secondly, we can pose the assumption that the jet power is correlated to jet radio luminosity, for all jet powers relevant to BeXRBs, in similar fashion to black hole jets: $L_R \propto L_J^{1.4}$ \citep{blandford79,markoff01,falcke1996}. With those scalings, Equation \ref{eq:L_J} can be written as

\begin{equation}
    L_R \propto P^{-14/5} B^{6/5} L_X^{4/5}
    \label{eq:L_R}
\end{equation}

\noindent This equation takes the same functional form as Equation \ref{eq:linmodel}, fitted to the $L_X$--$L_R$ relation, with $\beta = 0.8$ and $\xi~\propto~P^{-14/5} B^{6/5}$. With our measurements of both $\beta$ and $\xi$ for BeXRBs and the results from \citet{gallo18} for hard-state NS LMXBs, we can assess whether the magneto-rotational models are able to approximately explain the differences between both classes of accreting NSs.

Firstly, we can consider the full sample of giant BeXRB outbursts studied in this work. For these combined data sets, we measure a coupling index of $\beta = 0.84 \pm 0.06$, consistent with a slope of $\beta = 0.8$. However, as discussed, we should be careful when we apply Equation \ref{eq:L_R} to any sample of accreting NSs. When doing so, we implicitly assume a single spin and magnetic field for all sources within this sample. In other words, we neglect the dependence of the magneto-rotational models on spin and magnetic field. For 1A 0535+262 and J0243.6+6124 individually, we instead measured $\beta = 0.81 \pm 0.27$ and $\beta = 0.65^{+0.17}_{-0.15}$, respectively. Both slopes remain consistent with each other and with the $\beta=0.8$ value from Equation \ref{eq:L_R}.

We can also assess the normalization, and possible effects of the magnetic field and spin, which can be parameterized in this model as $\xi = \xi_0 (P/\text{1 sec})^{-14/5} (B/10^{12}\text{ G})^{6/5}$. For 1A 0535+262, using its measured $\xi$, $P=103$ sec, and $B=5\times10^{12}$ G, we can infer that $\xi_0 = 168.7^{+124.5}_{-62.2}$. For Swift J0243.6+6124, no cyclotron line has been detected \citep{jaisawal2017,jaisawal2019,zhang2019,tao2019}. Instead, we can apply the scaling with spin and magnetic field strength to infer the magnetic field, required to explain the difference between its measured $\log \xi$ and that of 1A 0535+262, given their known difference in spin. Considering the $1-\sigma$ ranges in $\log \xi$ for both sources, one then finds that the magnetic field of Swift J0243.6+6124 should lie between $2\times10^{10}$ G and $1.4\times10^{11}$ G. Indirect estimates of this field have been obtained via pulse frequency evolution modelling and through searches for the transitional propeller X-ray luminosity or the critical X-ray luminosity, yielding contrasting results: while several authors report evidence for a field strength of $B \geq 10^{13}$ G \citep{doroshenko17,vandeneijnden2019_reb,kong2020}, others find $B < 10^{13}$ G \citep{doroshenko2020,sugizaki2020}, and finally some conclude both ranges are possible \citep{wilson18,tsygankov18}. Therefore, with current evidence, we cannot rule out a magnetic field in the range required to explain the difference between 1A 0535+262 and Swift J0243.6+6124 in the magneto-rotational models. However, this range is significantly lower than a subset of estimates for Swift J0243.6+6124 and the magnetic field typically observed in BeXRBs. For such, more typical, BeXRB fields i.e. ($B\geq10^{12}$ G), the magneto-rotational models would have predicted a larger difference between the two targets.

It is worth briefly pointing out that the radio non-detections of two other BeXRBs, GRO J1008-57 and SAX J2103.5+4545, are not surprising in this model interpretation. The former target has similar spin and magnetic field to 1A 0535+262, but a significantly larger distance, while spin period of the latter is more than three times larger than that of 1A 0535+262. As both sources where observed at similar X-ray luminosities as the 1A 0535+262 outburst decay, we expect lower radio flux densities under the assumption of the magneto-rotational models. Combined with their higher radio luminosity limits, the non-detections are therefore consistent with this model.

Finally, we can conduct a similar comparison between NS BeXRB and the full sample of NS LMXBs (ignoring for simplicity the intrinsic variations in the latter sample), for which \citet{gallo18} measure $\log \xi = -0.17 \pm 0.05$. That value, in combination with the measurement of $\xi_0$ from 1A 0535+262, does not fit well with typical spin and magnetic field values assumed or measured for NS LMXBs. For instance, it implies a maximum spin frequency of $\sim 7$ Hz for a $10^{8}$ G magnetic field, an order of magnitude below the typical spin frequencies of accreting millisecond X-ray pulsars (AMXPs). At a field strength of $10^7$ G, at the low end of what is typically invoked for AMXPs, this maximum spin frequency only increases to $\sim 20$ Hz. We therefore find that the magneto-rotational models cannot reproduce both the slope and normalisation of the observed $L_X$--$L_R$ relationship in a consistent fashion for the NS LMXBs and BeXRBs. Instead, the measured difference in radio luminosity normalisation is smaller than predicted, leading to the low inferred spin frequency for NS LMXBs mentioned above\footnote{We note that even for an individual radio-bright AMXP \citep[for instance IGR J17591-2342; see e.g.][]{russell18,gusinskaia20_igrj17591}, the larger difference in $\log \xi$ is not sufficient to be consistent with magneto-rotational models. Even if it were, it would go against our initial assumption underlying the comparison: a single magneto-rotational model holds for \textit{all} NS X-ray binaries}. We conclude that the magneto-rotational models can qualitatively account for the differences seen between the samples of NS LMXBs and BeXRBs, and between the two considered BeXRBs, but currently fails to quantitatively explain these for reasonable spin and magnetic field values.

\subsubsection{Assessing the implicit assumptions}

In the analysis above, we make a number of assumptions to compare BeXRBs and LMXBs. Therefore, it is important to assess whether the difficulty to explain the quantitative differences between these sources in the magneto-rotational models, arises due to these assumptions. We can start by discussing the role of the X-ray luminosity in these calculations. For instance, we assumed that the X-ray luminosity scales in a linear fashion with mass accretion rate across all considered X-ray luminosities, including the super-Eddington ones reached by Swift J0243.6+6124. When considering the radio-detected BeXRBs individually, we find their $L_X$--$L_R$ slopes to be consistent despite the different (but overlapping) ranges in $L_X$ they span, as expected in this scenario. However, this does not imply that the same inflow-outflow coupling necessarily operates in the sub- and super-Eddington regime. Especially given the small range in X-ray luminosity and uncertainties on $\beta$, that conclusion cannot be made.

Another possible issue with the X-ray luminosity, as mentioned by \citet{chatzis2021}, may be the relatively low contribution of the accretion flow to the total X-ray emission. If the emission is dominated by the accretion column, we may need to consider instead only a fraction of the X-ray luminosity as input for Equation \ref{eq:L_R}. If this fraction is independent of the total luminosity, such a change only affects the normalisation, increasing the inferred value of $\xi_0$. As the accretion column emission is only expected to play such a significant role in BeXRBs, and not LMXBs\footnote{Fractional variabilities in accreting millisecond pulsars are typically of the order of a few per cent or less \citep{patruno18}.}, an increased value of $\xi_0$ exacerbates the issue that we measure a smaller normalisation difference between these source classes than expected in the magneto-rotational models. We note, on the other hand, that a substantial fraction of X-ray lumionosity of LMXBs can originate from a boundary layer, which is not present in BeXRBs. Moreover, both the accretion column and boundary layer luminosity, fundamentally, scale with accretion rate. A systematic exploration of these X-ray spectral decompositions on the tracks of NSs in the $L_X$--$L_R$ plane would help disentangle these effects.

Another assumption, especially relevant for 1A 0535+262 and NS LMXBs without a spectral index measurements, is that the observed radio fluxes can be extrapolated to $5$ GHz radio luminosities without loss of information. Changes in spectral index are likely occurring throughout outbursts of most sources, based on the monitoring of sources where spectral index measurements were made \citep{vandeneijnden2019_reb,gusinskaia20,russell2021}. By ignoring or not measuring such changes, the radio flux density to luminosity conversion will introduce scatter into the relationship between X-ray and radio luminosity, and possible affect its slope. In addition, using the $5$-GHz radio luminosity ingores changes in spectral break frequency and the optically thin slope, which strongly affect the total jet power \citep{russell14}.

\subsection{The radio properties of BeXRBs in the context of all X-ray binaries}

\label{sec:biggerpicture}

\begin{figure*}
    \centering
    \includegraphics[width=\textwidth]{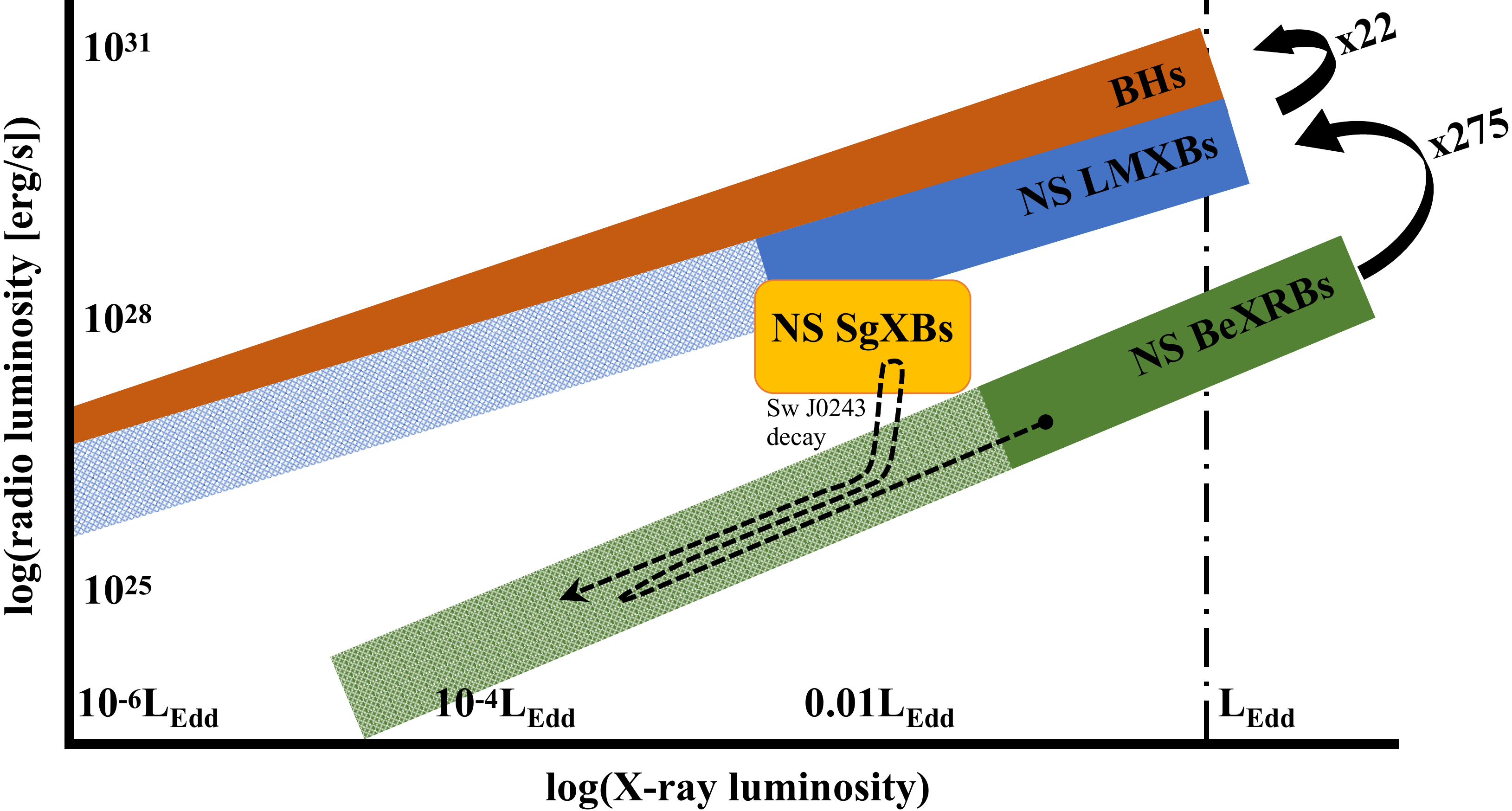}
    \caption{A schematic overview of X-ray binaries in the X-ray -- radio luminosity diagram. For each region, the shaded region indicates where a correlation has been extrapolated; these correlations have been observed over the range indicated by the filled regions. For the black hole systems, we treat the population as a single category and do not show a radio-loud and radio-quiet track. For the NS SgXBs, we only indicate the location of radio-detected sources. However, several sources in this class are not detected with radio upper limits below this region. The dashed line shows the simplest trajectory of Swift J0243.6+6124 during its outburst decay, consistent with radio detections and upper limits. However, the shape of this track when the source is not detected, may differ. Finally, the factors $22$ and $275$ shown on the right refer to the difference in normalisation of the X-ray --  radio correlations, measured at $L_X = 2\times10^{36}$ erg/s (i.e. $0.01$ $L_{\rm Edd}$, where $L_{\rm Edd}$ is the Eddington luminosity of a NS). See the main text for all caveats.}
    \label{fig:schematic}
\end{figure*}

Having focused on the radio behaviour of BeXRBs in the previous two sections, we will now turn to a comparison with the broader class of X-ray binaries. Based on the assumption that giant BeXRB outbursts show a single $L_X$--$L_R$ correlation, we have drawn a schematic to summarize the $L_X$--$L_R$ plane for various types of X-ray binaries in Figure \ref{fig:schematic}. In this Figure, the solid regions indicate measured correlations between the two luminosities, while the dashed regions indicate extrapolated behaviour. The hard state black hole systems, regardless of donor mass, are radio-brightest \citep{fender01,migliari06}; in this schematic, we follow \citet{gallo18} and treat the entire black hole population as one and do not distinguish a radio-loud and radio-quiet track \citep[see e.g.][for more discussion]{soleri11,gallo2014,dincer14,meyer14,drappeau15}. Different types of NS X-ray binaries populate different regions in this diagram: combined into one class, the low-mass systems approximately trace the black hole correlation with similar coupling index \citep{gallo18}, although the sample's radio luminosity normalisation at $2\times10^{36}$ erg /s (i.e. $0.01L_{\rm Edd}$, where $L_{\rm Edd}$ is defined assuming a NS) is a factor $\sim 22$ lower and individual sources can show deviating behavior \citep[e.g.][]{migliari06,gusinskaia20}. 

As argued in this paper, the NS BeXRBs are even radio fainter, by a further factor of $\sim 275$ at $0.01L_{\rm Edd}$ compared to the NS LMXBs, while showing a slightly steeper index. These two NS classes show a large range of extrapolation, particularly at low X-ray luminosity; systematic radio detections have only been obtained down to $\sim 0.01L_{\rm Edd}$ for NS LMXBs, while this limit is $\sim 0.03L_{\rm Edd}$ for the NS BeXRBs. Therefore, we make the simplest assumption of a single powerlaw coupling down to low luminosities, although this remains to be confirmed observationally. The latter class does extend to significantly super-Eddington luminosities, due to the inclusion of Swift J0243.6+6124. Finally, we note that the hard boundaries drawn between classes in Figure \ref{fig:schematic} are, in reality, not clear-cut. Variations between individual sources mean that, e.g., compact object type cannot be determined beyond doubt from the position in this diagram. 

This simplistic picture is complicated by the inclusion of two additional pieces of information. Firstly, the outburst decay and especially re-flares of Swift J0243.6+6124 are significantly radio-brighter than the extrapolated NS BeXRB correlation, as indicated by the dashed arrow in the schematic. We stress that the exact path of this source during its outburst has not been fully constrained, due to several radio non-detections -- the dashed arrow shows the simplest route consistent with the observations. Secondly, the NS supergiant X-ray binaries (SgXRBs), where a NS in a tight orbit persistently accretes from the strong stellar wind of a supergiant donor star, are wedged in between the two aforementioned NS correlations \citep{vandeneijnden2021}. An important caveat to that statement is, however, that this only holds for the radio-detected NS SgXBs -- several of such sources are not radio detected and fall below the yellow area \citep{vandeneijnden2021}. This complicating behaviour of Swift J0243.6+6124 and the NS SgXBs, both radio-bright compared to the NS BeXRB correlation, is strongly suggestive of an additional radio-emission mechanism. This raises the question `What could such a mechanism be?'

For the radio behaviour of Swift J0243.6+6124 during its outburst decay and X-ray re-flares, \citet{vandeneijnden2019_reb} suggested a two-fold explanation. The initial radio flaring might have originated in large-scale shocks, as jet material interacts with the ISM, while the radio properties during the X-ray re-flare could represent a rapidly re-establishing jet. While the latter scenario proposes a jet that would be, in terms of radio luminosity, remarkably similar to NS LMXBs, its inferred similarity to the jet observed in the super-Eddington outburst phase remained puzzling. The possible presence of an ultra-fast disk outflow during the super-Eddington phases has been proposed to play a role in regulating the maximum radio luminosity of the super-Eddington jet; however, such an explanation is quite speculative \citep{vandeneijnden2019_chandra}.

An alternative answer may therefore lie in the model by \citet{chatzis2021} discussed previously, which is able to describe these sub-Eddington radio observations satisfactorily. A challenge for this model would be, then, to explain the launch of a (roughly) spherical outflow from the accretion disk at X-ray luminosities between the super-Eddington and propeller regimes. In addition, if this shock model indeed explains the outlying behaviour of Swift J0243.6+6124, it should similarly predict no or fainter radio emission in the two considered outbursts of GRO J1008-57, the giant outburst of SAX J2103.5+4545, and in the late giant outburst decay of 1A 0535+262. As the particle acceleration and shock emission properties in this model depend heavily on the system's geometry (i.e. orbital separation and viewing angle) and wind properties (mass loss rate and velocity), such differences may be expected. For instance, the wind properties of Be stars are very poorly constrained in BeXRBs, and could differ strongly, causing differences in the location and energetics of a shock. Another factor to consider is the difference between these transient states. An X-ray re-flare of Swift J0243.6+6124, Type-I outburst of GRO J1008-57, and giant outburst of 1A 0535+262 or SAX J2103.5+4545 do not necessarily respresent the exact same accretion flow state despite similar $L_X$. 

Turning briefly to the NS SgXBs, a logical next question is then whether a stellar wind or the \citet{chatzis2021} shock model could be the inferred additional radio emission process. As discussed in detail in \citet{vandeneijnden2021}, thermal stellar wind emission may play a role in a subset of targets. However, it is not expected to be the driving factor of this enhanced radio luminosity, as not all radio-detected NS SgXBs launch a stellar wind capable of explaining the radio emission, while some non-detected targets should have been detected in this scenario. Several lines of reasoning also argue against the model by \citet{chatzis2021}. Firstly, it is unclear whether an accretion disk, capable of launching one of the two shocking outflows, is present in all NS SgXBs \citep[see e.g.][for a recent discussion]{elmellah2019}. Secondly, the stellar wind of the massive star is significantly denser than those in BeXRBs, which makes it unlikely that emission from a shock deep in the massive stellar wind can be observed. However, the fundamental idea of this model -- shocks occur between the stellar wind and some other structure, causing the acceleration of relativistic electrons -- may still contribute. For instance, the presence of large scale accretion and photo-ionisation wakes in some NS SgXBs \citep{blondin1991,kaper1994} may provide sites for shocks with the stellar wind to develop on larger physical scales that are less affected by effects suppressing the radio emission \citep[i.e. free-free absorption and the Razin effect;][]{hornby1966}. A more detailed model, as well as further observations of radio NS SgXBs and a better understanding of the circumstances (e.g. binary and stellar wind properties) where accretion and photo-ionisation wakes are formed, are necessary to further consider such a scenario.

With regards to the SgXBs, we will make two final comments. Firstly, in the above discussion, we have assumed that strongly-magnetized NSs in SgXBs are equally capable of launching jets as NS BeXRBs, and would do so via the same mechanism. While that may be a reasonable assumption in terms of the NS properties, the accretion flow itself differs significantly between these two source classes. For instance, if a smaller disk, or no disk at all, is present in a NS SgXB, magneto-rotational models may or may not operate. However, whether that predicts a lower radio luminosity, or instead allows for another (possibly radio-brighter) jet launch mechanism to take over, cannot be determined without adjusting strong-B jet launch models for spherical accretion flows or focused winds. Secondly, the above discussion regarding additional radio emission mechanisms, especially shocks, does not require a NS primary. However, for systems with BH primaries, such as Cyg X-1, any resulting radio emission is significantly fainter than the jet and would be virtually undetectable; for the radio-detected BH system MWC 656, on the other hand, this scenario does not apply, as it hosts a Be-star instead of supergiant donor.

\subsection{Future Galactic and extragalactic prospects}

In our own Galaxy, the advent of the next-generation VLA (ngVLA), as well as the SKA and SKA precursors in the Southern hemisphere, with their enhanced sensitivity, would greatly extend the range of X-ray luminosity and distances where BeXRB radio emission and jets may be probed with observed of reasonable length (see the blue lines in Figure \ref{fig:lxlr_fitted}). Given the typical range of radio luminosities observed in isolated Be stars, such future observations may probe down to the regime where this emission cannot simply be ignored. For instance, for sources within $\sim 2$ kpc, a one-hour ngVLA observation is sensitive down to $\sim 2\times10^{25}$ erg/s, reaching far into the range of isolated Be star radio emission (see Section \ref{sec:whatdowesee}). Not all Be stars are, however, detected at radio frequencies. Therefore, with coordinated X-ray observations and sufficiently dense radio monitoring, transient radio emission could still be tracked down to low accretion rates for those BeXRBs hosting radio-faint Be stars.

In this work, we present evidence for the existence of an X-ray -- radio luminosity coupling for BeXRBs. If we assume that this holds more generally for strongly-magnetized accreting NSs, we can use this correlation to briefly move focus to ultra-luminous X-ray sources (ULXs). ULXs are extragalactic X-ray sources with X-ray luminosities exceeding the Eddington luminosity of a $\sim 10$ $M_{\odot}$ black hole \citep[i.e. $\sim 10^{39}$ erg/s;][]{kaaret17}. While the exact nature of ULX compact objects long remained unclear, with both stellar-mass compact objects and intermediate mass black holes considered as options, the detection of pulsations from multiple ULXs \citep{bachetti14,furst16b,israel17b} has unambiguously shown that at least a fraction of them host accreting NSs. The exact fraction remains unknown, although both observational \citep{walton18c} and theoretical considerations \citep{king16_ulx} are consistent with a significant proportion. ULXs also show evidence for outflows, both through the detection of resolved (feedback) structures \citep{pakull2003,kaaret03} and X-ray absorption lines from ultra-fast outflows \citep{pinto16}. However, unresolved radio counterparts from compact jets have not been detected unambiguously from ULX pulsars \citep{mezcua15,cseh15a,kaaret17}.

The identified NSs in ULXs rotate slowly, similar to their strongly-magnetized Galactic counterparts. Extending our suggested BeXRB X-ray--radio luminosity relation to a typical ULX luminosity, one might ask what the prospects are for detecting radio point source emission? If we assume an X-ray luminosity of $\sim 10^{41}$ erg/s \citep[on the high end of their luminosity distribution;][]{kaaret17}, the predicted radio luminosity would be of the order $\sim 10^{30}$ erg/s. At a Mpc distance and $5$ GHz, this is equivalent to a $0.2$ $\mu$Jy flux density. Such depths are out of reach for any current facilities in reasonable observing times, but are approached by the planned ngVLA sensitivity \citep[$\sim 0.23$ $\mu$Jy at 8 GHz in 1 hour of observing time;][]{selina18}. At such depths, confusion limits, host galaxy emission, diffuse feedback structures and other extended, close-by sources may complicate any searches for radio point source emission (especially at low frequencies). However, given the $>3$ orders of magnitude difference in radio luminosity normalisation compared to the black hole systems, the detection and flux density of radio emission may help to understand the nature of the compact object accretor.

\section{Acknowledgments}

The authors thank the referee for a constructive report. JvdE is supported by a Lee Hysan Junior Research Fellowship awarded by St. Hilda's College. TDR acknowledges financial contribution from the agreement ASI-INAF n.2017-14-H.0. GRS is supported by NSERC Discovery Grants RGPIN-2016-06569 and RGPIN-2021-04001. The authors acknowledge the use of public data from the Swift data archive. This research has made use of MAXI data provided by RIKEN, JAXA and the MAXI team. The National Radio Astronomy Observatory is a facility of the National Science Foundation operated under cooperative agreement by Associated Universities, Inc. The Australia Telescope Compact Array is part of the Australia Telescope National Facility which is funded by the Australian Government for operation as a National Facility managed by CSIRO. We acknowledge the Gomeroi people as the traditional owners of the ATCA observatory site. This research has made use of data and software provided by the High Energy Astrophysics Science Archive Research Center (HEASARC) and NASA's Astrophysics Data System Bibliographic Services. This work has made use of data from the European Space Agency (ESA) mission {\it Gaia} (\url{https://www.cosmos.esa.int/gaia}), processed by the {\it Gaia} Data Processing and Analysis Consortium (DPAC, \url{https://www.cosmos.esa.int/web/gaia/dpac/consortium}). Funding for the DPAC has been provided by national institutions, in particular the institutions participating in the {\it Gaia} Multilateral Agreement.

\section*{Data Availability}

All radio observations can be accessed via the VLA Data Archive (\url{https://archive.nrao.edu/archive/archiveproject.jsp}) or the Australia Telescope Online Archive (\url{https://atoa.atnf.csiro.au}). Relevant project codes are listed in the Online Supplementary Materials. All X-ray data is publicly available via the HEASARC (pointed observations) or \textit{Swift}/BAT Hard X-ray Transient Monitor (\url{https://swift.gsfc.nasa.gov/results/transients/}). A Jupyter notebook reproducing the Figures and \textsc{LinMix} fits in this paper, will be publically accessible upon acceptance and publication via \url{https://github.com/jvandeneijnden/LxLrCouplingInBeXRBs}.



\newcommand{\noop}[1]{}
\begin{thebibliography}{}
\makeatletter
\relax
\def\mn@urlcharsother{\let\do\@makeother \do\$\do\&\do\#\do\^\do\_\do\%\do\~}
\def\mn@doi{\begingroup\mn@urlcharsother \@ifnextchar [ {\mn@doi@}
  {\mn@doi@[]}}
\def\mn@doi@[#1]#2{\def\@tempa{#1}\ifx\@tempa\@empty \href
  {http://dx.doi.org/#2} {doi:#2}\else \href {http://dx.doi.org/#2} {#1}\fi
  \endgroup}
\def\mn@eprint#1#2{\mn@eprint@#1:#2::\@nil}
\def\mn@eprint@arXiv#1{\href {http://arxiv.org/abs/#1} {{\tt arXiv:#1}}}
\def\mn@eprint@dblp#1{\href {http://dblp.uni-trier.de/rec/bibtex/#1.xml}
  {dblp:#1}}
\def\mn@eprint@#1:#2:#3:#4\@nil{\def\@tempa {#1}\def\@tempb {#2}\def\@tempc
  {#3}\ifx \@tempc \@empty \let \@tempc \@tempb \let \@tempb \@tempa \fi \ifx
  \@tempb \@empty \def\@tempb {arXiv}\fi \@ifundefined
  {mn@eprint@\@tempb}{\@tempb:\@tempc}{\expandafter \expandafter \csname
  mn@eprint@\@tempb\endcsname \expandafter{\@tempc}}}

\bibitem[\protect\citeauthoryear{{Ambrosino} et~al.,}{{Ambrosino}
  et~al.}{2021}]{ambrosino2021}
{Ambrosino} F.,  et~al., 2021, \mn@doi [Nature Astronomy]
  {10.1038/s41550-021-01308-0}, \href
  {https://ui.adsabs.harvard.edu/abs/2021NatAs...5..552A} {5, 552}

\bibitem[\protect\citeauthoryear{{Arnason}, {Papei}, {Barmby}, {Bahramian}  \&
  {Gorski}}{{Arnason} et~al.}{2021}]{arnason2021}
{Arnason} R.~M.,  {Papei} H.,  {Barmby} P.,  {Bahramian} A.,   {Gorski} M.~D.,
  2021, \mn@doi [\mnras] {10.1093/mnras/stab345}, \href
  {https://ui.adsabs.harvard.edu/abs/2021MNRAS.502.5455A} {502, 5455}

\bibitem[\protect\citeauthoryear{{Atri} et~al.,}{{Atri}
  et~al.}{2019}]{atri2019}
{Atri} P.,  et~al., 2019, \mn@doi [\mnras] {10.1093/mnras/stz2335}, \href
  {https://ui.adsabs.harvard.edu/abs/2019MNRAS.489.3116A} {489, 3116}

\bibitem[\protect\citeauthoryear{{Bachetti} et~al.,}{{Bachetti}
  et~al.}{2014}]{bachetti14}
{Bachetti} M.,  et~al., 2014, \mn@doi [\nat] {10.1038/nature13791}, \href
  {http://cdsads.u-strasbg.fr/abs/2014Natur.514..202B} {514, 202}

\bibitem[\protect\citeauthoryear{{Bailer-Jones}, {Rybizki}, {Fouesneau},
  {Mantelet}  \& {Andrae}}{{Bailer-Jones} et~al.}{2018}]{bailerjones18}
{Bailer-Jones} C.~A.~L.,  {Rybizki} J.,  {Fouesneau} M.,  {Mantelet} G.,
  {Andrae} R.,  2018, \mn@doi [\aj] {10.3847/1538-3881/aacb21}, \href
  {https://ui.adsabs.harvard.edu/abs/2018AJ....156...58B} {156, 58}

\bibitem[\protect\citeauthoryear{{Bailer-Jones}, {Rybizki}, {Fouesneau},
  {Demleitner}  \& {Andrae}}{{Bailer-Jones} et~al.}{2020}]{bailerjones2020}
{Bailer-Jones} C.~A.~L.,  {Rybizki} J.,  {Fouesneau} M.,  {Demleitner} M.,
  {Andrae} R.,  2020, arXiv e-prints, \href
  {https://ui.adsabs.harvard.edu/abs/2020arXiv201205220B} {p. arXiv:2012.05220}

\bibitem[\protect\citeauthoryear{{Barthelmy} et~al.,}{{Barthelmy}
  et~al.}{2005}]{barthelmy05}
{Barthelmy} S.~D.,  et~al., 2005, \mn@doi [\ssr] {10.1007/s11214-005-5096-3},
  \href {https://ui.adsabs.harvard.edu/abs/2005SSRv..120..143B} {120, 143}

\bibitem[\protect\citeauthoryear{{Baykal}, {Inam}, {Stark}, {Heffner}, {Erkoca}
   \& {Swank}}{{Baykal} et~al.}{2007}]{baykal2007}
{Baykal} A.,  {Inam} S.~{\c{C}}.,  {Stark} M.~J.,  {Heffner} C.~M.,  {Erkoca}
  A.~E.,   {Swank} J.~H.,  2007, \mn@doi [\mnras]
  {10.1111/j.1365-2966.2006.11231.x}, \href
  {https://ui.adsabs.harvard.edu/abs/2007MNRAS.374.1108B} {374, 1108}

\bibitem[\protect\citeauthoryear{{Bell}}{{Bell}}{1978}]{bell1978}
{Bell} A.~R.,  1978, \mn@doi [\mnras] {10.1093/mnras/182.2.147}, \href
  {https://ui.adsabs.harvard.edu/abs/1978MNRAS.182..147B} {182, 147}

\bibitem[\protect\citeauthoryear{{Blandford} \& {K{\"o}nigl}}{{Blandford} \&
  {K{\"o}nigl}}{1979}]{blandford79}
{Blandford} R.~D.,  {K{\"o}nigl} A.,  1979, \mn@doi [\apj] {10.1086/157262},
  \href {https://ui.adsabs.harvard.edu/abs/1979ApJ...232...34B} {232, 34}

\bibitem[\protect\citeauthoryear{{Blondin}, {Stevens}  \& {Kallman}}{{Blondin}
  et~al.}{1991}]{blondin1991}
{Blondin} J.~M.,  {Stevens} I.~R.,   {Kallman} T.~R.,  1991, \mn@doi [\apj]
  {10.1086/169934}, \href
  {https://ui.adsabs.harvard.edu/abs/1991ApJ...371..684B} {371, 684}

\bibitem[\protect\citeauthoryear{{Bozzo}, {Falanga}  \& {Stella}}{{Bozzo}
  et~al.}{2008}]{bozzo2008}
{Bozzo} E.,  {Falanga} M.,   {Stella} L.,  2008, \mn@doi [\apj]
  {10.1086/589990}, \href
  {https://ui.adsabs.harvard.edu/abs/2008ApJ...683.1031B} {683, 1031}

\bibitem[\protect\citeauthoryear{{Brumback}, {Hickox}, {F{\"u}rst},
  {Pottschmidt}, {Hemphill}, {Tomsick}, {Wilms}  \& {Ballhausen}}{{Brumback}
  et~al.}{2018}]{brumback2018}
{Brumback} M.~C.,  {Hickox} R.~C.,  {F{\"u}rst} F.~S.,  {Pottschmidt} K.,
  {Hemphill} P.,  {Tomsick} J.~A.,  {Wilms} J.,   {Ballhausen} R.,  2018,
  \mn@doi [\apj] {10.3847/1538-4357/aa9e91}, \href
  {https://ui.adsabs.harvard.edu/abs/2018ApJ...852..132B} {852, 132}

\bibitem[\protect\citeauthoryear{{Burrows} et~al.,}{{Burrows}
  et~al.}{2004}]{burrows04}
{Burrows} D.~N.,  et~al., 2004, in {Flanagan} K.~A.,  {Siegmund} O.~H.~W.,
  eds,  Proceedings of the SPIE Vol. 5165, X-Ray and Gamma-Ray Instrumentation
  for Astronomy XIII. pp 201--216, \mn@doi{10.1117/12.504868}

\bibitem[\protect\citeauthoryear{{Camero Arranz}, {Wilson}, {Finger}  \&
  {Reglero}}{{Camero Arranz} et~al.}{2007}]{camero2007}
{Camero Arranz} A.,  {Wilson} C.~A.,  {Finger} M.~H.,   {Reglero} V.,  2007,
  \mn@doi [\aap] {10.1051/0004-6361:20077398}, \href
  {https://ui.adsabs.harvard.edu/abs/2007A&A...473..551C} {473, 551}

\bibitem[\protect\citeauthoryear{{Casares}, {Negueruela}, {Rib{\'o}}, {Ribas},
  {Paredes}, {Herrero}  \& {Sim{\'o}n-D{\'\i}az}}{{Casares}
  et~al.}{2014}]{casares2014}
{Casares} J.,  {Negueruela} I.,  {Rib{\'o}} M.,  {Ribas} I.,  {Paredes} J.~M.,
  {Herrero} A.,   {Sim{\'o}n-D{\'\i}az} S.,  2014, \mn@doi [\nat]
  {10.1038/nature12916}, \href
  {https://ui.adsabs.harvard.edu/abs/2014Natur.505..378C} {505, 378}

\bibitem[\protect\citeauthoryear{{Cash}}{{Cash}}{1979}]{cash1979}
{Cash} W.,  1979, \mn@doi [\apj] {10.1086/156922}, \href
  {http://adsabs.harvard.edu/abs/1979ApJ...228..939C} {228, 939}

\bibitem[\protect\citeauthoryear{{Chatzis}, {Petropoulou}  \&
  {Vasilopoulos}}{{Chatzis} et~al.}{2021}]{chatzis2021}
{Chatzis} M.,  {Petropoulou} M.,   {Vasilopoulos} G.,  2021, \mn@doi [\mnras]
  {10.1093/mnras/stab3098}, \href
  {https://ui.adsabs.harvard.edu/abs/2021MNRAS.tmp.2817C} {}

\bibitem[\protect\citeauthoryear{{Clark}, {Steele}  \& {Fender}}{{Clark}
  et~al.}{1998}]{clark1998}
{Clark} J.~S.,  {Steele} I.~A.,   {Fender} R.~P.,  1998, \mn@doi [\mnras]
  {10.1046/j.1365-8711.1998.01848.x}, \href
  {https://ui.adsabs.harvard.edu/abs/1998MNRAS.299.1119C} {299, 1119}

\bibitem[\protect\citeauthoryear{{Coe} et~al.,}{{Coe} et~al.}{1994}]{coe1994}
{Coe} M.~J.,  et~al., 1994, \mn@doi [\mnras] {10.1093/mnras/270.1.L57}, \href
  {https://ui.adsabs.harvard.edu/abs/1994MNRAS.270L..57C} {270, L57}

\bibitem[\protect\citeauthoryear{{Corbel}, {Fender}, {Tzioumis}, {Nowak},
  {McIntyre}, {Durouchoux}  \& {Sood}}{{Corbel} et~al.}{2000}]{corbel00}
{Corbel} S.,  {Fender} R.~P.,  {Tzioumis} A.~K.,  {Nowak} M.,  {McIntyre} V.,
  {Durouchoux} P.,   {Sood} R.,  2000, \aap, \href
  {https://ui.adsabs.harvard.edu/abs/2000A&A...359..251C} {359, 251}

\bibitem[\protect\citeauthoryear{{Corbel}, {Nowak}, {Fender}, {Tzioumis}  \&
  {Markoff}}{{Corbel} et~al.}{2003}]{corbel03}
{Corbel} S.,  {Nowak} M.~A.,  {Fender} R.~P.,  {Tzioumis} A.~K.,   {Markoff}
  S.,  2003, \mn@doi [\aap] {10.1051/0004-6361:20030090}, \href
  {http://cdsads.u-strasbg.fr/abs/2003A%26A...400.1007C} {400, 1007}

\bibitem[\protect\citeauthoryear{{Corbel}, {Coriat}, {Brocksopp}, {Tzioumis},
  {Fender}, {Tomsick}, {Buxton}  \& {Bailyn}}{{Corbel} et~al.}{2013}]{corbel13}
{Corbel} S.,  {Coriat} M.,  {Brocksopp} C.,  {Tzioumis} A.~K.,  {Fender} R.~P.,
   {Tomsick} J.~A.,  {Buxton} M.~M.,   {Bailyn} C.~D.,  2013, \mn@doi [\mnras]
  {10.1093/mnras/sts215}, \href
  {https://ui.adsabs.harvard.edu/abs/2013MNRAS.428.2500C} {428, 2500}

\bibitem[\protect\citeauthoryear{{Corbet}}{{Corbet}}{1986}]{corbet86}
{Corbet} R.~H.~D.,  1986, \mn@doi [\mnras] {10.1093/mnras/220.4.1047}, \href
  {http://cdsads.u-strasbg.fr/abs/1986MNRAS.220.1047C} {220, 1047}

\bibitem[\protect\citeauthoryear{{Cseh} et~al.,}{{Cseh} et~al.}{2015}]{cseh15a}
{Cseh} D.,  et~al., 2015, \mn@doi [\mnras] {10.1093/mnras/stu2363}, \href
  {http://cdsads.u-strasbg.fr/abs/2015MNRAS.446.3268C} {446, 3268}

\bibitem[\protect\citeauthoryear{{Das}, {Porth}  \& {Watts}}{{Das}
  et~al.}{2022}]{das2022}
{Das} P.,  {Porth} O.,   {Watts} A.,  2022, arXiv e-prints, \href
  {https://ui.adsabs.harvard.edu/abs/2022arXiv220400249D} {p. arXiv:2204.00249}

\bibitem[\protect\citeauthoryear{{Degenaar}, {Pinto}, {Miller}, {Wijnands},
  {Altamirano}, {Paerels}, {Fabian}  \& {Chakrabarty}}{{Degenaar}
  et~al.}{2017}]{degenaar17}
{Degenaar} N.,  {Pinto} C.,  {Miller} J.~M.,  {Wijnands} R.,  {Altamirano} D.,
  {Paerels} F.,  {Fabian} A.~C.,   {Chakrabarty} D.,  2017, \mn@doi [\mnras]
  {10.1093/mnras/stw2355}, \href
  {http://adsabs.harvard.edu/abs/2017MNRAS.464..398D} {464, 398}

\bibitem[\protect\citeauthoryear{{Din{\c{c}}er}, {Kalemci}, {Tomsick}, {Buxton}
   \& {Bailyn}}{{Din{\c{c}}er} et~al.}{2014}]{dincer14}
{Din{\c{c}}er} T.,  {Kalemci} E.,  {Tomsick} J.~A.,  {Buxton} M.~M.,   {Bailyn}
  C.~D.,  2014, \mn@doi [\apj] {10.1088/0004-637X/795/1/74}, \href
  {https://ui.adsabs.harvard.edu/abs/2014ApJ...795...74D} {795, 74}

\bibitem[\protect\citeauthoryear{{Doroshenko}, {Tsygankov}, {Mushtukov},
  {Lutovinov}, {Santangelo}, {Suleimanov}  \& {Poutanen}}{{Doroshenko}
  et~al.}{2017}]{doroshenko17}
{Doroshenko} V.,  {Tsygankov} S.~S.,  {Mushtukov} A. e.~A.,  {Lutovinov} A.~A.,
   {Santangelo} A.,  {Suleimanov} V.~F.,   {Poutanen} J.,  2017, \mn@doi
  [\mnras] {10.1093/mnras/stw3236}, \href
  {https://ui.adsabs.harvard.edu/abs/2017MNRAS.466.2143D} {466, 2143}

\bibitem[\protect\citeauthoryear{{Doroshenko} et~al.,}{{Doroshenko}
  et~al.}{2020}]{doroshenko2020}
{Doroshenko} V.,  et~al., 2020, \mn@doi [\mnras] {10.1093/mnras/stz2879}, \href
  {https://ui.adsabs.harvard.edu/abs/2020MNRAS.491.1857D} {491, 1857}

\bibitem[\protect\citeauthoryear{{Dougherty}, {Taylor}  \&
  {Waters}}{{Dougherty} et~al.}{1991}]{dougherty1991}
{Dougherty} S.~M.,  {Taylor} A.~R.,   {Waters} L.~B.~F.~M.,  1991, \aap, \href
  {https://ui.adsabs.harvard.edu/abs/1991A&A...248..175D} {248, 175}

\bibitem[\protect\citeauthoryear{{Drake}}{{Drake}}{1990}]{drake1990}
{Drake} S.~A.,  1990, \mn@doi [\aj] {10.1086/115541}, \href
  {https://ui.adsabs.harvard.edu/abs/1990AJ....100..572D} {100, 572}

\bibitem[\protect\citeauthoryear{{Drappeau}, {Malzac}, {Belmont}, {Gandhi}  \&
  {Corbel}}{{Drappeau} et~al.}{2015}]{drappeau15}
{Drappeau} S.,  {Malzac} J.,  {Belmont} R.,  {Gandhi} P.,   {Corbel} S.,  2015,
  \mn@doi [\mnras] {10.1093/mnras/stu2711}, \href
  {https://ui.adsabs.harvard.edu/abs/2015MNRAS.447.3832D} {447, 3832}

\bibitem[\protect\citeauthoryear{{El Mellah}, {Sander}, {Sundqvist}  \&
  {Keppens}}{{El Mellah} et~al.}{2019}]{elmellah2019}
{El Mellah} I.,  {Sander} A.~A.~C.,  {Sundqvist} J.~O.,   {Keppens} R.,  2019,
  \mn@doi [\aap] {10.1051/0004-6361/201834498}, \href
  {https://ui.adsabs.harvard.edu/abs/2019A&A...622A.189E} {622, A189}

\bibitem[\protect\citeauthoryear{{Espinasse} \& {Fender}}{{Espinasse} \&
  {Fender}}{2018}]{espinasse18}
{Espinasse} M.,  {Fender} R.,  2018, \mn@doi [\mnras] {10.1093/mnras/stx2467},
  \href {https://ui.adsabs.harvard.edu/abs/2018MNRAS.473.4122E} {473, 4122}

\bibitem[\protect\citeauthoryear{{Evans} et~al.,}{{Evans}
  et~al.}{2009}]{evans09}
{Evans} P.~A.,  et~al., 2009, \mn@doi [\mnras]
  {10.1111/j.1365-2966.2009.14913.x}, \href
  {https://ui.adsabs.harvard.edu/abs/2009MNRAS.397.1177E} {397, 1177}

\bibitem[\protect\citeauthoryear{{Fabrika}}{{Fabrika}}{2004}]{fabrika04}
{Fabrika} S.,  2004, Astrophysics and Space Physics Reviews, \href
  {http://cdsads.u-strasbg.fr/abs/2004ASPRv..12....1F} {12, 1}

\bibitem[\protect\citeauthoryear{{Falcke} \& {Biermann}}{{Falcke} \&
  {Biermann}}{1996}]{falcke1996}
{Falcke} H.,  {Biermann} P.~L.,  1996, \aap, \href
  {https://ui.adsabs.harvard.edu/abs/1996A&A...308..321F} {308, 321}

\bibitem[\protect\citeauthoryear{{Fender} \& {Kuulkers}}{{Fender} \&
  {Kuulkers}}{2001}]{fender01}
{Fender} R.~P.,  {Kuulkers} E.,  2001, \mn@doi [\mnras]
  {10.1046/j.1365-8711.2001.04345.x}, \href
  {https://ui.adsabs.harvard.edu/abs/2001MNRAS.324..923F} {324, 923}

\bibitem[\protect\citeauthoryear{{Fender}, {Belloni}  \& {Gallo}}{{Fender}
  et~al.}{2004}]{fender04}
{Fender} R.~P.,  {Belloni} T.~M.,   {Gallo} E.,  2004, \mn@doi [\mnras]
  {10.1111/j.1365-2966.2004.08384.x}, \href
  {http://cdsads.u-strasbg.fr/abs/2004MNRAS.355.1105F} {355, 1105}

\bibitem[\protect\citeauthoryear{{Finger}, {Cominsky}, {Wilson}, {Harmon}  \&
  {Fishman}}{{Finger} et~al.}{1994}]{finger1994}
{Finger} M.~H.,  {Cominsky} L.~R.,  {Wilson} R.~B.,  {Harmon} B.~A.,
  {Fishman} G.~J.,  1994, in {Holt} S.,  {Day} C.~S.,  eds,  American Institute
  of Physics Conference Series Vol. 308, The Evolution of X-ray Binariese.
  p.~459, \mn@doi{10.1063/1.46032}

\bibitem[\protect\citeauthoryear{{F{\"u}rst} et~al.,}{{F{\"u}rst}
  et~al.}{2016}]{furst16b}
{F{\"u}rst} F.,  et~al., 2016, \mn@doi [\apjl] {10.3847/2041-8205/831/2/L14},
  \href {http://cdsads.u-strasbg.fr/abs/2016ApJ...831L..14F} {831, L14}

\bibitem[\protect\citeauthoryear{{Gallo}, {Fender}  \& {Pooley}}{{Gallo}
  et~al.}{2003}]{gallo03}
{Gallo} E.,  {Fender} R.~P.,   {Pooley} G.~G.,  2003, \mn@doi [\mnras]
  {10.1046/j.1365-8711.2003.06791.x}, \href
  {http://adsabs.harvard.edu/abs/2003MNRAS.344...60G} {344, 60}

\bibitem[\protect\citeauthoryear{{Gallo} et~al.,}{{Gallo}
  et~al.}{2014}]{gallo2014}
{Gallo} E.,  et~al., 2014, \mn@doi [\mnras] {10.1093/mnras/stu1599}, \href
  {http://adsabs.harvard.edu/abs/2014MNRAS.445..290G} {445, 290}

\bibitem[\protect\citeauthoryear{{Gallo}, {Degenaar}  \& {van den
  Eijnden}}{{Gallo} et~al.}{2018}]{gallo18}
{Gallo} E.,  {Degenaar} N.,   {van den Eijnden} J.,  2018, \mn@doi [\mnras]
  {10.1093/mnrasl/sly083}, \href
  {http://cdsads.u-strasbg.fr/abs/2018MNRAS.478L.132G} {478, L132}

\bibitem[\protect\citeauthoryear{{Gehrels} et~al.,}{{Gehrels}
  et~al.}{2004}]{gehrels04}
{Gehrels} N.,  et~al., 2004, \mn@doi [\apj] {10.1086/422091}, \href
  {http://adsabs.harvard.edu/abs/2004ApJ...611.1005G} {611, 1005}

\bibitem[\protect\citeauthoryear{{Gendreau} et~al.,}{{Gendreau}
  et~al.}{2016}]{gendreau2016}
{Gendreau} K.~C.,  et~al., 2016, in Space Telescopes and Instrumentation 2016:
  Ultraviolet to Gamma Ray. p. 99051H, \mn@doi{10.1117/12.2231304}

\bibitem[\protect\citeauthoryear{{Grishina}, {Kopatskaya}  \&
  {Larionov}}{{Grishina} et~al.}{2020}]{grishina2020}
{Grishina} T.~S.,  {Kopatskaya} E.~N.,   {Larionov} V.~M.,  2020, The
  Astronomer's Telegram, \href
  {https://ui.adsabs.harvard.edu/abs/2020ATel13938....1G} {13938, 1}

\bibitem[\protect\citeauthoryear{{Gusinskaia} et~al.,}{{Gusinskaia}
  et~al.}{2017}]{gusinskaia17}
{Gusinskaia} N.~V.,  et~al., 2017, \mn@doi [\mnras] {10.1093/mnras/stx1235},
  \href {http://adsabs.harvard.edu/abs/2017MNRAS.470.1871G} {470, 1871}

\bibitem[\protect\citeauthoryear{{Gusinskaia} et~al.,}{{Gusinskaia}
  et~al.}{2020a}]{gusinskaia20_igrj17591}
{Gusinskaia} N.~V.,  et~al., 2020a, \mn@doi [\mnras] {10.1093/mnras/stz3460},
  \href {https://ui.adsabs.harvard.edu/abs/2020MNRAS.492.1091G} {492, 1091}

\bibitem[\protect\citeauthoryear{{Gusinskaia} et~al.,}{{Gusinskaia}
  et~al.}{2020b}]{gusinskaia20}
{Gusinskaia} N.~V.,  et~al., 2020b, \mn@doi [\mnras] {10.1093/mnras/stz3420},
  \href {https://ui.adsabs.harvard.edu/abs/2020MNRAS.492.2858G} {492, 2858}

\bibitem[\protect\citeauthoryear{{Hannikainen}, {Hunstead}, {Campbell-Wilson}
  \& {Sood}}{{Hannikainen} et~al.}{1998}]{hannikainen98}
{Hannikainen} D.~C.,  {Hunstead} R.~W.,  {Campbell-Wilson} D.,   {Sood} R.~K.,
  1998, \aap, \href {https://ui.adsabs.harvard.edu/abs/1998A&A...337..460H}
  {337, 460}

\bibitem[\protect\citeauthoryear{{Hornby} \& {Williams}}{{Hornby} \&
  {Williams}}{1966}]{hornby1966}
{Hornby} J.~M.,  {Williams} P.~J.~S.,  1966, \mn@doi [\mnras]
  {10.1093/mnras/131.2.237}, \href
  {https://ui.adsabs.harvard.edu/abs/1966MNRAS.131..237H} {131, 237}

\bibitem[\protect\citeauthoryear{{Hulleman}, {in 't Zand}  \&
  {Heise}}{{Hulleman} et~al.}{1998}]{hulleman1998}
{Hulleman} F.,  {in 't Zand} J.~J.~M.,   {Heise} J.,  1998, \aap, \href
  {https://ui.adsabs.harvard.edu/abs/1998A&A...337L..25H} {337, L25}

\bibitem[\protect\citeauthoryear{{Israel} et~al.,}{{Israel}
  et~al.}{2017}]{israel17b}
{Israel} G.~L.,  et~al., 2017, \mn@doi [\mnras] {10.1093/mnrasl/slw218}, \href
  {http://cdsads.u-strasbg.fr/abs/2017MNRAS.466L..48I} {466, L48}

\bibitem[\protect\citeauthoryear{{Jaisawal} \& {Naik}}{{Jaisawal} \&
  {Naik}}{2017}]{jaisawal2017}
{Jaisawal} G.~K.,  {Naik} S.,  2017, in {Serino} M.,  {Shidatsu} M.,  {Iwakiri}
  W.,   {Mihara} T.,  eds, 7 years of MAXI: monitoring X-ray Transients. p.~153
  (\mn@eprint {arXiv} {1705.05536})

\bibitem[\protect\citeauthoryear{{Jaisawal} et~al.,}{{Jaisawal}
  et~al.}{2019}]{jaisawal2019}
{Jaisawal} G.~K.,  et~al., 2019, \mn@doi [\apj] {10.3847/1538-4357/ab4595},
  \href {https://ui.adsabs.harvard.edu/abs/2019ApJ...885...18J} {885, 18}

\bibitem[\protect\citeauthoryear{{Johnston}, {Soria}  \& {Gibson}}{{Johnston}
  et~al.}{2016}]{johnston16}
{Johnston} H.~M.,  {Soria} R.,   {Gibson} J.,  2016, \mn@doi [\mnras]
  {10.1093/mnras/stv2669}, \href
  {http://adsabs.harvard.edu/abs/2016MNRAS.456..347J} {456, 347}

\bibitem[\protect\citeauthoryear{{Kaaret}, {Corbel}, {Prestwich}  \&
  {Zezas}}{{Kaaret} et~al.}{2003}]{kaaret03}
{Kaaret} P.,  {Corbel} S.,  {Prestwich} A.~H.,   {Zezas} A.,  2003, \mn@doi
  [Science] {10.1126/science.1079610}, \href
  {http://adsabs.harvard.edu/abs/2003Sci...299..365K} {299, 365}

\bibitem[\protect\citeauthoryear{{Kaaret}, {Feng}  \& {Roberts}}{{Kaaret}
  et~al.}{2017}]{kaaret17}
{Kaaret} P.,  {Feng} H.,   {Roberts} T.~P.,  2017, \mn@doi [\araa]
  {10.1146/annurev-astro-091916-055259}, \href
  {https://ui.adsabs.harvard.edu/abs/2017ARA&A..55..303K} {55, 303}

\bibitem[\protect\citeauthoryear{{Kaper}, {Hammerschlag-Hensberge}  \&
  {Zuiderwijk}}{{Kaper} et~al.}{1994}]{kaper1994}
{Kaper} L.,  {Hammerschlag-Hensberge} G.,   {Zuiderwijk} E.~J.,  1994, \aap,
  \href {https://ui.adsabs.harvard.edu/abs/1994A&A...289..846K} {289, 846}

\bibitem[\protect\citeauthoryear{{Kelly}}{{Kelly}}{2007}]{kelly2007}
{Kelly} B.~C.,  2007, \mn@doi [\apj] {10.1086/519947}, \href
  {https://ui.adsabs.harvard.edu/abs/2007ApJ...665.1489K} {665, 1489}

\bibitem[\protect\citeauthoryear{{King} \& {Lasota}}{{King} \&
  {Lasota}}{2016}]{king16_ulx}
{King} A.,  {Lasota} J.-P.,  2016, \mn@doi [\mnras] {10.1093/mnrasl/slw011},
  \href {https://ui.adsabs.harvard.edu/abs/2016MNRAS.458L..10K} {458, L10}

\bibitem[\protect\citeauthoryear{{Kong} et~al.,}{{Kong}
  et~al.}{2020}]{kong2020}
{Kong} L.~D.,  et~al., 2020, \mn@doi [\apj] {10.3847/1538-4357/abb241}, \href
  {https://ui.adsabs.harvard.edu/abs/2020ApJ...902...18K} {902, 18}

\bibitem[\protect\citeauthoryear{{K{\"u}hnel} et~al.,}{{K{\"u}hnel}
  et~al.}{2013}]{kuhnel2013}
{K{\"u}hnel} M.,  et~al., 2013, \mn@doi [\aap] {10.1051/0004-6361/201321203},
  \href {https://ui.adsabs.harvard.edu/abs/2013A&A...555A..95K} {555, A95}

\bibitem[\protect\citeauthoryear{{Kylafis}, {Contopoulos}, {Kazanas}  \&
  {Christodoulou}}{{Kylafis} et~al.}{2012}]{kylafis2012}
{Kylafis} N.~D.,  {Contopoulos} I.,  {Kazanas} D.,   {Christodoulou} D.~M.,
  2012, \mn@doi [\aap] {10.1051/0004-6361/201117052}, \href
  {https://ui.adsabs.harvard.edu/abs/2012A&A...538A...5K} {538, A5}

\bibitem[\protect\citeauthoryear{{Laplace}, {Mihara}, {Moritani}, {Nakajima},
  {Takagi}, {Makishima}  \& {Santangelo}}{{Laplace} et~al.}{2017}]{laplace17}
{Laplace} E.,  {Mihara} T.,  {Moritani} Y.,  {Nakajima} M.,  {Takagi} T.,
  {Makishima} K.,   {Santangelo} A.,  2017, \mn@doi [\aap]
  {10.1051/0004-6361/201629373}, \href
  {http://cdsads.u-strasbg.fr/abs/2017A%26A...597A.124L} {597, A124}

\bibitem[\protect\citeauthoryear{{Ludlam} et~al.,}{{Ludlam}
  et~al.}{2016}]{ludlam16}
{Ludlam} R.~M.,  et~al., 2016, \mn@doi [\apj] {10.3847/0004-637X/824/1/37},
  \href {http://adsabs.harvard.edu/abs/2016ApJ...824...37L} {824, 37}

\bibitem[\protect\citeauthoryear{{Ludlam} et~al.,}{{Ludlam}
  et~al.}{2017a}]{ludlam17a}
{Ludlam} R.~M.,  et~al., 2017a, \mn@doi [\apj] {10.3847/1538-4357/836/1/140},
  \href {http://adsabs.harvard.edu/abs/2017ApJ...836..140L} {836, 140}

\bibitem[\protect\citeauthoryear{{Ludlam}, {Miller}, {Cackett}, {Degenaar}  \&
  {Bostrom}}{{Ludlam} et~al.}{2017b}]{ludlam17b}
{Ludlam} R.~M.,  {Miller} J.~M.,  {Cackett} E.~M.,  {Degenaar} N.,   {Bostrom}
  A.~C.,  2017b, \mn@doi [\apj] {10.3847/1538-4357/aa661a}, \href
  {http://adsabs.harvard.edu/abs/2017ApJ...838...79L} {838, 79}

\bibitem[\protect\citeauthoryear{{Mandal}, {Pal}, {Hazra}, {Jana}, {Bhunia}  \&
  {Ghanta}}{{Mandal} et~al.}{2020}]{mandal2020}
{Mandal} M.,  {Pal} S.,  {Hazra} M.,  {Jana} A.,  {Bhunia} B.,   {Ghanta} A.,
  2020, The Astronomer's Telegram, \href
  {https://ui.adsabs.harvard.edu/abs/2020ATel14157....1M} {14157, 1}

\bibitem[\protect\citeauthoryear{{Markoff}, {Falcke}  \& {Fender}}{{Markoff}
  et~al.}{2001}]{markoff01}
{Markoff} S.,  {Falcke} H.,   {Fender} R.,  2001, \mn@doi [\aap]
  {10.1051/0004-6361:20010420}, \href
  {https://ui.adsabs.harvard.edu/abs/2001A&A...372L..25M} {372, L25}

\bibitem[\protect\citeauthoryear{{Martin}, {Nixon}, {Armitage}, {Lubow}  \&
  {Price}}{{Martin} et~al.}{2014}]{martin14}
{Martin} R.~G.,  {Nixon} C.,  {Armitage} P.~J.,  {Lubow} S.~H.,   {Price}
  D.~J.,  2014, \mn@doi [\apjl] {10.1088/2041-8205/790/2/L34}, \href
  {http://cdsads.u-strasbg.fr/abs/2014ApJ...790L..34M} {790, L34}

\bibitem[\protect\citeauthoryear{{Massi} \& {Kaufman Bernad{\'o}}}{{Massi} \&
  {Kaufman Bernad{\'o}}}{2008}]{massi08}
{Massi} M.,  {Kaufman Bernad{\'o}} M.,  2008, \mn@doi [\aap]
  {10.1051/0004-6361:20077567}, \href
  {http://adsabs.harvard.edu/abs/2008A%26A...477....1M} {477, 1}

\bibitem[\protect\citeauthoryear{{Matsuoka} et~al.,}{{Matsuoka}
  et~al.}{2009}]{matsuoka09}
{Matsuoka} M.,  et~al., 2009, \mn@doi [\pasj] {10.1093/pasj/61.5.999}, \href
  {http://adsabs.harvard.edu/abs/2009PASJ...61..999M} {61, 999}

\bibitem[\protect\citeauthoryear{{Matthews}, {Bell}  \& {Blundell}}{{Matthews}
  et~al.}{2020}]{matthews2020}
{Matthews} J.~H.,  {Bell} A.~R.,   {Blundell} K.~M.,  2020, \mn@doi [\nar]
  {10.1016/j.newar.2020.101543}, \href
  {https://ui.adsabs.harvard.edu/abs/2020NewAR..8901543M} {89, 101543}

\bibitem[\protect\citeauthoryear{{McMullin}, {Waters}, {Schiebel}, {Young}  \&
  {Golap}}{{McMullin} et~al.}{2007}]{mcmullin07}
{McMullin} J.~P.,  {Waters} B.,  {Schiebel} D.,  {Young} W.,   {Golap} K.,
  2007, in {Shaw} R.~A.,  {Hill} F.,   {Bell} D.~J.,  eds,  Astronomical
  Society of the Pacific Conference Series Vol. 376, Astronomical Data Analysis
  Software and Systems XVI. p.~127

\bibitem[\protect\citeauthoryear{{Merloni}, {Heinz}  \& {di Matteo}}{{Merloni}
  et~al.}{2003}]{merloni03}
{Merloni} A.,  {Heinz} S.,   {di Matteo} T.,  2003, \mn@doi [\mnras]
  {10.1046/j.1365-2966.2003.07017.x}, \href
  {http://adsabs.harvard.edu/abs/2003MNRAS.345.1057M} {345, 1057}

\bibitem[\protect\citeauthoryear{{Meyer-Hofmeister} \&
  {Meyer}}{{Meyer-Hofmeister} \& {Meyer}}{2014}]{meyer14}
{Meyer-Hofmeister} E.,  {Meyer} F.,  2014, \mn@doi [\aap]
  {10.1051/0004-6361/201322423}, \href
  {https://ui.adsabs.harvard.edu/abs/2014A&A...562A.142M} {562, A142}

\bibitem[\protect\citeauthoryear{{Mezcua}, {Roberts}, {Lobanov}  \&
  {Sutton}}{{Mezcua} et~al.}{2015}]{mezcua15}
{Mezcua} M.,  {Roberts} T.~P.,  {Lobanov} A.~P.,   {Sutton} A.~D.,  2015,
  \mn@doi [\mnras] {10.1093/mnras/stv143}, \href
  {http://cdsads.u-strasbg.fr/abs/2015MNRAS.448.1893M} {448, 1893}

\bibitem[\protect\citeauthoryear{{Migliari} \& {Fender}}{{Migliari} \&
  {Fender}}{2006}]{migliari06}
{Migliari} S.,  {Fender} R.~P.,  2006, \mn@doi [\mnras]
  {10.1111/j.1365-2966.2005.09777.x}, \href
  {http://adsabs.harvard.edu/abs/2006MNRAS.366...79M} {366, 79}

\bibitem[\protect\citeauthoryear{{Migliari}, {Fender}, {Rupen}, {Jonker},
  {Klein-Wolt}, {Hjellming}  \& {van der Klis}}{{Migliari}
  et~al.}{2003}]{migliari03}
{Migliari} S.,  {Fender} R.~P.,  {Rupen} M.,  {Jonker} P.~G.,  {Klein-Wolt} M.,
   {Hjellming} R.~M.,   {van der Klis} M.,  2003, \mn@doi [\mnras]
  {10.1046/j.1365-8711.2003.06795.x}, \href
  {http://cdsads.u-strasbg.fr/abs/2003MNRAS.342L..67M} {342, L67}

\bibitem[\protect\citeauthoryear{{Migliari}, {Tudose}, {Miller-Jones},
  {Kuulkers}, {Nakajima}  \& {Yamaoka}}{{Migliari} et~al.}{2011}]{migliari11}
{Migliari} S.,  {Tudose} V.,  {Miller-Jones} J.~C.~A.,  {Kuulkers} E.,
  {Nakajima} M.,   {Yamaoka} K.,  2011, The Astronomer's Telegram, \href
  {http://adsabs.harvard.edu/abs/2011ATel.3198....1M} {3198}

\bibitem[\protect\citeauthoryear{{Monageng}, {McBride}, {Coe}, {Steele}  \&
  {Reig}}{{Monageng} et~al.}{2017}]{monageng17}
{Monageng} I.~M.,  {McBride} V.~A.,  {Coe} M.~J.,  {Steele} I.~A.,   {Reig} P.,
   2017, \mn@doi [\mnras] {10.1093/mnras/stw2354}, \href
  {http://cdsads.u-strasbg.fr/abs/2017MNRAS.464..572M} {464, 572}

\bibitem[\protect\citeauthoryear{{Moritani} et~al.,}{{Moritani}
  et~al.}{2013}]{moritani13}
{Moritani} Y.,  et~al., 2013, \mn@doi [\pasj] {10.1093/pasj/65.4.83}, \href
  {http://cdsads.u-strasbg.fr/abs/2013PASJ...65...83M} {65, 83}

\bibitem[\protect\citeauthoryear{{Motch}, {Stella}, {Janot-Pacheco}  \&
  {Mouchet}}{{Motch} et~al.}{1991}]{motch1991}
{Motch} C.,  {Stella} L.,  {Janot-Pacheco} E.,   {Mouchet} M.,  1991, \mn@doi
  [\apj] {10.1086/169779}, \href
  {https://ui.adsabs.harvard.edu/abs/1991ApJ...369..490M} {369, 490}

\bibitem[\protect\citeauthoryear{{Nakajima} et~al.,}{{Nakajima}
  et~al.}{2020}]{nakajima2020}
{Nakajima} M.,  et~al., 2020, The Astronomer's Telegram, \href
  {https://ui.adsabs.harvard.edu/abs/2020ATel14173....1N} {14173, 1}

\bibitem[\protect\citeauthoryear{{Okazaki} \& {Negueruela}}{{Okazaki} \&
  {Negueruela}}{2001}]{okazaki01}
{Okazaki} A.~T.,  {Negueruela} I.,  2001, \mn@doi [\aap]
  {10.1051/0004-6361:20011083}, \href
  {http://cdsads.u-strasbg.fr/abs/2001A%26A...377..161O} {377, 161}

\bibitem[\protect\citeauthoryear{{Pakull} \& {Mirioni}}{{Pakull} \&
  {Mirioni}}{2003}]{pakull2003}
{Pakull} M.~W.,  {Mirioni} L.,  2003, in {Arthur} J.,  {Henney} W.~J.,  eds,
  Revista Mexicana de Astronomia y Astrofisica Conference Series Vol. 15,
  Revista Mexicana de Astronomia y Astrofisica Conference Series. pp 197--199

\bibitem[\protect\citeauthoryear{{Parfrey}, {Spitkovsky}  \&
  {Beloborodov}}{{Parfrey} et~al.}{2016}]{parfrey16}
{Parfrey} K.,  {Spitkovsky} A.,   {Beloborodov} A.~M.,  2016, \mn@doi [\apj]
  {10.3847/0004-637X/822/1/33}, \href
  {http://cdsads.u-strasbg.fr/abs/2016ApJ...822...33P} {822, 33}

\bibitem[\protect\citeauthoryear{{Patruno}, {Haskell}  \&
  {Andersson}}{{Patruno} et~al.}{2017}]{patruno17}
{Patruno} A.,  {Haskell} B.,   {Andersson} N.,  2017, preprint, \href
  {http://adsabs.harvard.edu/abs/2017arXiv170507669P} {} (\mn@eprint {arXiv}
  {1705.07669})

\bibitem[\protect\citeauthoryear{{Patruno}, {Wette}  \& {Messenger}}{{Patruno}
  et~al.}{2018}]{patruno18}
{Patruno} A.,  {Wette} K.,   {Messenger} C.,  2018, \mn@doi [\apj]
  {10.3847/1538-4357/aabf89}, \href
  {https://ui.adsabs.harvard.edu/abs/2018ApJ...859..112P} {859, 112}

\bibitem[\protect\citeauthoryear{{Pestalozzi}, {Torkelsson}, {Hobbs}  \&
  {L{\'o}pez-S{\'a}nchez}}{{Pestalozzi} et~al.}{2009}]{pestalozzi09}
{Pestalozzi} M.,  {Torkelsson} U.,  {Hobbs} G.,   {L{\'o}pez-S{\'a}nchez}
  {\'A}.~R.,  2009, \mn@doi [\aap] {10.1051/0004-6361/200913151}, \href
  {http://adsabs.harvard.edu/abs/2009A%26A...506L..21P} {506, L21}

\bibitem[\protect\citeauthoryear{{Pinto}, {Middleton}  \& {Fabian}}{{Pinto}
  et~al.}{2016}]{pinto16}
{Pinto} C.,  {Middleton} M.~J.,   {Fabian} A.~C.,  2016, \mn@doi [\nat]
  {10.1038/nature17417}, \href
  {http://adsabs.harvard.edu/abs/2016Natur.533...64P} {533, 64}

\bibitem[\protect\citeauthoryear{{Porter} \& {Rivinius}}{{Porter} \&
  {Rivinius}}{2003}]{porter03}
{Porter} J.~M.,  {Rivinius} T.,  2003, \mn@doi [\pasp] {10.1086/378307}, \href
  {http://cdsads.u-strasbg.fr/abs/2003PASP..115.1153P} {115, 1153}

\bibitem[\protect\citeauthoryear{{Reig}}{{Reig}}{2011}]{reig11}
{Reig} P.,  2011, \mn@doi [\apss] {10.1007/s10509-010-0575-8}, \href
  {http://cdsads.u-strasbg.fr/abs/2011Ap%26SS.332....1R} {332, 1}

\bibitem[\protect\citeauthoryear{{Reig}, {Negueruela}, {Fabregat}, {Chato},
  {Blay}  \& {Mavromatakis}}{{Reig} et~al.}{2004}]{reig2004}
{Reig} P.,  {Negueruela} I.,  {Fabregat} J.,  {Chato} R.,  {Blay} P.,
  {Mavromatakis} F.,  2004, \mn@doi [\aap] {10.1051/0004-6361:20035786}, \href
  {https://ui.adsabs.harvard.edu/abs/2004A&A...421..673R} {421, 673}

\bibitem[\protect\citeauthoryear{{Reig}, {S{\l}owikowska}, {Zezas}  \&
  {Blay}}{{Reig} et~al.}{2010}]{reig2010}
{Reig} P.,  {S{\l}owikowska} A.,  {Zezas} A.,   {Blay} P.,  2010, \mn@doi
  [\mnras] {10.1111/j.1365-2966.2009.15656.x}, \href
  {https://ui.adsabs.harvard.edu/abs/2010MNRAS.401...55R} {401, 55}

\bibitem[\protect\citeauthoryear{{Reig}, {Doroshenko}  \& {Zezas}}{{Reig}
  et~al.}{2014}]{reig2014}
{Reig} P.,  {Doroshenko} V.,   {Zezas} A.,  2014, \mn@doi [\mnras]
  {10.1093/mnras/stu1840}, \href
  {https://ui.adsabs.harvard.edu/abs/2014MNRAS.445.1314R} {445, 1314}

\bibitem[\protect\citeauthoryear{{Remillard} et~al.,}{{Remillard}
  et~al.}{2021}]{remillard2021}
{Remillard} R.~A.,  et~al., 2021, arXiv e-prints, \href
  {https://ui.adsabs.harvard.edu/abs/2021arXiv210509901R} {p. arXiv:2105.09901}

\bibitem[\protect\citeauthoryear{{Reynolds}, {Wolff}, {Miller}, {Arzoumanian},
  {Gendreau}, {Chakrabarty}  \& {Jenke}}{{Reynolds}
  et~al.}{2020}]{reynolds2020}
{Reynolds} M.,  {Wolff} M.,  {Miller} J.,  {Arzoumanian} Z.,  {Gendreau} K.,
  {Chakrabarty} D.,   {Jenke} P.,  2020, The Astronomer's Telegram, \href
  {https://ui.adsabs.harvard.edu/abs/2020ATel13749....1R} {13749, 1}

\bibitem[\protect\citeauthoryear{{Rib{\'o}} et~al.,}{{Rib{\'o}}
  et~al.}{2017}]{ribo17}
{Rib{\'o}} M.,  et~al., 2017, \mn@doi [\apjl] {10.3847/2041-8213/835/2/L33},
  \href {https://ui.adsabs.harvard.edu/abs/2017ApJ...835L..33R} {835, L33}

\bibitem[\protect\citeauthoryear{{Riley} et~al.,}{{Riley}
  et~al.}{2019}]{riley19}
{Riley} T.~E.,  et~al., 2019, \mn@doi [\apjl] {10.3847/2041-8213/ab481c}, \href
  {https://ui.adsabs.harvard.edu/abs/2019ApJ...887L..21R} {887, L21}

\bibitem[\protect\citeauthoryear{{Ruderman} \& {Sutherland}}{{Ruderman} \&
  {Sutherland}}{1975}]{ruderman1975}
{Ruderman} M.~A.,  {Sutherland} P.~G.,  1975, \mn@doi [\apj] {10.1086/153393},
  \href {https://ui.adsabs.harvard.edu/abs/1975ApJ...196...51R} {196, 51}

\bibitem[\protect\citeauthoryear{{Russell} et~al.,}{{Russell}
  et~al.}{2013}]{russell13}
{Russell} D.~M.,  et~al., 2013, \mn@doi [\mnras] {10.1093/mnras/sts377}, \href
  {http://cdsads.u-strasbg.fr/abs/2013MNRAS.429..815R} {429, 815}

\bibitem[\protect\citeauthoryear{{Russell}, {Soria}, {Miller-Jones}, {Curran},
  {Markoff}, {Russell}  \& {Sivakoff}}{{Russell} et~al.}{2014}]{russell14}
{Russell} T.~D.,  {Soria} R.,  {Miller-Jones} J.~C.~A.,  {Curran} P.~A.,
  {Markoff} S.,  {Russell} D.~M.,   {Sivakoff} G.~R.,  2014, \mn@doi [\mnras]
  {10.1093/mnras/stt2498}, \href
  {http://cdsads.u-strasbg.fr/abs/2014MNRAS.439.1390R} {439, 1390}

\bibitem[\protect\citeauthoryear{{Russell}, {Degenaar}, {Wijnands}, {van den
  Eijnden}, {Gusinskaia}, {Hessels}  \& {Miller-Jones}}{{Russell}
  et~al.}{2018}]{russell18}
{Russell} T.~D.,  {Degenaar} N.,  {Wijnands} R.,  {van den Eijnden} J.,
  {Gusinskaia} N.~V.,  {Hessels} J.~W.~T.,   {Miller-Jones} J.~C.~A.,  2018,
  \mn@doi [\apjl] {10.3847/2041-8213/aaf4f9}, \href
  {https://ui.adsabs.harvard.edu/abs/2018ApJ...869L..16R} {869, L16}

\bibitem[\protect\citeauthoryear{{Russell} et~al.,}{{Russell}
  et~al.}{2021}]{russell2021}
{Russell} T.~D.,  et~al., 2021, \mn@doi [\mnras] {10.1093/mnrasl/slab087},
  \href {https://ui.adsabs.harvard.edu/abs/2021MNRAS.508L...6R} {508, L6}

\bibitem[\protect\citeauthoryear{{Selina} et~al.,}{{Selina}
  et~al.}{2018}]{selina18}
{Selina} R.~J.,  et~al., 2018, {The ngVLA Reference Design}.
p.~15

\bibitem[\protect\citeauthoryear{{Shakura}, {Postnov}, {Kochetkova}  \&
  {Hjalmarsdotter}}{{Shakura} et~al.}{2012}]{shakura2012}
{Shakura} N.,  {Postnov} K.,  {Kochetkova} A.,   {Hjalmarsdotter} L.,  2012,
  \mn@doi [\mnras] {10.1111/j.1365-2966.2011.20026.x}, \href
  {https://ui.adsabs.harvard.edu/abs/2012MNRAS.420..216S} {420, 216}

\bibitem[\protect\citeauthoryear{{Shrader}, {Sutaria}, {Singh}  \&
  {Macomb}}{{Shrader} et~al.}{1999}]{shrader1999}
{Shrader} C.~R.,  {Sutaria} F.~K.,  {Singh} K.~P.,   {Macomb} D.~J.,  1999,
  \mn@doi [\apj] {10.1086/306785}, \href
  {https://ui.adsabs.harvard.edu/abs/1999ApJ...512..920S} {512, 920}

\bibitem[\protect\citeauthoryear{{Sidoli} \& {Paizis}}{{Sidoli} \&
  {Paizis}}{2018}]{sidoli2018}
{Sidoli} L.,  {Paizis} A.,  2018, \mn@doi [\mnras] {10.1093/mnras/sty2428},
  \href {https://ui.adsabs.harvard.edu/abs/2018MNRAS.481.2779S} {481, 2779}

\bibitem[\protect\citeauthoryear{{Sidoli}, {Israel}, {Esposito},
  {Rodr{\'\i}guez Castillo}  \& {Postnov}}{{Sidoli} et~al.}{2017}]{sidoli17}
{Sidoli} L.,  {Israel} G.~L.,  {Esposito} P.,  {Rodr{\'\i}guez Castillo} G.~A.,
    {Postnov} K.,  2017, \mn@doi [\mnras] {10.1093/mnras/stx1105}, \href
  {https://ui.adsabs.harvard.edu/abs/2017MNRAS.469.3056S} {469, 3056}

\bibitem[\protect\citeauthoryear{{Soleri} \& {Fender}}{{Soleri} \&
  {Fender}}{2011}]{soleri11}
{Soleri} P.,  {Fender} R.,  2011, \mn@doi [\mnras]
  {10.1111/j.1365-2966.2011.18303.x}, \href
  {https://ui.adsabs.harvard.edu/abs/2011MNRAS.413.2269S} {413, 2269}

\bibitem[\protect\citeauthoryear{{Staubert} et~al.,}{{Staubert}
  et~al.}{2019}]{staubert19}
{Staubert} R.,  et~al., 2019, \mn@doi [\aap] {10.1051/0004-6361/201834479},
  \href {https://ui.adsabs.harvard.edu/abs/2019A&A...622A..61S} {622, A61}

\bibitem[\protect\citeauthoryear{{Sugizaki}, {Oeda}, {Kawai}, {Mihara},
  {Makishima}  \& {Nakajima}}{{Sugizaki} et~al.}{2020}]{sugizaki2020}
{Sugizaki} M.,  {Oeda} M.,  {Kawai} N.,  {Mihara} T.,  {Makishima} K.,
  {Nakajima} M.,  2020, \mn@doi [\apj] {10.3847/1538-4357/ab93c7}, \href
  {https://ui.adsabs.harvard.edu/abs/2020ApJ...896..124S} {896, 124}

\bibitem[\protect\citeauthoryear{{Tao}, {Feng}, {Zhang}, {Bu}, {Zhang}, {Qu}
  \& {Zhang}}{{Tao} et~al.}{2019}]{tao2019}
{Tao} L.,  {Feng} H.,  {Zhang} S.,  {Bu} Q.,  {Zhang} S.,  {Qu} J.,   {Zhang}
  Y.,  2019, \mn@doi [\apj] {10.3847/1538-4357/ab0211}, \href
  {https://ui.adsabs.harvard.edu/abs/2019ApJ...873...19T} {873, 19}

\bibitem[\protect\citeauthoryear{{Taylor}, {Waters}, {Lamers}, {Persi}  \&
  {Bjorkman}}{{Taylor} et~al.}{1987}]{taylor1987}
{Taylor} A.~R.,  {Waters} L.~B.~F.~M.,  {Lamers} H.~J.~G.~L.~M.,  {Persi} P.,
  {Bjorkman} K.~S.,  1987, \mn@doi [\mnras] {10.1093/mnras/228.4.811}, \href
  {https://ui.adsabs.harvard.edu/abs/1987MNRAS.228..811T} {228, 811}

\bibitem[\protect\citeauthoryear{{Taylor}, {Waters}, {Bjorkman}  \&
  {Dougherty}}{{Taylor} et~al.}{1990}]{taylor1990}
{Taylor} A.~R.,  {Waters} L.~B.~F.~M.,  {Bjorkman} K.~S.,   {Dougherty} S.~M.,
  1990, \aap, \href {https://ui.adsabs.harvard.edu/abs/1990A&A...231..453T}
  {231, 453}

\bibitem[\protect\citeauthoryear{{Tsygankov}, {Mushtukov}, {Suleimanov},
  {Doroshenko}, {Abolmasov}, {Lutovinov}  \& {Poutanen}}{{Tsygankov}
  et~al.}{2017}]{tsygankov17}
{Tsygankov} S.~S.,  {Mushtukov} A.~A.,  {Suleimanov} V.~F.,  {Doroshenko} V.,
  {Abolmasov} P.~K.,  {Lutovinov} A.~A.,   {Poutanen} J.,  2017, \mn@doi [\aap]
  {10.1051/0004-6361/201630248}, \href
  {http://cdsads.u-strasbg.fr/abs/2017A%26A...608A..17T} {608, A17}

\bibitem[\protect\citeauthoryear{{Tsygankov}, {Doroshenko}, {Mushtukov},
  {Lutovinov}  \& {Poutanen}}{{Tsygankov} et~al.}{2018}]{tsygankov18}
{Tsygankov} S.~S.,  {Doroshenko} V.,  {Mushtukov} A.~A.,  {Lutovinov} A.~A.,
  {Poutanen} J.,  2018, \mn@doi [\mnras] {10.1093/mnrasl/sly116}, \href
  {http://cdsads.u-strasbg.fr/abs/2018MNRAS.479L.134T} {479, L134}

\bibitem[\protect\citeauthoryear{{Tudor} et~al.,}{{Tudor}
  et~al.}{2017}]{tudor17}
{Tudor} V.,  et~al., 2017, \mn@doi [\mnras] {10.1093/mnras/stx1168}, \href
  {http://adsabs.harvard.edu/abs/2017MNRAS.470..324T} {470, 324}

\bibitem[\protect\citeauthoryear{{Tudose}, {Migliari}, {Miller-Jones},
  {Nakajima}, {Yamaoka}  \& {Kuulkers}}{{Tudose} et~al.}{2010}]{tudose10}
{Tudose} V.,  {Migliari} S.,  {Miller-Jones} J.~C.~A.,  {Nakajima} M.,
  {Yamaoka} K.,   {Kuulkers} E.,  2010, The Astronomer's Telegram, \href
  {http://adsabs.harvard.edu/abs/2010ATel.2798....1T} {2798}

\bibitem[\protect\citeauthoryear{{Verner}, {Ferland}, {Korista}  \&
  {Yakovlev}}{{Verner} et~al.}{1996}]{verner96}
{Verner} D.~A.,  {Ferland} G.~J.,  {Korista} K.~T.,   {Yakovlev} D.~G.,  1996,
  \mn@doi [\apj] {10.1086/177435}, \href
  {http://adsabs.harvard.edu/abs/1996ApJ...465..487V} {465, 487}

\bibitem[\protect\citeauthoryear{{Walton} et~al.,}{{Walton}
  et~al.}{2018}]{walton18c}
{Walton} D.~J.,  et~al., 2018, \mn@doi [\apj] {10.3847/1538-4357/aab610}, \href
  {http://cdsads.u-strasbg.fr/abs/2018ApJ...856..128W} {856, 128}

\bibitem[\protect\citeauthoryear{{Williams}, {Gies}, {Matson}, {Touhami},
  {Grundstrom}, {Huang}  \& {McSwain}}{{Williams} et~al.}{2010}]{williams2010}
{Williams} S.~J.,  {Gies} D.~R.,  {Matson} R.~A.,  {Touhami} Y.,  {Grundstrom}
  E.~D.,  {Huang} W.,   {McSwain} M.~V.,  2010, \mn@doi [\apjl]
  {10.1088/2041-8205/723/1/L93}, \href
  {https://ui.adsabs.harvard.edu/abs/2010ApJ...723L..93W} {723, L93}

\bibitem[\protect\citeauthoryear{{Wilms}, {Allen}  \& {McCray}}{{Wilms}
  et~al.}{2000}]{wilms00}
{Wilms} J.,  {Allen} A.,   {McCray} R.,  2000, \mn@doi [\apj] {10.1086/317016},
  \href {https://ui.adsabs.harvard.edu/abs/2000ApJ...542..914W} {542, 914}

\bibitem[\protect\citeauthoryear{{Wilson-Hodge} et~al.,}{{Wilson-Hodge}
  et~al.}{2018}]{wilson18}
{Wilson-Hodge} C.~A.,  et~al., 2018, \mn@doi [\apj] {10.3847/1538-4357/aace60},
  \href {http://cdsads.u-strasbg.fr/abs/2018ApJ...863....9W} {863, 9}

\bibitem[\protect\citeauthoryear{{Yamamoto}, {Mihara}, {Sugizaki}, {Nakajima},
  {Makishima}  \& {Sasano}}{{Yamamoto} et~al.}{2014}]{yamamoto2014}
{Yamamoto} T.,  {Mihara} T.,  {Sugizaki} M.,  {Nakajima} M.,  {Makishima} K.,
  {Sasano} M.,  2014, \mn@doi [\pasj] {10.1093/pasj/psu028}, \href
  {https://ui.adsabs.harvard.edu/abs/2014PASJ...66...59Y} {66, 59}

\bibitem[\protect\citeauthoryear{{Zhang}, {Harding}  \& {Muslimov}}{{Zhang}
  et~al.}{2000}]{zhang2000}
{Zhang} B.,  {Harding} A.~K.,   {Muslimov} A.~G.,  2000, \mn@doi [\apjl]
  {10.1086/312542}, \href
  {https://ui.adsabs.harvard.edu/abs/2000ApJ...531L.135Z} {531, L135}

\bibitem[\protect\citeauthoryear{{Zhang} et~al.,}{{Zhang}
  et~al.}{2019}]{zhang2019}
{Zhang} Y.,  et~al., 2019, \mn@doi [\apj] {10.3847/1538-4357/ab22b1}, \href
  {https://ui.adsabs.harvard.edu/abs/2019ApJ...879...61Z} {879, 61}

\bibitem[\protect\citeauthoryear{{van den Eijnden}, {Degenaar}, {Russell},
  {Wijnand s}, {Miller-Jones}, {Sivakoff}  \& {Hern{\'a}ndez Santisteban}}{{van
  den Eijnden} et~al.}{2018a}]{vandeneijnden2018_swj0243}
{van den Eijnden} J.,  {Degenaar} N.,  {Russell} T.~D.,  {Wijnand s} R.,
  {Miller-Jones} J.~C.~A.,  {Sivakoff} G.~R.,   {Hern{\'a}ndez Santisteban}
  J.~V.,  2018a, \mn@doi [\nat] {10.1038/s41586-018-0524-1}, \href
  {https://ui.adsabs.harvard.edu/abs/2018Natur.562..233V} {562, 233}

\bibitem[\protect\citeauthoryear{{van den Eijnden} et~al.,}{{van den Eijnden}
  et~al.}{2018b}]{vandeneijnden2018_igr17379}
{van den Eijnden} J.,  et~al., 2018b, The Astronomer's Telegram, \href
  {http://cdsads.u-strasbg.fr/abs/2018ATel11520....1V} {11520}

\bibitem[\protect\citeauthoryear{{van den Eijnden}, {Degenaar}, {Russell},
  {Hern{\'a}ndez Santisteban}, {Wijnands}, {Miller-Jones}, {Rouco Escorial}  \&
  {Sivakoff}}{{van den Eijnden} et~al.}{2019a}]{vandeneijnden2019_reb}
{van den Eijnden} J.,  {Degenaar} N.,  {Russell} T.~D.,  {Hern{\'a}ndez
  Santisteban} J.~V.,  {Wijnands} R.,  {Miller-Jones} J.~C.~A.,  {Rouco
  Escorial} A.,   {Sivakoff} G.~R.,  2019a, \mn@doi [\mnras]
  {10.1093/mnras/sty3479}, \href
  {http://cdsads.u-strasbg.fr/abs/2019MNRAS.483.4628V} {483, 4628}

\bibitem[\protect\citeauthoryear{{van den Eijnden} et~al.,}{{van den Eijnden}
  et~al.}{2019b}]{vandeneijnden2019_chandra}
{van den Eijnden} J.,  et~al., 2019b, \mn@doi [\mnras] {10.1093/mnras/stz1548},
  \href {https://ui.adsabs.harvard.edu/abs/2019MNRAS.487.4355V} {487, 4355}

\bibitem[\protect\citeauthoryear{{van den Eijnden} et~al.,}{{van den Eijnden}
  et~al.}{2020}]{vandeneijnden2020_a0535atel}
{van den Eijnden} J.,  et~al., 2020, The Astronomer's Telegram, \href
  {https://ui.adsabs.harvard.edu/abs/2020ATel14193....1V} {14193, 1}

\bibitem[\protect\citeauthoryear{{van den Eijnden} et~al.,}{{van den Eijnden}
  et~al.}{2021}]{vandeneijnden2021}
{van den Eijnden} J.,  et~al., 2021, \mn@doi [\mnras] {10.1093/mnras/stab1995},
  \href {https://ui.adsabs.harvard.edu/abs/2021MNRAS.507.3899V} {507, 3899}

\makeatother
\end{thebibliography}

\input{main.bbl}


\appendix

\section{Online Supplementary Materials}

\section{Observational details and analysis}
\label{app:observations_fluxes}

\begin{figure*}
    \centering
    \includegraphics[width=\textwidth]{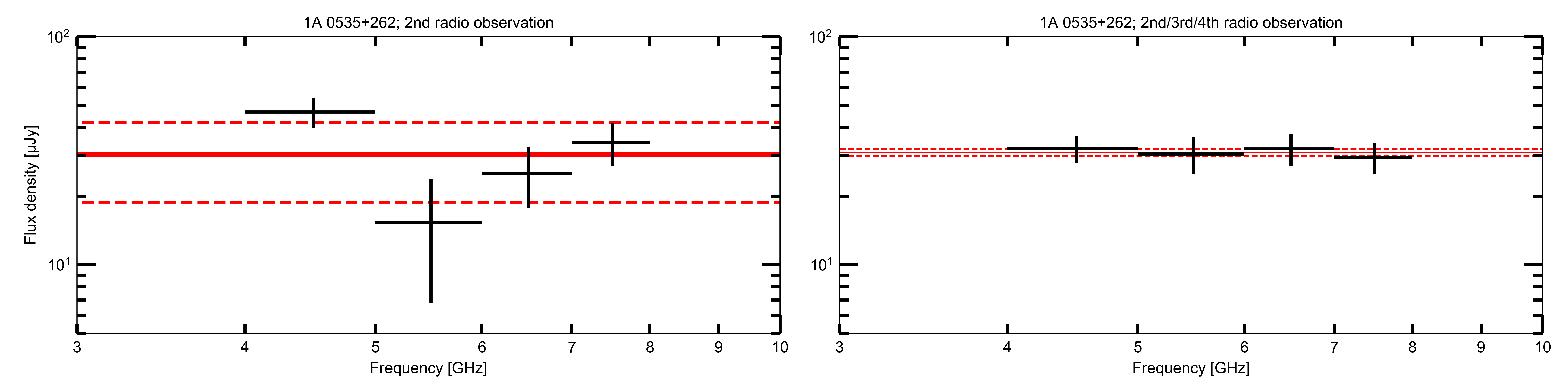}
    \caption{\textit{Left:} the VLA radio spectrum of 1A 0535+262, taken during its brightest (i.e. second) radio epoch. The source shows a large level of scatter due to its faintness. The red line and dashed line show the mean flux and its $1-\sigma$ confidence interval, calculated from these four bands. The spectral index, fitted to these data points, is $\alpha=-0.9\pm 1.1$. \textit{Right:} same as left panel, calculated from the combined data of observations 2, 3, and 4. The spectral index is measured to be $\alpha=-0.1\pm 0.1$.}
    \label{fig:spectrum}
\end{figure*}

In Tables \ref{tab:radio_obs} and \ref{tab:xrays_j1008_saxJ2103}, we list further details on the analysed radio and X-ray observations. In Figure \ref{fig:spectrum}, we show the radio spectra of 1A 0535+262 in the second and combined second to fourth observations. In Tables \ref{tab:NICER_0535_1} and \ref{tab:NICER_0535_2}, we list details regarding the analysed \textit{Swift} and \textit{NICER} observations of 1A 0535+262, as well as regarding the spectral fit. Note that we only analyse \textit{Swift} data up to MJD 59300, as the final radio observation was taken three weeks prior to that date. We refer to the main paper for full details, especially regarding the model choice, fitted energy band, and the number of included narrow Gaussian lines in the iron line complex. In two observations, the fitted \textit{NICER} band is highlighted by an asterisk. For those observations, even restricting the band to $2$--$10$ keV did not alleviate the issues with instrumental residuals sufficiently to yield $\chi^2_\nu < 2$. Therefore, in those two observations, we additionally added a $1$\% systematic uncertainty to the spectrum. As a result, the uncertainty on the flux and parameters may be enhanced. 

\begin{figure*}
    \centering
    \includegraphics[width=\textwidth]{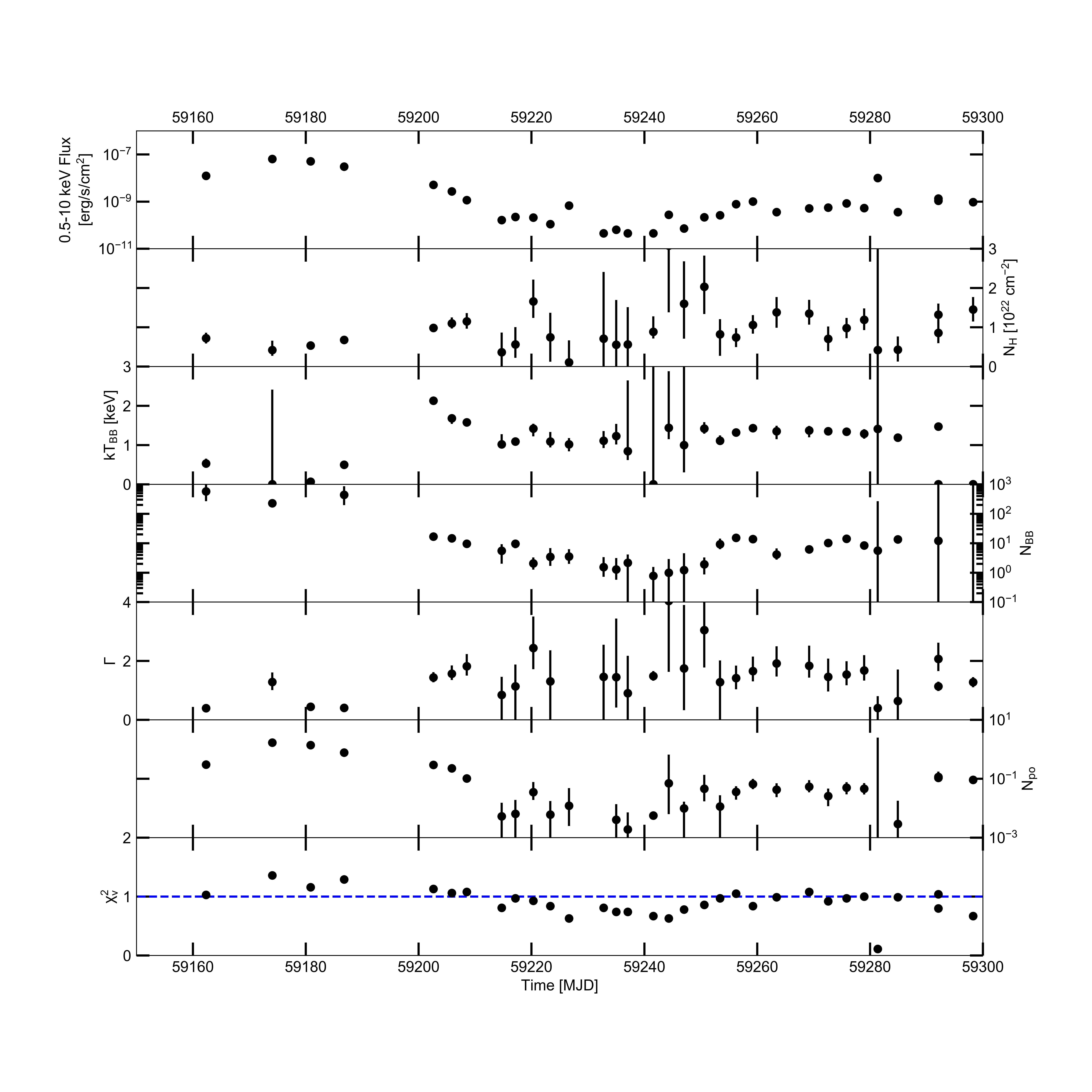}
    \caption{Light curves of the parameters fitted to the \textit{Swift} spectra of 1A 0535+262 up to MJD 59300. In order from top to bottom, the panels show: the X-ray flux, absorption column; black body temperature and normalisation; power law index and normalisation; and $\chi^2_\nu$ of the fit. Several parameters show larger error bars, often cased by a relatively short exposure (cf. MJD 59281.35). All parameter values are available in the machine-readable parameter file.} 
    \label{fig:swift_params_a0535}
\end{figure*}

\begin{figure*}
    \centering
    \includegraphics[width=\textwidth]{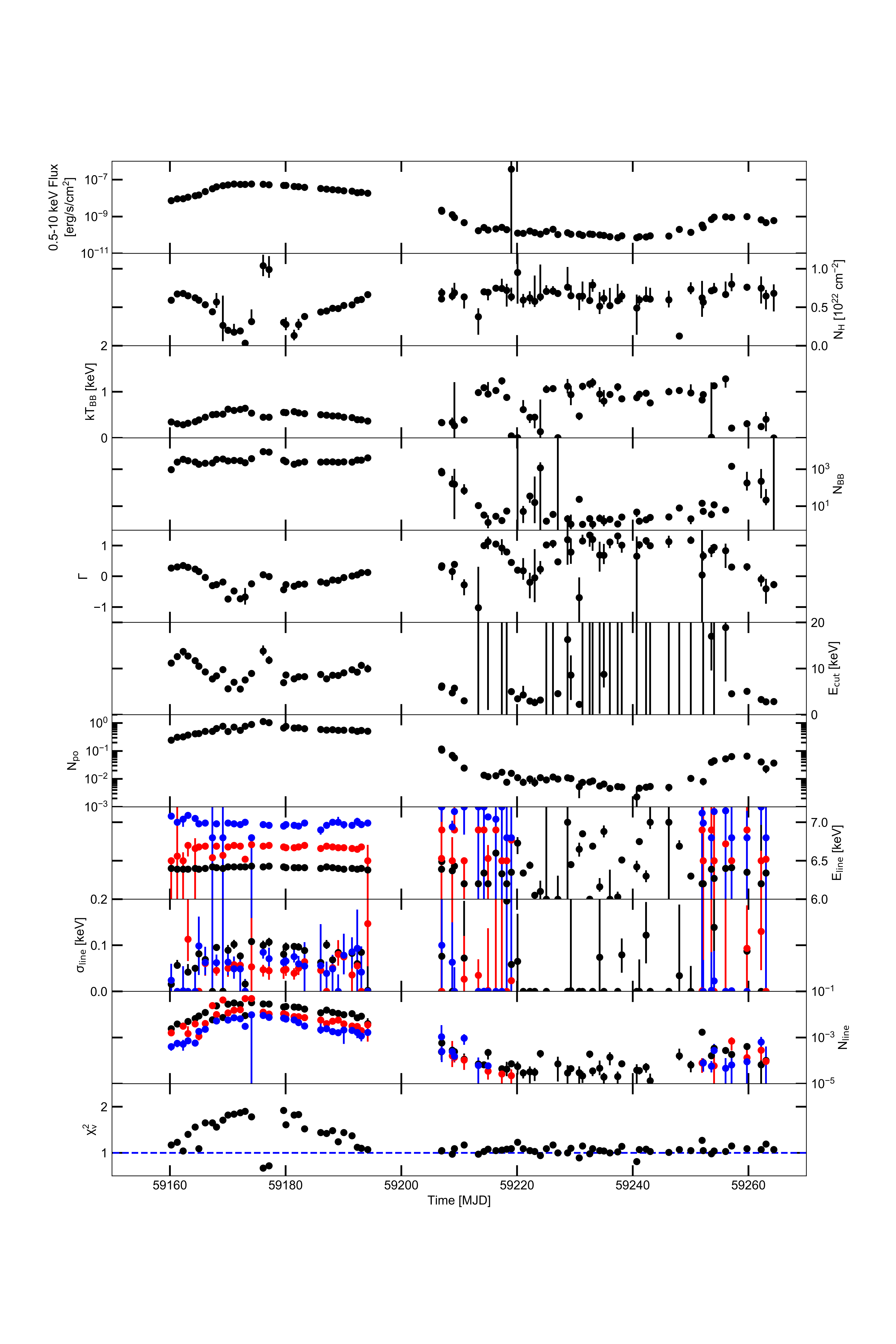}
    \caption{Light curves of the parameters fitted to all \textit{NICER} spectra of 1A 0535+262. In order from top to bottom, the panels show: the X-ray flux, absorption column; black body temperature and normalisation; power law index, exponential cut-off energy, and normalisation; Gaussian energy, width, and normalisation of all three lines (in black, red, and blue); and $\chi^2_\nu$ of the fit. Several parameters show larger error bars, often cased by a relatively short exposure At the highest fluxes, early in the outburst, instrumental residuals increase $\chi^2_\nu$ systematically. At lower flux (i.e. after MJD 59200), several parameters are poorly constrained: especially the Gaussian line parameters and cutoff energy show significant uncertainties. All parameter values are available in the machine-readable parameter file.}
    \label{fig:NICER_params_a0535}
\end{figure*}

In Figures \ref{fig:swift_params_a0535} and \ref{fig:NICER_params_a0535}, we plot the evolution of the parameters fitted to the \textit{Swift} and \textit{NICER} spectra, respectively. In both data sets, the parameter uncertainties increase systematically towards lower X-ray flux, as expected. However, other effects cause large uncertainties in several observations as well: short exposure times, for instance, or the complexity of the \textit{NICER} model, especially regarding the Gaussian lines and cutoff energy. In the \textit{NICER} light curves, one can see how the Gaussian line energies are fitted at reasonably stable values of $\sim 6.4$, $\sim 6.67$, and $\sim 6.96$ keV at high X-ray flux, as expected for the iron K$\alpha$ complex. At lower fluxes, however, the energies, as well as their widths, are more poorly constrained. Finally, the $\chi^2_\nu$ of the \textit{NICER} fits clearly peaks at high flux, due to the appearance of significant instrumental residuals below $3$ keV -- we stress that the $\chi^2_\nu$ values in both Table and Figure are calculated \textit{after} applying the energy band restriction. 

All parameters values, with uncertainties, as well as the information from Tables \ref{tab:NICER_0535_1} and \ref{tab:NICER_0535_2}, are available in machine-readable format in the other files in these Online Supplementary Materials. Additionally, we include machine-readable files with the fluxes in three energy bands (0.5-10 keV, 1-10 keV, 2-10 keV) for all sources. Alternatively, all machine-readable table files can be accessed via the \textsc{github} repository of this paper at \url{https://github.com/jvandeneijnden/RadioMonitoringOfTransientBeXRBs}.

\begin{figure*}
    \centering
    \includegraphics[width=\textwidth]{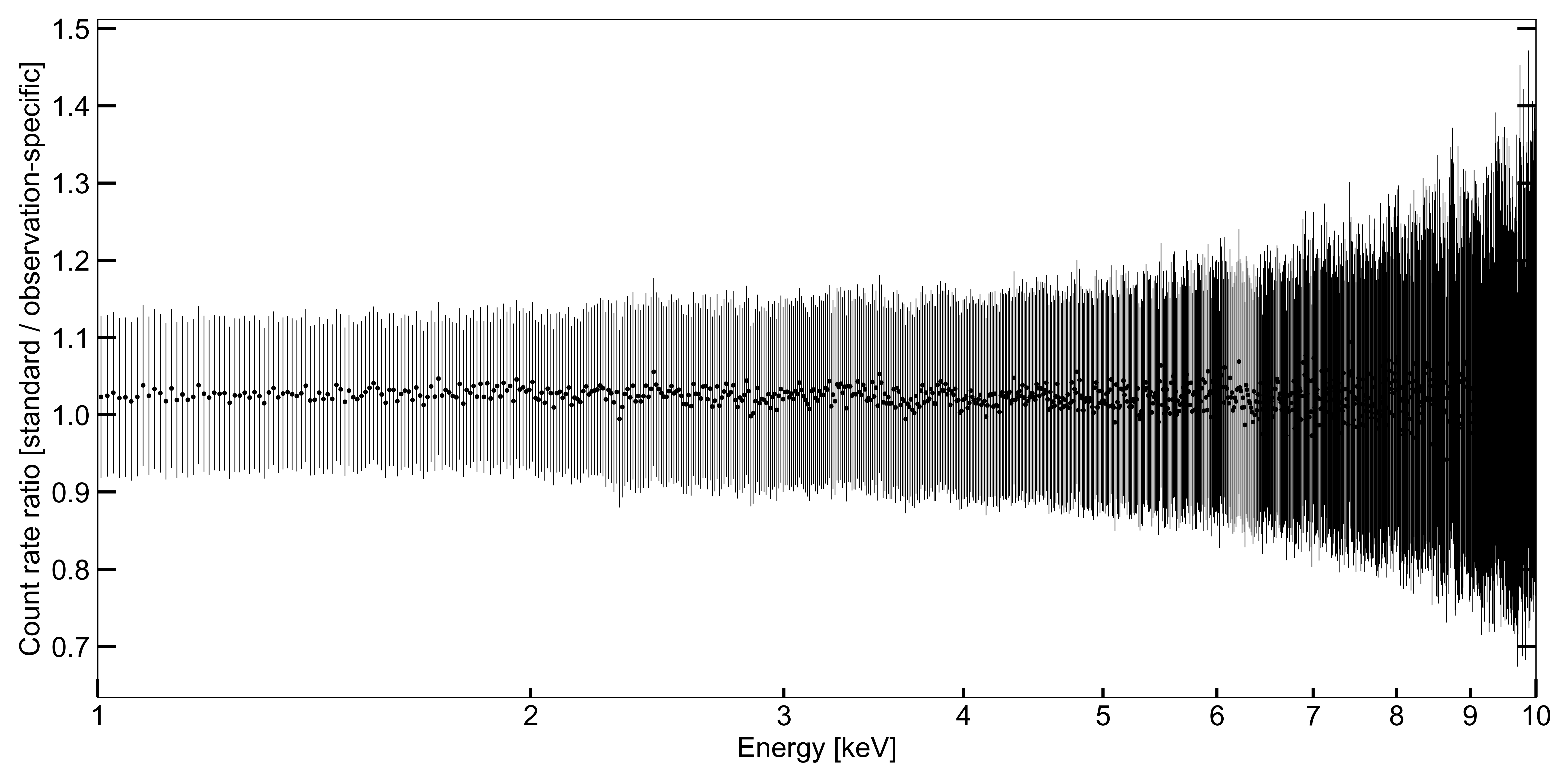}
    \caption{The ratio between the \textit{NICER} spectra measured for ObsID 3200360135, modelled with standard and observation-specific response and background files.} 
    \label{fig:resp_comp}
\end{figure*}

As detailed in the main paper, the \textit{NICER} spectra were modelled using the standard, pre-calculated reponse files for the \textsc{rmf} and \textsc{arf}. However, we explicitly tested the use of \textsc{nicerarf} and \textsc{nicerrmf} to generate observation-specific response files, following the \textit{NICER} analysis threads\footnote{\url{https://heasarc.gsfc.nasa.gov/docs/nicer/analysis\_threads/arf-rmf/}}. In addition, we also generated background files using the pre-release version v0p6 of \textsc{nicer\_bkg\_estimator}\footnote{\url{https://heasarc.gsfc.nasa.gov/docs/nicer/tools/nicer\_bkg\_est\_tools.html}}. However, we found that this does not lead to significant differences in the analysed spectra. In Figure \ref{fig:resp_comp}, we show the ratio between the \textit{NICER} spectra, measured in counts/s/keV from ObsID 3200360135, plotted with the pre-calculated and observation-specific instrument response. While the energy-averaged ratio is slightly offset from unity, due to the subtraction of the background in the latter spectrum, no energy-dependent structures can be identified. In fact, at all energies, the ratio is consistent with unity. This observation, shown as the blue spectrum in Figure 1 of the main paper, shows strong instrumental effects below $2$ keV; evidently, these effects are not reduced by the alternative response and background approach.

Finally, we also tested whether the binsize used in the analysis affected the presence of these residual features and the measured parameters. Rebinning spectra by a factor 3 or 10 did not alleviate the issues with residuals, but did naturally lead to slightly enhanced uncertainties on the parameters and higher $\chi^2_\nu$ values. Importantly, the fitted parameters and derived fluxes did not change significantly. However, we note that the derived errors and $\chi^2_\nu$ listed in these supplementary materials are only valid for the default \textit{NICER} binning.

\section{MCMC run figures}
\label{sec:appendix_MCMC}

\begin{figure*}
    \centering
    \includegraphics[width=\textwidth]{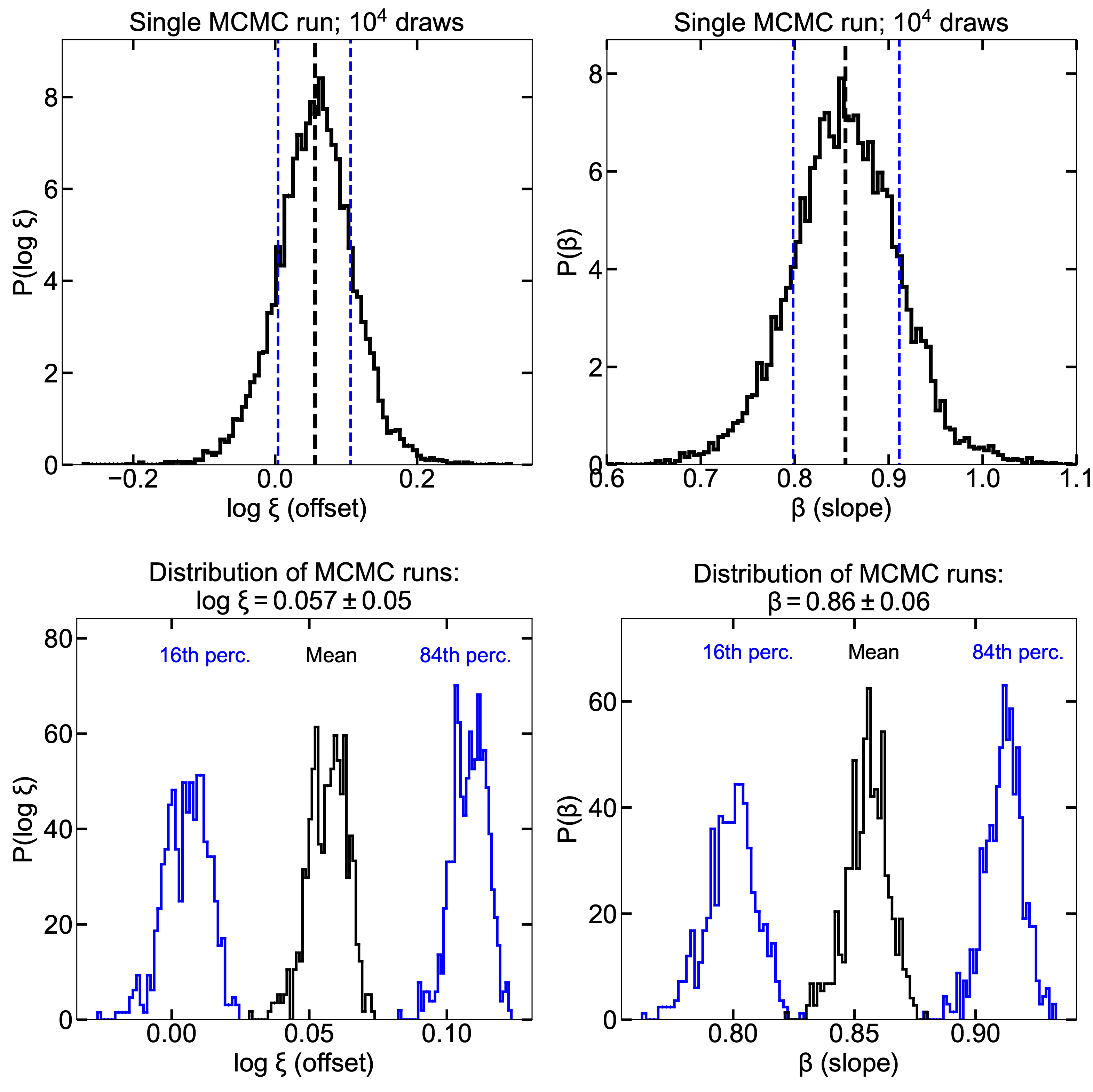}
    \caption{\textit{top panels:} posterior distributions of the offset $\log \xi$ and slope $\beta$ for a single MCMC run. \textit{Bottom panels:} distributions of the 16$^{\rm th}$ and 84$^{\rm th}$ percentile (blue), as well as the mean (black), of the offset and slope after 500 MCMC runs.}
    \label{fig:MCMC_4panels}
\end{figure*}

In Figure \ref{fig:MCMC_4panels}, we show the posterior distributions of the offset $\log \xi$ and slope $\beta$ of a single MCMC run to fit the behaviour of giant BeXRB outbursts in the $L_X$--$L_R$ plane (top left and right panels, respectively). In the bottom, we show the distribution of the 16$^{\rm th}$ and 84$^{\rm th}$ percentile (blue), as well as the mean (black), of the same two parameters, after 500 MCMC runs (bottom left and right panels, respectively). These percentiles and the mean are indicated in the top panels as well, for the single example run.

\begin{table*}
\caption{Details of the radio observations in this work. For each studied outburst, we list the observation number, date in MJD, observation length in hours, observing frequency $\nu$ and bandwidth $\Delta \nu$ in GHz, measured flux density or $3-\sigma$ upper limit in $\mu$Jy, the observatory and its configuration, and finally the program ID. These program IDs provide the most straightforward access to publicly available data in the observatory archives.}
\label{tab:radio_obs}
\begin{tabular}{lllllllll}
\hline
\multicolumn{9}{c}{\textit{1A 0535+262}} \\ 
\hline
Number & MJD & $\Delta T$ [hr] & $\nu$ [GHz] & $\Delta\nu$ [GHz] & Flux density [$\mu$Jy] & Observatory & Configuration & Program ID \\
\hline
1 & 59163.42 & 1 & 6 & 4 & $12.5 \pm 3.9$ & VLA & BnA & 20A-171 \\ 
2 & 59168.47 & 1 & 6 & 4 & $39.2 \pm 4.0$ & VLA & BnA$\rightarrow$A & 20A-171 \\ 
3 & 59170.25 & 1 & 6 & 4& $27.9 \pm 4.0$ & VLA & BnA$\rightarrow$A & 20A-171 \\ 
4 & 59172.23 & 1 & 6 & 4& $27.9 \pm 4.3$& VLA & BnA$\rightarrow$A & 20A-171 \\ 
5 & 59177.12 & 1 & 6 & 4& $20.3 \pm 3.3$& VLA & BnA$\rightarrow$A & 20A-171 \\ 
6 & 59178.4 & 1 & 6 & 4& $27.5 \pm 4.0$& VLA & BnA$\rightarrow$A & 20A-171 \\ 
7 & 59178.45 & 1 & 6 & 4& $20.6 \pm 4.0$& VLA & BnA$\rightarrow$A & 20A-171 \\ 
8 & 59185.09 & 1 & 6 & 4& $21.6 \pm 4.3$& VLA & BnA$\rightarrow$A & 20A-171 \\ 
9 & 59186.47 & 1 & 6 & 4& $16.2 \pm 4.2$& VLA & BnA$\rightarrow$A & SM0612 \\ 
10 & 59189.09 & 1 & 6 & 4& $9.8 \pm 4.1$& VLA & BnA$\rightarrow$A & 20A-171 \\ 
11 & 59191.40 & 1 & 6 & 4& $ < 11.7$ & VLA & BnA$\rightarrow$A & 20A-171 \\ 
12 & 59201.31 & 1 & 6 & 4& $ < 12.9 $ & VLA & A & SG9053\\ 
13 & 59202.36 & 1 & 6 & 4& $ < 12.6$ & VLA & A & SG9053\\ 
14 & 59220.01 & 1 & 6 & 4& $ < 18.0$ & VLA & A & SG9053\\ 
15 & 59235.07 & 1 & 6 & 4& $ < 16.5$ & VLA & A & SG9053\\ 
16 & 59247.93 & 1 & 6 & 4& $ < 14.0$ & VLA & A & SG9053\\ 
17 & 59262.05 & 1 & 6 & 4& $ < 13.0$ & VLA & A & SG9053\\ 
18 & 59266.89 & 1 & 6 & 4& $ < 13.0$ & VLA & A & SG9053\\ 
19 & 59278.95 & 1 & 6 & 4& $ < 15.0$ & VLA & A$\rightarrow$D & SG9053\\ 
20 & 59279.10 & 1 & 6 & 4& $ < 15.0$ & VLA & A$\rightarrow$D & SG9053\\ 
\hline
\multicolumn{9}{c}{\textit{GRO J1008-57: giant outburst}} \\ 
\hline
Number & MJD & $\Delta T$ [hr] & $\nu$ [GHz] & $\Delta\nu$ [GHz] & Flux density [$\mu$Jy] & Observatory & Configuration & Program ID \\
\hline
\multirow{2}{*}{1} & \multirow{2}{*}{58984.43} & \multirow{2}{*}{4} &  5.5 & 1 & $< 33.0$ & \multirow{2}{*}{ATCA} & \multirow{2}{*}{1.5C} & \multirow{2}{*}{C3299} \\
& & & 9.0 & 1  & $< 22.5$ & & & \\ \cline{3-6}
\multirow{2}{*}{2} & \multirow{2}{*}{58990.42} & \multirow{2}{*}{4}& 5.5 & 1 & $< 36.0$ & \multirow{2}{*}{ATCA} & \multirow{2}{*}{1.5C} & \multirow{2}{*}{C3299}\\
& & & 9.0 & 1  & $< 24.0$ & & & \\ \cline{3-6}
\multirow{2}{*}{3} & \multirow{2}{*}{59001.19} & \multirow{2}{*}{4}& 5.5 & 1 & $< 33.0$ & \multirow{2}{*}{ATCA} & \multirow{2}{*}{1.5C} & \multirow{2}{*}{C3299} \\
& & & 9.0 & 1  & $< 22.5$ & & & \\  \cline{3-6}
\multirow{2}{*}{4} & \multirow{2}{*}{59013.15} & \multirow{2}{*}{4}& 5.5 & 1 & $< 16.0$ & \multirow{2}{*}{ATCA} & \multirow{2}{*}{1.5C} & \multirow{2}{*}{C3299} \\
& & & 9.0 & 1  & $< 23.0$ & & & \\ \cline{3-6}
\multirow{2}{*}{5} & \multirow{2}{*}{59025.00} & \multirow{2}{*}{4}& 5.5 & 1 & $< 180.0$ & \multirow{2}{*}{ATCA} & \multirow{2}{*}{H214} & \multirow{2}{*}{C3299} \\
& & & 9.0 & 1  & $< 66.0$ & & & \\ \cline{3-6}
\multirow{2}{*}{\textit{1--4}} & \multirow{2}{*}{N/A} & \multirow{2}{*}{16}& 5.5 & 1 & $< 20.0$ & \multirow{2}{*}{ATCA} & \multirow{2}{*}{1.5C} & \multirow{2}{*}{C3299} \\
& & & 9.0 & 1  & $< 24.0$ & & & \\
\hline
\multicolumn{9}{c}{\textit{GRO J1008-57: periastron outburst}} \\ 
\hline
Number & MJD & $\Delta T$ [hr] & $\nu$ [GHz] & $\Delta\nu$ [GHz] & Flux density [$\mu$Jy] & Observatory & Configuration & Program ID \\
\hline
\multirow{2}{*}{1} & \multirow{2}{*}{58639.17} & \multirow{2}{*}{4}& 5.5 & 1  & $< 33.0$ & \multirow{2}{*}{ATCA} & \multirow{2}{*}{6A} & \multirow{2}{*}{C3298} \\
& & & 9.0 & 1  & $< 27.0$ & & & \\ \cline{3-6}
\multirow{2}{*}{2} & \multirow{2}{*}{58643.27} & \multirow{2}{*}{4}& 5.5 & 1  & $< 30.0$ & \multirow{2}{*}{ATCA} & \multirow{2}{*}{6A} & \multirow{2}{*}{C3298} \\
& & & 9.0 & 1  & $< 27.0$ & & & \\ \cline{3-6}
\multirow{2}{*}{3} & \multirow{2}{*}{58647.15} & \multirow{2}{*}{4}& 5.5 & 1  & $< 30.0$ & \multirow{2}{*}{ATCA} & \multirow{2}{*}{6A} & \multirow{2}{*}{C3298} \\
& & & 9.0 & 1  & $< 30.0$ & & & \\ \cline{3-6}
\multirow{2}{*}{4} & \multirow{2}{*}{58651.15} & \multirow{2}{*}{4}& 5.5 & 1  & $< 27.0$ & \multirow{2}{*}{ATCA} & \multirow{2}{*}{6A} & \multirow{2}{*}{C3298} \\
& & & 9.0 & 1  & $< 24.0$ & & & \\ \cline{3-6}
\multirow{2}{*}{5} & \multirow{2}{*}{58655.13} & \multirow{2}{*}{4}& 5.5 & 1  & $< 33.0$ & \multirow{2}{*}{ATCA} & \multirow{2}{*}{6A} & \multirow{2}{*}{C3298} \\
& & & 9.0 & 1  & $< 30.0$ & & & \\ \cline{3-6}
\multirow{2}{*}{6} & \multirow{2}{*}{58658.08} & \multirow{2}{*}{4}& 5.5 & 1  & $< 28.0$ & \multirow{2}{*}{ATCA} & \multirow{2}{*}{6A} & \multirow{2}{*}{C3298} \\
& & & 9.0 & 1  & $< 45.0$ & & & \\ \cline{3-6}
\multirow{2}{*}{\textit{1--6}} & \multirow{2}{*}{N/A} & \multirow{2}{*}{4}& 5.5 & 1  & $< 27.0$ & \multirow{2}{*}{ATCA} & \multirow{2}{*}{6A} & \multirow{2}{*}{C3298} \\
& & & 9.0 & 1  & $< 19.5$ & & & \\
\hline
\multicolumn{9}{c}{\textit{SAX J2103.5+4545: giant outburst}} \\ 
\hline
Number & MJD & $\Delta T$ [hr] & $\nu$ [GHz] & $\Delta\nu$ [GHz] & Flux density [$\mu$Jy] & Observatory & Configuration & Program ID \\
\hline
1 & 59101.02 & 2 & 6 & 4 & $ < 18.0$ &  VLA & B & 20A-171 \\
\hline
\end{tabular}\\
\end{table*}

\begin{table*}
\caption{Full details of the X-ray observations and spectral fits for the two outbursts of GRO J1008-57 and the giant outburst of SAX J2103.5+4545. We refer to the main text for full details on data selection and model choice. When a joint fit is performed to the \textit{Swift}/XRT PC and WT mode data, the exposure times of both instruments are listed. * As detailed in the text, we apply a correction based on the limited soft energy coverage of \textit{MAXI} and have divided the measured X-ray flux by a factor 1.7. Here, we list these corrected fluxes. ** Joint fit between all seventeen spectra, returning a single $N_H$ measurement and fit statistic.}
\label{tab:xrays_j1008_saxJ2103}
\begin{tabular}{llllllllll}
\hline
\multicolumn{10}{c}{\textit{GRO J1008-57: periastron outburst}} \\
\hline
Observatory/ & \multirow{2}{*}{\#} & \multirow{2}{*}{ObsID/mode} & \multirow{2}{*}{MJD} & Exposure & $N_H$ & \multirow{2}{*}{$\Gamma$} & 0.5-10 keV flux & \multirow{2}{*}{C-stat} & \multirow{2}{*}{DOF} \\
Instrument & & & & [seconds] & [$10^{22}$ cm$^{-2}$] & & [erg/s/cm$^2$] & & \\
\hline
\textit{Swift}/XRT & 1 & 0003103149/pc & 58639.91 & 951 & $2.2\pm0.4$ & $1.0\pm0.2$ & $(1.9\pm0.1)\times10^{-10}$ & 288 & 269 \\
\textit{Swift}/XRT & 2 & 0003103150/pc & 58643.24 & 1079 & $2.6\pm0.5$ & $1.1\pm0.2$ & $(6.3\pm0.4)\times10^{-10}$ & 241 & 315 \\
\textit{Swift}/XRT & 3 & 0003103151/pc+wt & 58647.02 & 659+136 & $2.3\pm0.3$ & $0.76\pm0.11$ & $(1.12\pm0.05)\times10^{-9}$ & 718 & 865 \\
\textit{Swift}/XRT & 4 & 0003103152/wt & 58652.21 & 829 & $3.0\pm0.1$ & $0.92\pm0.04$ & $(1.92\pm0.02)\times10^{-9}$ & 718 & 865 \\
\textit{Swift}/XRT & 5 & 0003103153/wt & 58655.80 & 362 & $3.5\pm0.3$ & $1.15\pm0.09$ & $(9.3\pm0.3)\times10^{-10}$ & 757 & 711 \\
\textit{Swift}/XRT & 6 & 0003103154/pc+wt & 58658.45 & 687+258 & $2.7\pm0.3$ & $0.9\pm0.1$ & $(1.07\pm0.04)\times10^{-9}$ & 888 & 1063 \\
\hline
\multicolumn{10}{c}{\textit{GRO J1008-57: giant outburst}} \\
\hline
Observatory/ & \multirow{2}{*}{\#} & \multirow{2}{*}{ObsID/mode} & \multirow{2}{*}{MJD} & Exposure & $N_H$ & \multirow{2}{*}{$\Gamma$} & 0.5-10 keV flux* & \multirow{2}{*}{$\chi^2_\nu$} & \multirow{2}{*}{DOF} \\
Instrument & & & & [seconds] & [$10^{22}$ cm$^{-2}$] & & [erg/s/cm$^2$] & & \\
\hline
\textit{MAXI}/GSC & 1 & -- & 58980 \& 58983 & 929 & \multirow{17}{*}{$4.5\pm0.8$**} & $1.6 \pm 0.3$ & $(1.2 \pm 0.5)\times10^{-9}$ & \multirow{17}{*}{1.036**} & \multirow{17}{*}{466**} \\
\textit{MAXI}/GSC & 2 & -- & 58986 & 310 & & $0.8 \pm 0.3$ & $(1.7 \pm 0.6)\times10^{-9}$ & & \\
\textit{MAXI}/GSC & 3 & -- & 58993 & 580 & & $1.1 \pm 0.1$ & $(4.0 \pm 0.7)\times10^{-9}$ & & \\
\textit{MAXI}/GSC & 4 & -- & 58994 & 581 & & $1.4 \pm 0.2$ & $(4.5 \pm 0.9)\times10^{-9}$ & & \\
\textit{MAXI}/GSC & 5 & -- & 58995 & 620 & & $1.6 \pm 0.2$ & $(5 \pm 1)\times10^{-9}$ & & \\
\textit{MAXI}/GSC & 6 & -- & 58996 & 1072 & & $1.3 \pm 0.1$ & $(5.0 \pm 0.7)\times10^{-9}$ & & \\
\textit{MAXI}/GSC & 7 & -- & 58997 & 1107 & & $1.2 \pm 0.1$ & $(4.3 \pm 0.6)\times10^{-9}$ & & \\
\textit{MAXI}/GSC & 8 & -- & 58998 & 2533 & & $1.3 \pm 0.1$ & $(4.1 \pm 0.5)\times10^{-9}$ & & \\
\textit{MAXI}/GSC & 9 & -- & 58999 & 2603 & & $1.5 \pm 0.1$ & $(4.4 \pm 0.6)\times10^{-9}$ & & \\
\textit{MAXI}/GSC & 10 & -- & 59000 & 1466 & & $1.4 \pm 0.1$ & $(3.8 \pm 0.6)\times10^{-9}$ & & \\
\textit{MAXI}/GSC & 11 & -- & 59012 & 2188 & & $1.5 \pm 0.1$ & $(1.6 \pm 0.3)\times10^{-9}$ & & \\
\textit{MAXI}/GSC & 12 & -- & 59013 & 2495 & & $1.5 \pm 0.1$ & $(1.5 \pm 0.3)\times10^{-9}$ & & \\
\textit{MAXI}/GSC & 13 & -- & 59014 & 2773 & & $1.4 \pm 0.1$ & $(1.2 \pm 0.3)\times10^{-9}$ & & \\
\textit{MAXI}/GSC & 14 & -- & 59015 & 2170 & & $1.5 \pm 0.2$ & $(1.0 \pm 0.3)\times10^{-9}$ & & \\
\textit{MAXI}/GSC & 15 & -- & 59016 & 1937 & & $1.6 \pm 0.2$ & $(8 \pm 3)\times10^{-10}$ & & \\
\textit{MAXI}/GSC & 16 & -- & 59020$\rightarrow$59022 & 7149 & & $2.1 \pm 0.4$ & $(5 \pm 3)\times10^{-10}$ & & \\
\textit{MAXI}/GSC & 17 & -- & 59023$\rightarrow$59025 & 10090 & & $2.1 \pm 0.4$ & $(4 \pm 2)\times10^{-10}$ & & \\
\hline
\multicolumn{10}{c}{\textit{SAX J2103.5+4545: giant outburst}} \\
\hline
\multirow{2}{*}{Observatory} & \multirow{2}{*}{\#} & \multirow{2}{*}{ObsID/mode} & \multirow{2}{*}{MJD} & Exposure & $N_H$ & \multirow{2}{*}{$\Gamma$} & 0.5-10 keV flux & \multirow{2}{*}{C-stat} & \multirow{2}{*}{DOF} \\
& & & & [seconds] & [$10^{22}$ cm$^{-2}$] & & [erg/s/cm$^2$] & & \\
\hline
\textit{Swift}/XRT & 1 & 00030922079/wt & 59106.25 & 1026 & $1.5\pm0.2$ & $0.8\pm0.1$ & $(2.5\pm0.1)\times10^{-10}$ & 599 & 654 \\
\textit{Swift}/XRT & 2 & 00030922080/wt & 59107.84 & 953 & $1.5\pm0.3$ &  $0.7\pm0.1$ & $(2.2\pm0.1)\times10^{-10}$ & 497 & 598 \\ \hline
\end{tabular}\\
\end{table*}

\begin{table*}
\caption{Details of the X-ray spectral analysis of \textit{NICER} and \textit{Swift} monitoring of 1A 0535+262. In this Table, we list information about these observations, such as observatory, ObsID and observing mode, and time and exposure of the observation. We also list details about the spectral analysis, such as the fitted band, the measured flux, the number of narrow Gaussian model components fitted, and the fit statistic. Continues in Table \ref{tab:NICER_0535_2}. The flux error is listed at the $1-\sigma$ level.}
\label{tab:NICER_0535_1}
\begin{tabular}{lllllllllll}
\hline
Observatory/ & \multirow{2}{*}{Obs. \#} & \multirow{2}{*}{ObsID/mode} & \multirow{2}{*}{MJD} & Exposure & Fitted band & Flux & Error & \multirow{2}{*}{Gaussians} & \multirow{2}{*}{$\chi^2_{\nu}$} & \multirow{2}{*}{DOF} \\
Instrument & & & & [second] & [keV] & \multicolumn{2}{c}{[0.5-10 keV; erg/s/cm$^2$]} & & &  \\
\hline
\textit{NICER} & 1 & 3200360123 & 59160.23 & 8288 & 2-10 & 7.25e-09 & 6.60e-12 & 3 & 1.17 & 884 \\
\textit{NICER} & 2 & 3200360124 & 59161.26 & 3609 & 2-10 & 9.15e-09 & 1.05e-11 & 3 & 1.23 & 884 \\
\textit{NICER} & 3 & 3200360125 & 59162.29 & 2169 & 2-10 & 9.22e-09 & 1.54e-11 & 3 & 1.04 & 884 \\
\textit{NICER} & 4 & 3200360126 & 59163.13 & 5231 & 2-10 & 1.13e-08 & 1.08e-11 & 3 & 1.40 & 884 \\
\textit{NICER} & 5 & 3200360127 & 59164.36 & 12166 & 2-10 & 1.34e-08 & 7.51e-12 & 3 & 1.56 & 884 \\
\textit{NICER} & 6 & 3200360128 & 59165.0 & 1510 & 2-10 & 1.47e-08 & 2.19e-11 & 3 & 1.09 & 884 \\
\textit{NICER} & 7 & 3200360129 & 59166.1 & 4591 & 2-10 & 2.27e-08 & 1.60e-11 & 3 & 1.65 & 884 \\
\textit{NICER} & 8 & 3200360130 & 59167.34 & 2569 & 2-10 & 3.22e-08 & 2.49e-11 & 3 & 1.65 & 884 \\
\textit{NICER} & 9 & 3200360131 & 59168.05 & 3608 & 2-10 & 4.07e-08 & 3.03e-11 & 3 & 1.56 & 784 \\
\textit{NICER} & 10 & 3200360132 & 59169.15 & 3352 & 2-10 & 4.71e-08 & 3.36e-11 & 3 & 1.71 & 784 \\
\textit{NICER} & 11 & 3200360133 & 59170.04 & 1333 & 2-10 & 5.21e-08 & 4.79e-11 & 3 & 1.82 & 884 \\
\textit{NICER} & 12 & 3200360134 & 59171.08 & 2918 & 2-10 & 5.77e-08 & 3.71e-11 & 3 & 1.84 & 784 \\
\textit{NICER} & 13 & 3200360135 & 59172.17 & 1814 & 2-10 & 5.49e-08 & 4.19e-11 & 3 & 1.87 & 884 \\
\textit{NICER} & 14 & 3200360136 & 59173.01 & 3451 & 2-10 & 5.57e-08 & 3.17e-11 & 3 & 1.90 & 784 \\
\textit{NICER} & 15 & 3200360137 & 59174.11 & 3397 & 2-10 & 5.80e-08 & 2.93e-11 & 3 & 1.78 & 784 \\
\textit{NICER} & 16 & 3200360139 & 59176.11 & 8572 & 2-10* & 5.63e-08 & 4.99e-11 & 3 & 0.67 & 784 \\
\textit{NICER} & 17 & 3200360140 & 59177.14 & 6892 & 2-10* & 5.31e-08 & 4.93e-11 & 3 & 0.72 & 784 \\
\textit{NICER} & 18 & 3200360141 & 59179.66 & 2447 & 1-10 & 4.88e-08 & 3.36e-11 & 3 & 1.92 & 884 \\
\textit{NICER} & 19 & 3200360142 & 59180.04 & 3683 & 2-10 & 4.87e-08 & 3.13e-11 & 3 & 1.61 & 784 \\
\textit{NICER} & 20 & 3200360143 & 59181.46 & 5783 & 2-10 & 4.32e-08 & 2.26e-11 & 3 & 1.82 & 784 \\
\textit{NICER} & 21 & 3200360144 & 59182.24 & 5438 & 2-10 & 4.17e-08 & 2.42e-11 & 3 & 1.83 & 784 \\
\textit{NICER} & 22 & 3200360145 & 59183.27 & 1718 & 1-10 & 3.82e-08 & 3.52e-11 & 3 & 1.52 & 884 \\
\textit{NICER} & 23 & 3200360147 & 59186.05 & 2559 & 1-10 & 3.30e-08 & 2.65e-11 & 3 & 1.44 & 884 \\
\textit{NICER} & 24 & 3200360148 & 59187.08 & 2518 & 1-10 & 3.07e-08 & 1.66e-11 & 3 & 1.42 & 884 \\
\textit{NICER} & 25 & 3200360149 & 59188.11 & 3086 & 1-10 & 2.86e-08 & 2.21e-11 & 3 & 1.48 & 884 \\
\textit{NICER} & 26 & 3200360150 & 59189.08 & 1786 & 1-10 & 2.76e-08 & 2.86e-11 & 3 & 1.24 & 884 \\
\textit{NICER} & 27 & 3200360151 & 59190.05 & 3234 & 1-10 & 2.48e-08 & 1.98e-11 & 3 & 1.44 & 884 \\
\textit{NICER} & 28 & 3200360152 & 59191.47 & 3560 & 1-10 & 2.39e-08 & 1.85e-11 & 3 & 1.37 & 884 \\
\textit{NICER} & 29 & 3200360153 & 59192.37 & 1662 & 1-10 & 1.96e-08 & 2.42e-11 & 3 & 1.12 & 884 \\
\textit{NICER} & 30 & 3200360154 & 59193.09 & 1445 & 1-10 & 2.05e-08 & 2.66e-11 & 3 & 1.10 & 884 \\
\textit{NICER} & 31 & 3200360155 & 59194.18 & 852 & 1-10 & 1.83e-08 & 3.15e-11 & 3 & 1.07 & 884 \\
\textit{NICER} & 32 & 3200360156 & 59206.93 & 949 & 1-10 & 2.22e-09 & 9.68e-12 & 3 & 1.04 & 884 \\
\textit{NICER} & 33 & 3200360157 & 59206.99 & 1635 & 1-10 & 1.94e-09 & 6.91e-12 & 3 & 1.05 & 884 \\
\textit{NICER} & 34 & 3200360158 & 59208.8 & 901 & 1-10 & 1.26e-09 & 7.12e-12 & 3 & 0.97 & 884 \\
\textit{NICER} & 35 & 3200360159 & 59209.17 & 1948 & 1-10 & 8.81e-10 & 4.44e-12 & 3 & 1.09 & 884 \\
\textit{NICER} & 36 & 3200360160 & 59210.85 & 1609 & 1-10 & 4.71e-10 & 3.08e-12 & 3 & 1.17 & 884 \\
\textit{NICER} & 37 & 3200360161 & 59213.3 & 378 & 1-10 & 1.71e-10 & 5.97e-12 & 3 & 0.97 & 884 \\
\textit{NICER} & 38 & 3200360162 & 59214.27 & 7263 & 1-10 & 2.57e-10 & 1.12e-12 & 3 & 1.03 & 884 \\
\textit{NICER} & 39 & 3200360163 & 59214.99 & 3617 & 1-10 & 1.86e-10 & 1.28e-12 & 3 & 1.08 & 884 \\
\textit{NICER} & 40 & 3200360164 & 59216.34 & 12302 & 1-10 & 2.16e-10 & 7.70e-13 & 3 & 1.05 & 884 \\
\textit{NICER} & 41 & 3200360165 & 59217.37 & 6975 & 1-10 & 2.60e-10 & 1.15e-12 & 3 & 1.06 & 884 \\
\textit{NICER} & 42 & 3200360166 & 59218.21 & 3513 & 1-10 & 1.96e-10 & 1.38e-12 & 3 & 1.08 & 884 \\
\textit{NICER} & 43 & 3200360167 & 59219.0 & 6149 & 1-10 & 3.67e-07 & 3.61e-05 & 3 & 1.09 & 884 \\
\textit{NICER} & 44 & 3200360168 & 59220.08 & 1881 & 1-10 & 1.27e-10 & 1.41e-12 & 1 & 1.23 & 890 \\
\textit{NICER} & 45 & 3200360169 & 59221.05 & 5712 & 1-10 & 1.23e-10 & 9.17e-13 & 1 & 1.09 & 890 \\
\textit{NICER} & 46 & 3200360170 & 59222.21 & 2535 & 1-10 & 1.63e-10 & 1.36e-12 & 1 & 1.05 & 890 \\
\textit{NICER} & 47 & 3200360171 & 59223.05 & 3151 & 1-10 & 1.34e-10 & 1.12e-12 & 1 & 1.03 & 890 \\
\textit{NICER} & 48 & 3200360172 & 59224.02 & 2134 & 1-10 & 1.12e-10 & 1.27e-12 & 1 & 0.94 & 890 \\
\textit{NICER} & 49 & 3200360173 & 59225.05 & 3180 & 1-10 & 1.56e-10 & 1.32e-12 & 1 & 1.09 & 890 \\
\textit{NICER} & 50 & 3200360174 & 59226.21 & 2187 & 1-10 & 2.05e-10 & 1.71e-12 & 0 & 1.17 & 893 \\
\textit{NICER} & 51 & 3200360175 & 59227.05 & 3519 & 1-10 & 1.07e-10 & 9.54e-13 & 1 & 1.00 & 890 \\
\textit{NICER} & 52 & 3200360176 & 59228.73 & 885 & 1-10 & 1.33e-10 & 2.29e-12 & 1 & 1.00 & 890 \\
\textit{NICER} & 53 & 3200360177 & 59229.31 & 4755 & 1-10 & 1.10e-10 & 9.29e-13 & 1 & 1.10 & 890 \\
\textit{NICER} & 54 & 3200360178 & 59230.73 & 1380 & 1-10 & 1.11e-10 & 1.58e-12 & 1 & 0.89 & 890 \\
\textit{NICER} & 55 & 3200360179 & 59231.31 & 5382 & 1-10 & 9.47e-11 & 7.24e-13 & 1 & 1.15 & 890 \\
\textit{NICER} & 56 & 3200360180 & 59232.54 & 2473 & 1-10 & 1.13e-10 & 1.16e-12 & 1 & 0.98 & 890 \\
\textit{NICER} & 57 & 3200360181 & 59233.06 & 4095 & 1-10 & 1.08e-10 & 8.86e-13 & 1 & 1.09 & 890 \\ \hline
\end{tabular}\\
\end{table*}

\begin{table*}
\caption{Continuation of Table \ref{tab:NICER_0535_1}. * The exceptionally low $\chi^2_\nu$ value is caused by the short, $15$ second exposure and resulting over-fitting of the spectrum.}
\label{tab:NICER_0535_2}
\begin{tabular}{lllllllllll}
\hline
\multirow{2}{*}{Observatory} & \multirow{2}{*}{Obs. \#} & \multirow{2}{*}{ObsID/mode} & \multirow{2}{*}{MJD} & Exposure & Fitted band & Flux & Error & \multirow{2}{*}{Gaussians} & \multirow{2}{*}{$\chi^2_{\nu}$} & \multirow{2}{*}{DOF} \\
& & & & [second] & [keV] & \multicolumn{2}{c}{[0.5-10 keV; erg/s/cm$^2$]} & & &  \\
\hline
\textit{NICER} & 58 & 3200360182 & 59234.28 & 2712 & 1-10 & 1.01e-10 & 1.15e-12 & 1 & 1.05 & 890 \\
\textit{NICER} & 59 & 3200360183 & 59234.99 & 8227 & 1-10 & 9.52e-11 & 6.94e-13 & 1 & 1.04 & 890 \\
\textit{NICER} & 60 & 3200360184 & 59236.03 & 4121 & 1-10 & 8.12e-11 & 7.73e-13 & 1 & 1.00 & 890 \\
\textit{NICER} & 61 & 3200360185 & 59237.38 & 2583 & 1-10 & 7.09e-11 & 9.16e-13 & 1 & 1.02 & 890 \\
\textit{NICER} & 62 & 3200360186 & 59238.09 & 3829 & 1-10 & 9.06e-11 & 8.76e-13 & 1 & 1.14 & 890 \\
\textit{NICER} & 63 & 3200360187 & 59240.67 & 703 & 1-10 & 7.09e-11 & 2.12e-12 & 1 & 0.81 & 890 \\
\textit{NICER} & 64 & 3200360188 & 59241.13 & 4247 & 1-10 & 8.09e-11 & 7.93e-13 & 1 & 1.07 & 890 \\
\textit{NICER} & 65 & 3200360189 & 59242.29 & 2938 & 1-10 & 7.87e-11 & 9.01e-13 & 1 & 1.08 & 890 \\
\textit{NICER} & 66 & 3200360190 & 59243.0 & 4707 & 1-10 & 9.09e-11 & 7.60e-13 & 1 & 1.03 & 890 \\
\textit{NICER} & 67 & 3200360191 & 59246.23 & 1420 & 1-10 & 8.77e-11 & 1.45e-12 & 1 & 1.01 & 890 \\
\textit{NICER} & 68 & 3200360192 & 59248.03 & 847 & 1-10 & 2.00e-10 & 3.74e-12 & 1 & 1.07 & 890 \\
\textit{NICER} & 69 & 3200360193 & 59250.03 & 989 & 1-10 & 1.39e-10 & 2.17e-12 & 1 & 1.05 & 890 \\
\textit{NICER} & 70 & 3200360194 & 59251.97 & 1410 & 1-10 & 3.44e-10 & 4.18e-12 & 3 & 1.27 & 884 \\
\textit{NICER} & 71 & 3200360195 & 59252.16 & 1176 & 1-10 & 2.50e-10 & 2.85e-12 & 3 & 1.05 & 884 \\
\textit{NICER} & 72 & 3200360196 & 59253.59 & 6149 & 1-10 & 6.96e-10 & 1.88e-12 & 3 & 0.98 & 884 \\
\textit{NICER} & 73 & 3200360197 & 59254.04 & 1441 & 1-10 & 9.24e-10 & 4.85e-12 & 3 & 1.04 & 884 \\
\textit{NICER} & 74 & 3200360198 & 59256.04 & 1413 & 1-10 & 9.65e-10 & 5.18e-12 & 3 & 1.03 & 884 \\
\textit{NICER} & 75 & 3200360199 & 59257.07 & 1496 & 1-10 & 8.99e-10 & 4.61e-12 & 3 & 1.15 & 884 \\
\textit{NICER} & 76 & 3200360201 & 59259.72 & 1402 & 1-10 & 1.00e-09 & 5.52e-12 & 3 & 1.09 & 884 \\
\textit{NICER} & 77 & 3200360202 & 59262.17 & 695 & 1-10 & 6.67e-10 & 6.05e-12 & 3 & 1.07 & 884 \\
\textit{NICER} & 78 & 3200360203 & 59263.01 & 1144 & 1-10 & 4.67e-10 & 3.46e-12 & 3 & 1.19 & 884 \\
\textit{NICER} & 79 & 3200360204 & 59264.37 & 298 & 1-10 & 6.03e-10 & 8.10e-12 & 0 & 1.07 & 893 \\
\hline
\textit{Swift} & 1 & 00035066077/pc & 59162.31 & 2964 & 1-10 & 1.25e-08 & 1.86e-10 & 0 & 1.03 & 811 \\
\textit{Swift} & 2 & 00035066078/wt & 59174.07 & 344 & 1-10 & 6.43e-08 & 6.42e-10 & 0 & 1.36 & 863 \\
\textit{Swift} & 3 & 00035066080/wt & 59180.84 & 908 & 1-10 & 5.11e-08 & 2.61e-10 & 0 & 1.16 & 891 \\
\textit{Swift} & 4 & 00089186001/wt & 59186.8 & 1380 & 1-10 & 3.06e-08 & 1.46e-10 & 0 & 1.29 & 894 \\
\textit{Swift} & 5 & 00013945001/wt & 59202.61 & 1053 & 1-10 & 5.12e-09 & 3.32e-11 & 0 & 1.13 & 884 \\
\textit{Swift} & 6 & 00013945002/wt & 59205.87 & 554 & 1-10 & 2.72e-09 & 2.78e-11 & 0 & 1.06 & 814 \\
\textit{Swift} & 7 & 00013945003/wt & 59208.53 & 898 & 1-10 & 1.15e-09 & 1.55e-11 & 0 & 1.08 & 774 \\
\textit{Swift} & 8 & 00013945004/wt & 59214.7 & 903 & 1-10 & 1.64e-10 & 7.34e-12 & 0 & 0.81 & 623 \\
\textit{Swift} & 9 & 00013945005/wt & 59217.15 & 1053 & 1-10 & 2.22e-10 & 7.15e-12 & 0 & 0.97 & 649 \\
\textit{Swift} & 10 & 00013945006/wt & 59220.33 & 1098 & 1-10 & 2.10e-10 & 1.14e-11 & 0 & 0.93 & 606 \\
\textit{Swift} & 11 & 00013945007/wt & 59223.33 & 953 & 1-10 & 1.11e-10 & 5.41e-12 & 0 & 0.84 & 549 \\
\textit{Swift} & 12 & 00013945008/wt & 59226.65 & 354 & 1-10 & 6.77e-10 & 6.86e-11 & 0 & 0.63 & 324 \\
\textit{Swift} & 13 & 00013945010/wt & 59232.76 & 1038 & 1-10 & 4.52e-11 & 1.88e-12 & 0 & 0.81 & 431 \\
\textit{Swift} & 14 & 00013945011/wt & 59235.01 & 1099 & 1-10 & 6.45e-11 & 3.27e-12 & 0 & 0.74 & 512 \\
\textit{Swift} & 15 & 00013945012/wt & 59237.06 & 1043 & 1-10 & 4.51e-11 & 3.72e-12 & 0 & 0.74 & 436 \\
\textit{Swift} & 16 & 00013945013/wt & 59241.58 & 1128 & 1-10 & 4.54e-11 & 2.21e-12 & 0 & 0.67 & 489 \\
\textit{Swift} & 17 & 00013945014/wt & 59244.31 & 963 & 1-10 & 2.75e-10 & 8.65e-11 & 0 & 0.63 & 559 \\
\textit{Swift} & 18 & 00013945015/wt & 59247.02 & 668 & 1-10 & 7.28e-11 & 4.50e-12 & 0 & 0.78 & 396 \\
\textit{Swift} & 19 & 00013945016/wt & 59250.61 & 1048 & 1-10 & 2.16e-10 & 2.28e-11 & 0 & 0.86 & 614 \\
\textit{Swift} & 20 & 00013945017/wt & 59253.4 & 858 & 1-10 & 2.65e-10 & 8.46e-12 & 0 & 0.97 & 647 \\
\textit{Swift} & 21 & 00013945018/wt & 59256.26 & 917 & 1-10 & 7.79e-10 & 1.21e-11 & 0 & 1.05 & 751 \\
\textit{Swift} & 22 & 00013945019/wt & 59259.24 & 933 & 1-10 & 1.01e-09 & 1.15e-11 & 0 & 0.84 & 772 \\
\textit{Swift} & 23 & 00013945020/wt & 59263.43 & 790 & 1-10 & 3.58e-10 & 8.73e-12 & 0 & 0.99 & 645 \\
\textit{Swift} & 24 & 00013945021/wt & 59269.2 & 950 & 1-10 & 5.18e-10 & 9.78e-12 & 0 & 1.08 & 699 \\
\textit{Swift} & 25 & 00013945022/wt & 59272.58 & 908 & 1-10 & 5.59e-10 & 9.99e-12 & 0 & 0.92 & 739 \\
\textit{Swift} & 26 & 00013945023/wt & 59275.84 & 883 & 1-10 & 8.46e-10 & 1.18e-11 & 0 & 0.97 & 749 \\
\textit{Swift} & 27 & 00013945024/wt & 59278.95 & 1043 & 1-10 & 5.36e-10 & 7.67e-12 & 0 & 1.00 & 714 \\
\textit{Swift} & 28 & 00013945025/wt & 59281.35 & 15 & 1-10 & 1.00e-08 & 2.93e-09 & 0 & 0.11* & 248 \\
\textit{Swift} & 29 & 00013945026/wt & 59284.93 & 928 & 1-10 & 3.56e-10 & 1.01e-11 & 0 & 0.99 & 699 \\
\textit{Swift} & 30 & 00013945027/pc & 59292.1 & 350 & 1-10 & 1.34e-09 & 6.90e-11 & 0 & 0.80 & 345 \\
\textit{Swift} & 31 & 00013945027/wt & 59292.1 & 608 & 1-10 & 1.09e-09 & 2.12e-11 & 0 & 1.04 & 730 \\
\textit{Swift} & 32 & 00013945028/pc & 59298.27 & 567 & 1-10 & 9.52e-10 & 4.72e-11 & 0 & 0.67 & 380 \\
\hline
\end{tabular}\\
\end{table*}

\bsp	
\label{lastpage}
\end{document}